\documentclass[12pt,a4paper]{article} 
\usepackage{t1enc}
\usepackage{mathrsfs}
\usepackage[utf8]{inputenc}
\usepackage[dvips]{graphicx}
\usepackage{afterpage}
\usepackage{amsmath}
\usepackage{amssymb}
\usepackage{bbm}
\usepackage{multicol}
\usepackage{multirow}
\usepackage[table]{xcolor}
\usepackage{booktabs}
\usepackage{subfig}
\usepackage{hyperref}
\usepackage{floatrow}
\newfloatcommand{capbtabbox}{table}[][\FBwidth]

\newcommand{\css}{C^{\rm VV}_{\rm sca}}
\def\fm{{\rm fm}}

\newcommand{\Oasq}{\mathcal{O}(a^2)}
\newcommand{\mps}{m_\pi}
\newcommand{\fps}{f_\pi}
\newcommand{\mpstm}{m_\pi^{\rm tm}}
\newcommand{\fpstm}{f_\pi^{\rm tm}}
\newcommand{\mpsov}{m_\pi^{\rm ov}}
\newcommand{\fpsov}{f_\pi^{\rm ov}}
\newcommand{\mvv}{M_{\rm VV}}
\newcommand{\mvs}{M_{\rm VS}}
\newcommand{\mss}{M_{\rm SS,\pm}}
\newcommand{\mssz}{M_{\rm SS,0,conn}}
\newcommand{\mov}{m_{\rm ov}}
\newcommand{\MSb}{\overline{\mathrm{MS}}}
\newcommand{\mev}{\mathrm{MeV}}
\newcommand{\gev}{\mathrm{GeV}}

\newcommand{\nft}{N_{\rm f}=2}

\newcommand{\nn}{\nonumber}

\newcommand{\norm}[1]{
	\ensuremath{\left\|#1\right\|}}

\newcommand{\abs}[1]{
	\ensuremath{\left|#1\right|}}

\newcommand{\ID}[1]{
	\mathds{1}}
\newcommand{\e}{\,\mathrm{e}}

\begin{document}
\begin{flushright}
 DESY 12-192\\
 HU-EP-12/32 \\
 SFB/CPP-12-80\\
FTUAM-12-99\\
IFT-UAM/CSIC-12-67

\end{flushright}

\begin{center}
\Large
Overlap valence quarks on a twisted mass sea: a case study 
for mixed action Lattice QCD

\normalsize

\vspace{0.6cm}

Krzysztof Cichy$^{1,2}$, Vincent Drach$^{1}$, Elena Garc\'ia-Ramos$^{1,3}$,\\
Gregorio Herdo\'iza$^{4}$,
Karl Jansen$^{1}$\\
\vspace{0.3cm}
$^{1}$\emph{NIC, DESY, Platanenallee 6, D-15738 Zeuthen, Germany}\\
$^2$ \emph{Adam Mickiewicz University, Faculty of Physics,\\
Umultowska 85, 61-614 Poznan, Poland}\\
$^3$\emph{Humboldt Universit\"at zu Berlin, Newtonstr. 15, 12489 Berlin}\\
$^4$\emph{Departamento de F\'isica Te\'orica and Instituto de F\'isica Te\'orica UAM/CSIC
Universidad Aut\'onoma de Madrid,\\ Cantoblanco E-28049 Madrid, Spain}

\begin{center}
\includegraphics
[width=0.2\textwidth,angle=0]
{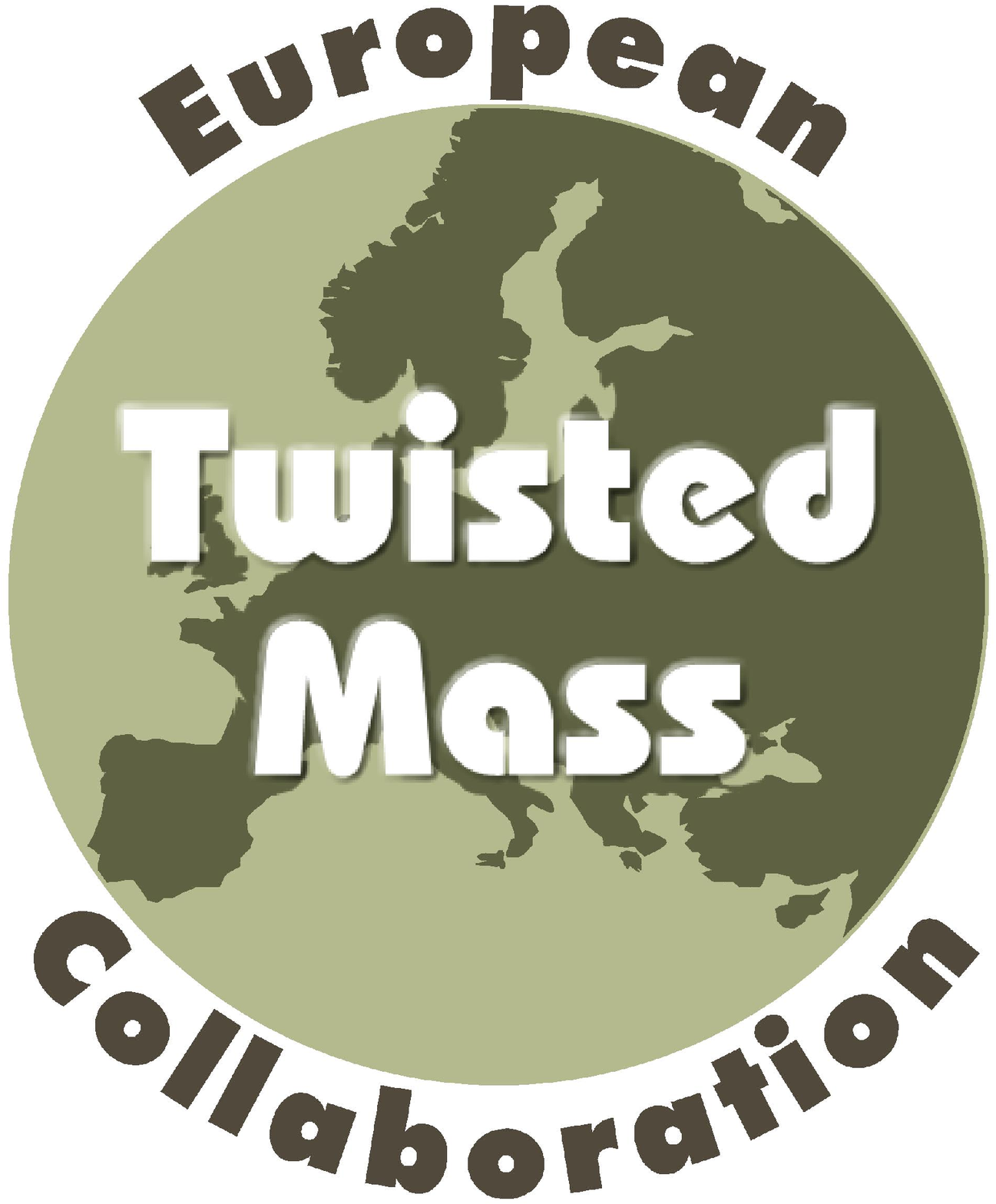}
\end{center}

\begin{abstract}
\noindent We discuss a Lattice QCD mixed action investigation employing
Wilson maximally twisted mass sea and overlap valence fermions.
Using four values of the lattice spacing, we demonstrate
that the overlap Dirac operator assumes a point-like locality
in the continuum limit. We also show that by adopting suitable
matching conditions for the sea and valence theories a
consistent continuum limit for the
pion decay constant and light baryon masses can be obtained.
Finally, we confront results for sea-valence
mixed meson masses and the valence scalar correlator
with corresponding expressions of chiral perturbation theory.
This allows us to extract
low energy constants of mixed action chiral perturbation
which characterize the strength
of unitarity violations in our mixed action setup. 
\end{abstract}


\end{center}

\section{Introduction}

The discovery that despite 
the no-go theorem of Ref.~\cite{RL-81-052},
an exact chiral symmetry can be established for  
lattice regulated field theories
without violating essential field theoretical 
conditions \cite{Luscher:1998pqa}, is certainly one of the major 
conceptual breakthroughs within 
lattice field theory in the last years. The fundamental theoretical development for this 
achievement is the Ginsparg-Wilson 
relation \cite{Ginsparg:1981bj}. The theoretical and 
conceptual advantages of a lattice Dirac operator 
obeying this relation have been pointed out in 
Refs.~\cite{hep-lat/9709110,Hasenfratz:1998ri,hep-lat/9802007,Niedermayer:1998bi}. 
A practical solution -- suitable for numerical simulations --
of an operator satisfying the Ginsparg-Wilson relation, 
termed overlap Neuberger-Dirac operator,   
was found in Refs.~\cite{hep-lat/9707022,hep-lat/9801031}.          

Unfortunately, soon after this exciting development in lattice field theory, it 
turned out that for numerical simulations overlap 
fermions are very expensive. In fact, the numerical cost is 
so demanding that many groups working in lattice QCD 
are nowadays still using lattice discretizations based
on computationally much cheaper 
Wilson or staggered like fermions\footnote{See, however, the work in 
a chiral invariant Higgs-Yukawa model, where extensive use 
of the overlap operator is made \cite{Gerhold:2009ub,Gerhold:2010bh,Gerhold:2010wv}.}. 
Although with such much simpler lattice fermions,
presently, simulations very close to or even at the physical value of the pion mass
can be performed, see e.g. Ref.~\cite{Hoelbling:2011kk}, 
it would be nevertheless highly desirable to take advantage 
of the conceptually much cleaner overlap fermions, or other Ginsparg-Wilson 
fermions that can be applied in practice, to investigate
problems where chiral symmetry or topological aspects play 
a significant role. Given the high numerical demand of overlap 
fermions, such simulations are, however, restricted so far   
to the future, when supercomputers of 
much higher performance than available today, in combination with, hopefully, 
much improved algorithms are developed. 

An intermediate step towards the goal to employ overlap fermions is to 
use them only in the valence quark sector, while 
generating gauge field configurations with much cheaper sea quarks, e.g. 
Wilson or staggered like quarks. 
Such a situation is called {\em mixed action setup}.
The basic idea in such a setup is to relate the sea and valence quark actions 
by a suitable matching condition, e.g. by demanding that at a given value 
of the lattice spacing the pion mass 
is the same in both discretizations. With such a condition, in principle, the 
continuum limit of a mixed action setup is well defined and should give 
correct answers for all physical observables of interest. 

However, at any non-zero value of the lattice spacing care has to be taken when 
using  mixed actions. Since the sea and valence actions are 
different, there are unitarity violations the size of which 
is unknown a priori and  
should be determined. Also, the eigenvalue spectra of the 
sea and valence lattice Dirac operators are not matched, which can lead to problems 
especially for the zero mode contributions of the chiral invariant 
overlap Dirac operator. In addition, 
the locality property of the overlap Dirac operator 
\cite{Hernandez:1998et}
needs to be determined in order to see whether hadronic quantities 
could be affected by finite values of the exponential decay rate. 
All these aspects demand therefore a principal 
investigation of a mixed action before large scale simulations can be started.  

In this paper, we want to report on one such basic 
mixed action analysis\footnote{Mixed actions have been used by various groups.
Here we mention studies using domain wall sea and overlap valence
fermions~\cite{Allton:2006nu,Li:2010pw}, Wilson clover sea
and overlap valence quarks~\cite{Durr:2007ef,Bernardoni:2010nf}, domain wall valence fermions on
a staggered sea~\cite{Renner:2004ck,Beane:2006kx,Aubin:2008wk}. Moreover, another class of
mixed actions involves using variants of the sea action in the valence
sector, such as Osterwalder-Seiler quarks on a twisted mass
sea~\cite{Constantinou:2010qv,Farchioni:2010tb,Herdoiza:2011gp}.}, namely using
maximally twisted mass 
fermions \cite{Frezzotti:2003ni} in the sea sector and overlap fermions in the valence sector. 
In a previous work \cite{Cichy:2010yd,arXiv:1012.4412}, a first study of this mixed action 
setup has been performed and particular difficulties with such an approach 
have been reported, which arise
from the fact that the exact zero modes of the overlap valence Dirac operator 
are not matched exactly by the sea quark twisted mass Dirac operator. This phenomenon  
appears at typical values of the lattice spacing currently used in simulations, i.e. 
$0.04\mathrm{fm}  \lesssim a \lesssim 0.1 \mathrm{fm}$. In such a situation, special 
care must be taken, in particular in the way the sea and valence quark theories
are matched. 

Here we want to extend the analysis of Ref.~\cite{arXiv:1012.4412} in several directions. The first is 
that we add a new and finer value of the lattice spacing than available in 
Ref.~\cite{arXiv:1012.4412} to the analysis
of the continuum limit scaling test of the pion decay constant. 
As we will discuss below, the results of the analysis 
with this new value of the lattice spacing confirm the findings of Ref.~\cite{arXiv:1012.4412} 
and strengthen the conclusion 
that there are critical values for the 
pion mass and the physical volume below which simulations can strongly suffer from the effects of
the mismatch of the  zero modes between sea and valence sectors.
We will also demonstrate that the improved matching
condition, suggested
in Ref.~\cite{arXiv:1012.4412}, indeed works well for the continuum limit 
of the pion decay constant and the nucleon and $\Delta$ masses 
which are new observables added to the study of the properties of the mixed action we are
considering here.

As a second aspect, we want to show that the overlap Dirac operator, which is only 
exponentially localized at non-zero lattice spacing \cite{Hernandez:1998et}, reaches the expected 
point-like locality property in the continuum limit. We will use four values of the 
lattice spacing to perform such a continuum limit extrapolation and, in particular, compare 
the localization range of the overlap Dirac operator 
with hadronic scales, i.e. the pion and the nucleon masses. To our 
knowledge, this is the first unquenched continuum limit study of the locality properties of 
the overlap Dirac operator.      

As a third and also new investigation, we study the meson masses constructed in the mixed action 
theory and confront them with expressions of mixed action chiral perturbation 
theory. This allows us to compute a number of low energy constants 
of the mixed action effective chiral Lagrangian. 
In addition, we will look at the non-singlet scalar correlator to test
for possible unitarity violation which are inherent in mixed action simulations, as mentioned above.

The main goal of the paper is to provide a basic and principal investigation of a mixed action 
employing overlap fermions in the valence sector. The results of this investigation 
can and will help any further calculation aiming at computing physical 
quantities in that it provides important information on the properties
of mixed actions which can help to perform safe simulations in the future.  

The outline of the paper is as follows. 
Sec.~2 discusses some theoretical aspects of our setup and gives parameters of our lattices.
In Sec.~3, we look at the continuum limit behaviour of the locality of the
overlap Dirac operator. Sec.~4 reports the results of our new and finest lattice spacing by
including them in a continuum limit scaling test 
of the pion decay constant. In Sec.~5, we discuss mixed meson masses and unitarity violations in
the non-singlet scalar correlator.
In Sec.~6, we employ our mixed action setup in the baryon sector and perform a continuum limit
scaling test for nucleon and $\Delta$ masses.
Sec.~7 concludes and summarizes our main findings.

\section{Simulation setup}
\label{section-setup}
\subsection{Overlap Fermions}
The fundamental Nielsen-Ninomiya theorem \cite{RL-81-052,Friedan:1982nk}, 
formulated in 1981,
apparently excluded the possibility of simulating chiral fermions on the
lattice, without violating essential properties, such as the absence of
doubler modes or the locality of the theory. 
However, already in 1982 it was shown by Ginsparg and Wilson
\cite{Ginsparg:1981bj} that a remnant of chiral symmetry is present on the
lattice without the doublers, if the corresponding lattice Dirac operator $\hat D$
obeys an equation now called the Ginsparg-Wilson relation:
\begin{equation}
\label{gw}
\gamma_5 \hat D + \hat D\gamma_5 = a\hat D\gamma_5\hat D.
\end{equation} 
For many years, though, no practical solution to this equation had been known.
The situation changed in the second part of the 1990s when the Ginsparg-Wilson 
relation was reconsidered and its conceptional advantages pointed out
\cite{hep-lat/9709110,hep-lat/9802007,hep-lat/9707022,hep-lat/9801031}. 
Moreover, in Ref.~\cite{Luscher:1998pqa} the key observation was made that any lattice 
Dirac operator satisfying the Ginsparg-Wilson relation 
leads to an exact chiral symmetry
at non-vanishing value of the lattice spacing.
Neuberger found a particular closed 
form of a lattice Dirac operator $D$ -- called overlap Dirac operator --
that obeys the Ginsparg-Wilson
relation and that can be employed in practical simulations. 
The massless overlap Dirac operator is given
by\,\footnote{For an early review of the overlap fermions formalism, see
  e.g. Ref.~\cite{Niedermayer:1998bi}.} \cite{hep-lat/9707022,hep-lat/9801031}:
\begin{equation}
  \label{massless_overlap}
  \hat D_{\rm ov}(0)=\frac{1}{a}\Big(1-A(A^\dagger A)^{-1/2}\Big).
\end{equation}
In the kernel operator $A$ we choose the standard Wilson-Dirac operator:
\begin{equation}
  \label{overlap-A}
  A=1+s-a\hat D_{\rm Wilson}(0),
\end{equation}
where $s$ is a parameter which satisfies $|s|<1$ and can be tuned to optimize
locality properties \cite{Hernandez:1998et} (see Sec.~\ref{section-locality} for
a detailed test of
locality in our setup). The Wilson-Dirac operator is defined by:
\begin{equation}
  \label{eq:DW}
  \hat D_{\rm
Wilson}(m_0)=\frac{1}{2}
\left(\gamma_\mu(\nabla_\mu^*+\nabla_\mu)-a\nabla_\mu^*\nabla_\mu\right)+m_0,
\end{equation}
where $m_0$ is the bare Wilson quark mass and $\nabla_\mu$ ($\nabla^*_\mu$) are
the forward (backward) covariant derivatives.
The massive overlap Dirac operator with the bare overlap
quark mass $\mov$ reads:
\begin{equation}
  \label{overlap-massive}
  \hat D_{\rm ov}(\mov) = \left(1-\frac{a\mov}{2}\right)\hat
D_{\rm ov}(0)+\mov.
\end{equation} 

The most important property of overlap fermions
is the fact that chiral
symmetry in a lattice modified form can be established 
with very important consequences, among them the absence 
of $\mathcal{O}(a)$ lattice
artefacts.

From the technical side, an important feature of overlap fermions is that the
explicit construction of the overlap Dirac operator involves an
approximation (in our case the Chebyshev polynomial approximation) 
of the operator $(A^\dagger A)^{-1/2}$ that needs
to be realized up to machine precision, in order to consider the lattice chiral
symmetry to be exact. This leads to a large computational cost of overlap
fermions, around two orders of magnitude larger than of e.g. Wilson twisted mass
fermions. For a comparison of overlap and Wilson twisted mass fermions
in terms of computing cost, we refer
to~\cite{Chiarappa:2006hz}.

Further problems occur in simulations with {\em dynamical} overlap fermions.
The overlap Dirac operator develops discontinuities when changing
topological sectors and the proposed solutions to alleviate this problem
~\cite{Fodor:2003bh,Cundy:2005mr,Schaefer:2006bk,Cundy:2008zc} lead to a further
growth of the computational cost.
The discontinuity problem can also be avoided by fixing
topology ~\cite{Fukaya:2006vs,Aoki:2008tq}, which, however, still needs
very computer time expensive simulations and introduces
additional $\mathcal{O}(1/V)$ finite volume effects.
The mixed action approach,  which uses computationally 
cheap sea quark simulations on which then valence overlap
quarks are evaluated, is another way of
avoiding the large computational costs of dynamical overlap simulations. 
This is the setup considered in this paper. 
Of course, in this approach exact chiral symmetry is only 
realized in the valence sector. 

Since the pseudoscalar decay constant will play an important role in the 
following, we discuss this observable here. 
For overlap fermions, the pseudoscalar decay
constant $\fpsov$ can be extracted using 
the PCAC relation and does not require the computation of
any renormalization constants (see e.g. Ref.~\cite{Bietenholz:2004wv}):
\begin{equation}
 \label{fpsov}
 \fpsov = \frac{2\mov}{ ( \mpsov )^2} \, |\langle 0| P |\pi \rangle_{\rm ov} |
\,,
\end{equation}
where $\mpsov$ is the mass of the charged pseudoscalar meson and $|\langle
0| P |\pi \rangle_{\rm ov}|$ the matrix element of the pseudoscalar current,
both extracted from the two-point pseudoscalar correlation function $C_{\rm
PP}(t)$, built of two mass-degenerate overlap valence quarks.

\subsection{Wilson Twisted Mass Fermions}
Wilson Twisted Mass (tm) fermions~\cite{Frezzotti:2000nk} were
originally introduced to deal with the problem of unphysically small eigenvalues (zero
modes) of the Wilson-Dirac operator.
As pointed out in Ref.~\cite{Frezzotti:2003ni} 
an essential advantage of this formulation of lattice QCD is also the
possibility to obtain automatic $\mathcal{O}(a)$-improvement, by tuning just
one parameter -- the bare Wilson quark mass to its 
critical value. This
property has been confirmed in detailed continuum limit scaling studies in the
quenched approximation~\cite{Jansen:2005kk} and with two dynamical
quarks~\cite{Baron:2009wt}.
What is more, the twisted mass discretization can reduce the effects of
explicit chiral symmetry breaking by the Wilson term in the 
renormalization process. In fact, the 
problem of operators belonging to different chiral representations can be 
suppressed or even avoided in this formulation.
Among the disadvantages of this formulation are the explicit breaking of parity
and isospin symmetry, being, however, $\Oasq$ cut-off
effects 
\cite{Frezzotti:2003ni,Jansen:2005cg,Frezzotti:2007qv,Dimopoulos:2009qv,Baron:2009wt}, which were
observed
to be substantial only in the neutral pion mass. 

Twisted mass fermions are obtained by adding a chirally rotated mass
term to the Wilson-Dirac operator in Eq.~(\ref{eq:DW}) in the following way:
\begin{equation}
  \label{eq:Dtm}
  \hat D_{\rm tm} = \hat D_{\rm Wilson}(m_0)+i\mu\gamma_5\tau_3\,,
\end{equation} 
where $\mu$ is the twisted mass parameter and $\tau_3$ is the third
Pauli matrix acting in flavour space.

A comprehensive simulation programme has been undertaken by the
European Twisted Mass (ETM) collaboration, including simulations with dynamical
up and down quarks ($N_{\rm f}=2$)
\cite{Boucaud:2007uk,Boucaud:2008xu,Baron:2009wt}
and with dynamical up, down, strange and charm quarks 
($N_{\rm f}=2+1+1$)~\cite{Baron:2010bv,Baron:2010th}.

The charged pseudoscalar meson decay constant can be obtained in a similar way
as in the case of overlap fermions, without the need to compute any
renormalization factors:
\begin{equation}
  \label{fpstm}
  \fpstm = \frac{2\mu}{(\mpstm)^2} \,  |\langle 0| P|\pi^\pm \rangle_{\rm tm}
|\,, 
\end{equation}
where $\mpstm$ is the pseudoscalar meson mass and $|\langle 0| P|\pi^\pm
\rangle_{\rm tm}|$ the matrix element of the pseudoscalar current,
both extracted from the two-point pseudoscalar correlation function $C_{\rm
PP}(t)$, built of two mass-degenerate tm valence quarks.

\subsection{Mixed action setup}
Our mixed action setup consists of two mass-degenerate
flavours of Wilson twisted mass quarks at maximal twist in the sea sector and
overlap valence fermions. The parameters of our dynamical
ensembles are provided in Tab.~\ref{tab:setup}. The ensembles were generated by
the ETM Collaboration~\cite{Baron:2009wt}, using the tree-level Symanzik
improved gauge action and $\nft$ flavours of twisted mass
fermions, tuned to maximal twist by setting the hopping
parameter $\kappa = 1/(8+ 2am_0)$ to its critical value, at which the PCAC
quark mass vanishes \cite{Boucaud:2007uk}.

The light-quark ensembles, labeled in Tab.~\ref{tab:setup} with a
subscript $\ell$, correspond to a fixed physical situation with roughly fixed
pseudoscalar meson mass $\mps r_0 \approx 0.8$ (where $r_0$ is the Sommer
parameter~\cite{Sommer:1993ce}) and
lattice size $L
\approx 1.3\,\fm$. The non-perturbatively renormalized
~\cite{Constantinou:2010gr} quark mass $\mu_{q}^{\MSb}(\mu=2\,\gev)\approx
20\,\mev$, which in infinite volume gives a pseudoscalar meson mass of
around $300\,\mev$. The chirally extrapolated values
    of the Sommer parameter for our ensembles are $r_0/a=5.25(2)$ at
$\beta=3.90$,
    $r_0/a=6.61(2)$ at $\beta=4.05$ and $r_0/a=8.33(5)$ at
    $\beta=4.20$, which corresponds to lattice spacings $a\approx0.079\,
    \mathrm{fm}$, $a\approx0.063\, \mathrm{fm}$ and $a\approx0.051\,
    \mathrm{fm}$, respectively~\cite{Baron:2009wt}. At $\beta=4.35$, we only
have one sea quark mass available, which yields $r_0/a=9.82(4)$ and a lattice
spacing $a\approx0.042\,\fm$.

For our investigations of locality (Sec.~\ref{section-locality}), we use ensembles with $L/a=32$
at: $\beta=3.9$, $a\mu=0.004$ (ensemble $B_{\ell,32}$), $\beta=4.05$, $a\mu=0.003$ ($C_{\ell,32}$),
$\beta=4.2$, $a\mu=0.0065$ (with a heavier quark mass, ensemble $D_{h,32}$) and $\beta=4.35$,
$a\mu=0.00175$ ($E_\ell$). In addition, we use two additional volumes at $\beta=3.9$, $a\mu=0.004$
(with linear lattice extent of 0.6 fm ($B_{\ell,8}$) and 1.6 fm ($B_{\ell,20}$)), to check finite
size effects in the context of locality.

We have also used one small-volume ensemble with a heavier quark mass, labeled
as $B_h$. It corresponds to the same lattice size and lattice spacing as
$B_\ell$, but the renormalized quark mass is $\mu_{q}^{\MSb}(\mu=2\,\gev)\approx
40\,\mev$, which corresponds to $\mps r_0 \approx 1.0$ and a
pseudoscalar meson mass $\mps \approx 450\,\mev$ in infinite volume.

Finally, to explicitly check whether one is safe against the effects of 
zero modes of the overlap operator, discussed in Ref.~\cite{arXiv:1012.4412},
we have also considered an ensemble $B_s$ that corresponds to a physical extent
of the lattice of $L\approx2\,\fm$ and a rather heavy renormalized quark mass
of $\mu_{q}^{\MSb}(\mu=2\,\gev)\approx
45\,\mev$, which yields $\mps r_0 \approx 1.0$ and a
pseudoscalar meson mass of $\mps \approx 480\,\mev$ in infinite volume. For
this ensemble, $\mps L \approx 4.7$ and hence the effects of zero modes should be very
much suppressed \cite{arXiv:1012.4412}.

%%%%%%%%%%%%%%%%%%%%%%%%%%%%%%%%%%%%%%%%%%%%%%%%%%%%%%%%%%%%%%%%%%%%%%%%%%%%%% 
\begin{table}[t!]
  \centering
  \begin{tabular*}{1.0\textwidth}{@{\extracolsep{\fill}}lccccccc}
    \hline\hline
    Label & $\beta$ & $(L/a)^3\times T/a$  & $a\mu$ &
$\kappa_\mathrm{crit}$ &
    $\mpstm r_0$ & $L/r_0$ & $\mpstm L$ \\
    \hline\hline
    $B_{\ell}$  & $3.90$ & $16^3\times 32$ & $0.0040$ &  $0.160856$  & $0.84$  &
$3.0$ & $2.5$  \\
    $C_{\ell}$          & $4.05$ & $20^3\times 40$ & $0.0030$ & $0.157010$ &
$0.83$ & $3.0$ & $2.4$  \\
    $D_{\ell}$          & $4.20$ & $24^3\times 48$ & $0.0020$ & $0.154073$ &
$0.82$ &     $2.9$  & $2.4$ \\
    $E_{\ell}$          & $4.35$ & $32^3\times 64$ & $0.00175$ & $0.151740$ &
$0.74$ & $3.3$ & $2.4$ \\
\hline
$B_h$      &  $3.90$ & $16^3\times 32$ & $0.0074$ & $0.160856$ & $1.03$  &
$3.0$ & $3.1$ \\ 
\hline
$B_s$         &  $3.90$ & $24^3\times 48$ & $0.0085$ & $0.160856$ & $1.02$  &
$4.6$ & $4.7$ \\ 
\hline
$B_{\ell,8}$      &  $3.90$ & $8^3\times 16$ & $0.0040$ & $0.160856$ & --  &
$1.5$ & -- \\ 
$B_{\ell,20}$      &  $3.90$ & $20^3\times 40$ & $0.0040$ & $0.160856$ & $0.73$  &
$3.8$ & $2.8$ \\ 
$B_{\ell,32}$      &  $3.90$ & $32^3\times 64$ & $0.0040$ & $0.160856$ & $0.70$  &
$6.1$ & $4.3$ \\ 
    $C_{\ell,32}$          & $4.05$ & $32^3\times 64$ & $0.0030$ & $0.157010$ &
$0.69$ & $4.8$ & $3.3$  \\
    $D_{h,32}$          & $4.20$ & $32^3\times 64$ & $0.0065$ & $0.154073$ &
$1.10$ &     $3.8$  & $4.2$ \\
    \hline\hline\\
  \end{tabular*}
  \caption{The parameters of dynamical $\nft$ maximally twisted mass
ensembles used in this work. We give the label, the values of the
    inverse coupling $\beta$, the lattice volume $L^3\times T$, the
    twisted mass parameter $a\mu$, the critical hopping parameter
    $\kappa_\mathrm{crit} = 1/(8+ 2am_{\rm crit})$, the approximate
    values of the pseudoscalar meson mass and of the lattice size in
    units of the Sommer scale $r_0$ and the product $\mpstm L$. 
    As demonstrated in Ref.~\cite{arXiv:1012.4412}, the values of $\mpstm L$ 
    provide a very good estimator for the 
    the role of the zero modes of the overlap operator: they can be large (if $\mpstm L<3$),
potentially large in some observables (if $3<\mpstm L<4$) or almost negligibly small
(if $\mpstm L>4$).}
  \label{tab:setup}
\end{table}
%%%%%%%%%%%%%%%%%%%%%%%%%%%%%%%%%%%%%%%%%%%%%%%%%%%%%%%%%%%%%%%%%%%%%%%%%%%%%% 

Our setup for the overlap fermion valence sector has been described in detail
in Ref.~\cite{Cichy:2009dy}. 
Here we only briefly mention that we performed one iteration of HYP smearing
\cite{hep-lat/0103029} of gauge links
to reduce the condition number of $A^\dagger A$.
To assure the best locality properties, we
set $s=0$ in Eq.~\eqref{overlap-A} (for details, see the next section). 

We have produced all-to-all overlap and twisted mass propagators for a wide
range of quark masses, which allowed us to compute three types of
correlation functions: sea-sea (unitary tm), valence-valence (overlap)
and valence-sea (mixed overlap-tm).

\section{Locality of the overlap operator}
\label{section-locality}

The aim of this section is a comprehensive investigation of the 
locality of the overlap lattice Dirac operator. 
We will use ensembles at four values of the 
lattice spacing, which will allow us 
to perform a continuum limit study of the effective decay rate of the 
overlap operator to see whether indeed point-like locality is 
recovered in the continuum limit. 
We will also investigate the influence of the parameter $s$ in 
Eq.~(\ref{overlap-A}), as well as finite size effects, for which we use four different physical volumes. 
Finally, we will perform a 
comparison of the decay rate with hadronic scales
of the theory, in particular the pseudoscalar 
and the nucleon masses, in order to 
achieve a quantitative measure for the decay rate relative 
to physical scales.
Such a comparison is clearly of importance: as long as 
the decay rate is smaller than, say, a hadron mass,
computations with the overlap operator are not very 
useful. If, as expected, the overlap operator converges 
in the continuum limit to the Dirac operator with point-like
locality, there must exist a value of the lattice spacing, 
where the decay rate is much larger than the hadronic scale. 
It is therefore one goal of this paper to determine the 
ratio of the hadronic mass and the decay rate and to determine 
this ratio 
at the values of the lattice spacings used here. 

In this section, we will closely follow the investigation 
of the locality property of the overlap operator of Ref.~\cite{Hernandez:1998et}. 
While for small enough values of the lattice spacing, the locality 
of the overlap operator can be analytically proven, 
for coarser values the locality property needs to be determined
numerically. 
In fact, in Ref.~\cite{Hernandez:1998et} it has been demonstrated
that the locality property can be maintained to rather coarse 
values of the lattice spacing of about $a\approx 0.1$ fm, 
which has been pushed to even coarser values of $a\approx 0.2$ fm
in \cite{Draper:2006wb}.
Since the values of the lattice spacings employed in this work 
are all below $0.1$ fm, we expect that the valence 
overlap operator is exponentially localized for all ensembles
considered, an expectation which will be confirmed below. 

\subsection{Effective decay rate}

It is clear that the overlap Dirac operator (denoted $D$ in this section) is not local in a 
geometrical 
sense, which would mean that there exists a finite $\xi$ such that the following requirement is
fulfilled:
\begin{eqnarray}
  \label{eq:1}
  &&D\Psi(n)=\sum_mD(n,m)\Psi(m),\\
 && D(n,m)=0 \quad \textrm{for all points } n,\,m \textrm{ such that} \quad \norm{n-m}> \xi, \nonumber
\end{eqnarray}
for some norm $\norm{\cdot}$ to be defined below. However, from the point of view of continuum
limit behaviour of physical quantities, such geometrical locality is not necessary. 
It is sufficient if the
 Dirac operator $D(n,m)$ decays exponentially fast, i.e.:
\begin{equation}
  \label{eq:2}
 \norm{D(n,m)}\leq C\e^{-\hat\rho\norm{n-m}}\, 
\end{equation}
with some decay rate $\hat\rho$ (in lattice units, with corresponding 
physical decay rate $\rho=\hat\rho/a$) and normalization factor $C$. 

For the norm in Eq.~\eqref{eq:1}, we use the taxi-driver distance
for periodic
boundary conditions:
\begin{equation}
  \label{eq:4}
  \norm{n}_1=\sum_{\mu}\min\{\abs{n_{\mu}},\abs{N_\mu-n_{\mu}}\}, \quad 0\leq n_{\mu} < N_\mu,
\end{equation}
where $N_\mu$ is the number of lattice sites in direction $\mu$.
We also define the norm of the operator as the row sum norm:
\begin{equation}
  \label{eq:5}
    \norm{D(n,m)}\equiv \max_{\substack{1\leq\mu\leq
    4\\1\leq a \leq 3}} \sum^4_{\nu=1}\sum^3_{b=1}\abs{D(n,m)_{a,b}^{\mu,\nu}}\,   
\end{equation}
where $\mu,\nu$ denote the Lorentz indices and $a,b$ the colour indices.
For the same value of the taxi-driver distance, the operator norm can take several values. Hence,
we also define the maximum norm:
\begin{equation}
  \norm{D}_{\textrm{max}}(d)= \displaystyle\max_{\norm{n-m}_{1}=d}\norm{D(n-m,0)}.
\end{equation}
From now on, we will only consider the maximum norm, since it corresponds to the
most restrictive case, i.e. proving locality using the maximum norm implies locality for any other
choice of the norm. 

\subsection{Continuum limit analysis}

In this subsection, we will perform a continuum limit scaling test 
of the decay rate of the overlap operator using ensembles $B_{\ell,32}, C_{\ell,32}, D_{h,32},
E_{\ell}$  of Tab.~\ref{tab:setup}. 
These ensembles correspond to four lattice spacings,
at different physical volumes, but keeping constant the lattice size $L/a=32$.

The Dirac operator is local if it fulfills
Eq.~\eqref{eq:2}. We define the effective decay rate $\hat\rho_{\textrm{eff}}$ as
follows: 
\begin{equation}
\hat \rho_{\textrm{eff}} = \ln\left(\frac{\norm{D}_{\rm
        max}(n_{i})}{\norm{D}_{\rm max}(n_{i}+1)}\right).
\end{equation}
Larger values of the decay rate imply better locality properties.

The full decay property of the overlap operator, including small 
distances, is determined by a sum of exponential decays 
with corresponding decay rates: 
\begin{equation}
\norm{D}_{\rm max}=\sum_iC_i \e^{-\rho_i\norm{x_i}}
\xrightarrow{\norm{x}\rightarrow\infty}
\norm{D}_{\rm max}=C\e^{-\rho\norm{x}}\,.
\end{equation}
Therefore, the desired decay rate is only attained at asymptotically
large distances, where the effective decay rate $\hat \rho_{\textrm{eff}}$ 
will be constant as a function of the taxi-driver distance. 

\begin{figure}[t!]
\begin{center}
\includegraphics[width=0.495\textwidth]
{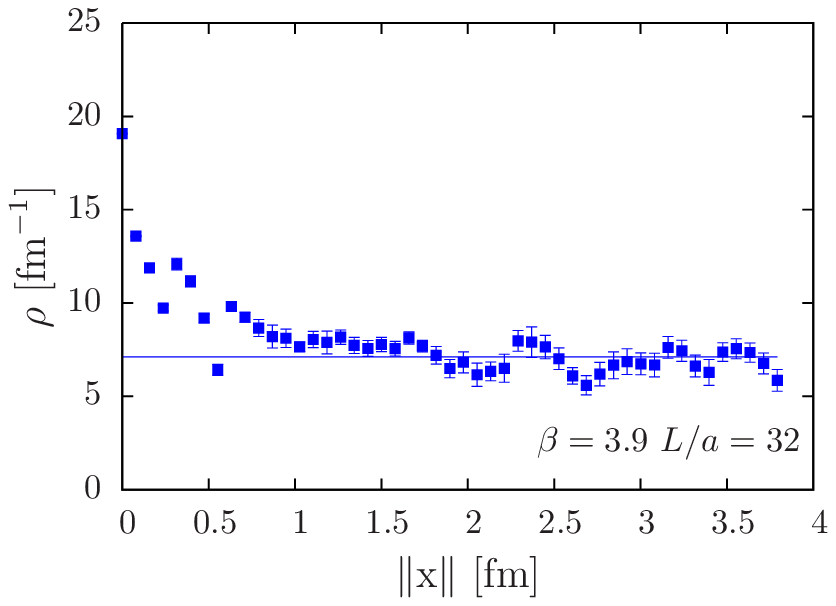}
\includegraphics[width=0.495\textwidth]
{ind_b405_L32}
\includegraphics[width=0.495\textwidth]
{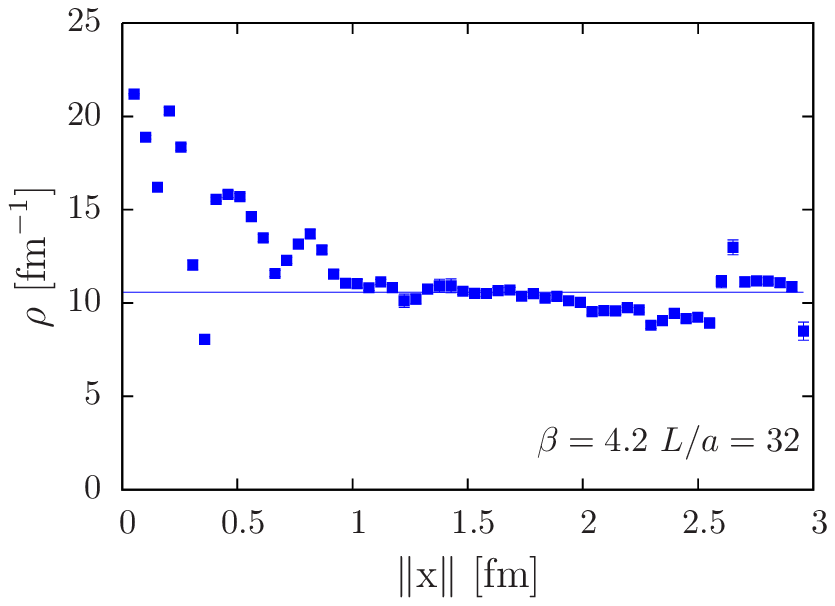}
\includegraphics[width=0.495\textwidth]
{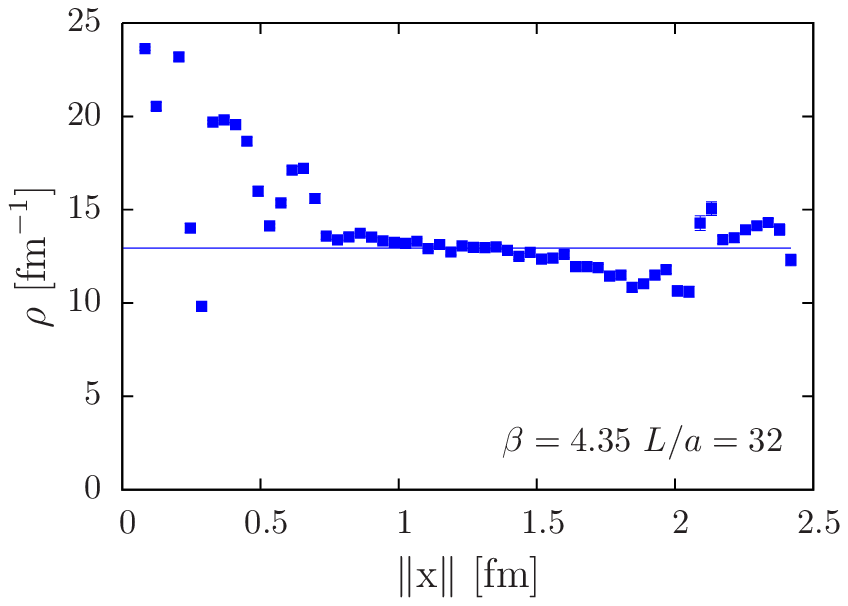}
\end{center}
\caption{\label{individual}Effective decay rate $\rho$ [fm$^{-1}$] as a function of the taxi driver
    distance [fm] for four different values of the lattice spacing.
 The blue line is the final fit obtained following
  the method explained in the text. The fitting intervals [fm] are
  indicated in Tab.~\ref{cont_lim_table}. The error bars are often
smaller than the symbol size; therefore the large fluctuations induce
significant systematic uncertainties in the determination of the
effective decay rate.} 
\label{rhos}
\end{figure}

In Fig.$\;$\ref{rhos}, we show the effective decay rate as a function
of the taxi-driver distance for four different values of the lattice
spacing.  Our typical statistics was around 50 configurations well separated in
Monte Carlo time.
In order to observe the expected plateau-like behaviour 
of $\hat \rho_{\textrm{eff}}$, it is necessary to have a large value for
$L/a$. If $L/a$ is too small, the signal is dominated
by hypercubic artefacts and the extracted values of $\rho$ are not reliable. 
We have studied this effect in free-field theory, for different values of $L/a$ and we have
concluded that $L/a=32$ is necessary to observe a stable plateau of the effective decay rate (see
Sec.~\ref{sec:locality_fve} for details).

Moreover, from Fig.$\;$\ref{rhos} it is clear that the determination of the 
effective decay rate is not straightforward, i.e. the observed large 
fluctuations have to be taken into account in the analysis as a systematic 
error. 
To this end, we follow the strategy of 
Refs.~\cite{Baron:2009wt,Durr:2010aw}.
As a first step, we fit the effective decay rates obtained at 
different values of the taxi-driver distance to a constant, 
including at least 4 consecutive points.
For this, we fit the
value of the decay rate and the corresponding statistical error using a jackknife
procedure. For each fit, we also compute the corresponding value 
of $\chi^2$ and the confidence level CL, defined as:
\begin{equation}
  \label{eq:6}
  \textrm{CL}(q=\chi^2,n=\textrm{d.o.f.})=1-\gamma(q/2,n/2),
\end{equation}
where d.o.f. denotes the number of degrees of freedom for a given fit and $\gamma(q/2,n/2)$ is the
lower incomplete Gamma function. 

\begin{figure}[t!]
\begin{floatrow}
\ffigbox{%
\includegraphics[width=0.485\textwidth]{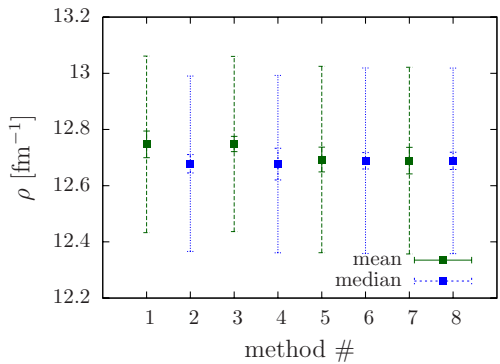}
}{%
  \caption{\label{methods_fig} Comparison of results for the ensemble
    $E_\ell$ obtained through different methods of extracting the
effective decay rate. The method numbering is explained in Tab.~\ref{methods_table}. The errors are
statistical (smaller error) and systematic (larger error).}
}
\capbtabbox{%
%\begin{footnotesize}
\begin{tabular}{c c c}
    \cline{1-3}
      $ \#$  &  $\rho$ and $\delta\rho_{\mbox{{\tiny stat}}}$ &  weight \\
      \cline{1-3} \cline{1-3}
      1 &  mean  & CL \\
      \cline{1-3}
      2 &  median & CL \\
      \cline{1-3}
      3 &  fit(mean)  & CL  \\
      \cline{1-3}
      4 &   \cellcolor[gray]{0.8} fit (median)  & \cellcolor[gray]{0.8} CL  \\
      \cline{1-3}
      5 &  mean & $\mbox{exp}(\chi^2/\mbox{{\tiny dof}}))$  \\
      \cline{1-3}
      6 &  median & $\mbox{exp}(\chi^2/\mbox{{\tiny dof}}))$  \\
     \cline{1-3}
     7 &  fit(mean) &  $\mbox{exp}(\chi^2/\mbox{{\tiny dof}}))$  \\
     \cline{1-3}
     8 &   fit (median)  & $\mbox{exp}(\chi^2/\mbox{{\tiny dof}}))$  \\
     \cline{1-3}
\end{tabular}
%\end{footnotesize}
}
{\caption{\label{methods_table} Different methods used to calculate the value of the
      effective decay rate $\rho$ and the corresponding error.} 
}

\end{floatrow}
\end{figure}

In this way, we obtain a large sample of fitted values of 
the effective decay rate from which we can construct weighted 
distributions
using two kinds of weights, the confidence level (CL) or a function of
$\chi^2$ which we choose as $\exp(\chi^2)/\textrm{d.o.f.}$ -- thus damping the
influence of bad fits. 
From this weighted distribution, we extract  
the mean and the median of the distribution. 
In practice, for our final values of the effective decay rates (shaded
entry in Tab.~\ref{methods_table}), 
we look for
the ``best fits'', i.e. the ones closest to the mean and median of the weighted distribution and
satisfying the constraint that their confidence level is at least 80\%.
The statistical error is then the jackknife error of the best fit, while the systematic error
is given by the $68.3\%$ confidence interval of the weighted distribution. The final error is
computed by adding both errors in quadrature.
We also analyzed the effect of adding a cut in the x axis in order to avoid possible
finite volume effects at large values of the taxi driver distance
(cut at $\|x\|=L$ or $\frac{\sqrt{7}L}{2}$, i.e. the maximum distance in a $L^3\times(2L)$ box in
the continuum, with periodic boundary conditions), and we found results that are fully compatible with
the ones without the cut, however, with a decrease of the systematic error.

Combining all these possibilities to extract the effective decay rate
by using the mean and median from the weighted distribution 
(labeled \emph{mean} and \emph{median} in Tab.~\ref{methods_table}) using 
either the confidence level or $\chi^2/{\rm d.o.f.}$, 
mean and median from ''best fit`` labeled \emph{fit(mean)} and \emph{fit(median)} in
Tab.~\ref{methods_table}), we 
thus have in total 8 different ways of computing the effective decay 
rate and its statistical \footnote{The statistical error quoted for
  the values of the mean and the median extracted from the
  distributions were obtained by analyzing the weighted distribution of all
the statistical errors of the different fits computed with jackknife.} and systematic error, see
Tab.~\ref{methods_table}.

In Fig.~\ref{methods_table}, we show the comparison of the final
results for the effective decay rate obtained using these 
8 different methods, for the case of the ensemble
$E_\ell$.
All methods lead to compatible results both in the central 
value, as well as in the total error, which gives us confidence that 
we indeed have the systematic errors of the effective decay rate 
under control.

For the continuum limit scaling test of the effective decay rate, we decided to use
method 4 of Tab.~\ref{methods_table}, 
taking the value for $\rho$ 
that yields a fitted $\rho$ as close as possible to the median of
the CL-weighted histogram.
However, we want to emphasize that all other variants 
would be equally good choices.

\begin{figure}[t!]
\begin{center}
\includegraphics[width=0.7\textwidth]{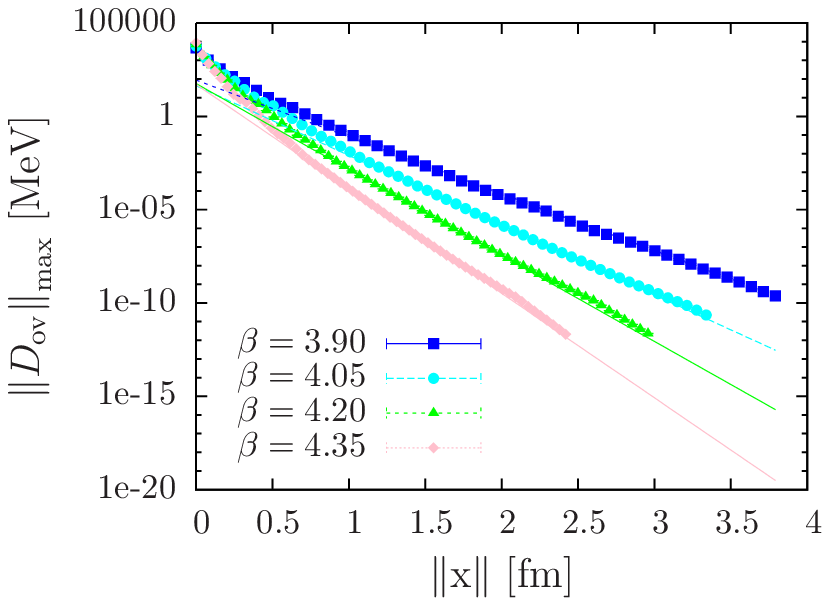}
\caption{Exponential decay of the maximum norm of the overlap Dirac operator for ensembles
$B_{\ell,32}$,
$C_{\ell,32}$, $D_{h,32}$, $E_\ell$.}
\label{norm}
\end{center}
\end{figure}

Before discussing the numerical results for the effective decay rate,  
we show in 
Fig. \ref{norm} the exponential decay of the maximum norm of the overlap
Dirac operator as a function of the taxi-driver distance. We use 
four different lattice spacings,
ranging from 0.079 to 0.042 fm, while keeping the 
value of $L/a=32$ fixed, as discussed above. 
We observe that decreasing the lattice spacing leads to an increasing slope of 
$\norm{D_{\rm ov}}_{\rm max}$.
This signals the continuum limit restoration of point-like locality.

\begin{table}[t!]{
    \begin{center}
      $\begin{array}{c c c c c c c c }
        \hline
       & \rho & \delta_{\text{stat}}\rho & \delta_{\text{syst}}\rho &  \delta\rho & 1/\rho & $interval$  &\\
        & [$fm$^{-1}] &[$fm$^{-1}] &[$fm$^{-1}] &[$fm$^{-1}]     &[$fm$] & [$fm$] \\
        \hline
        \hline
        B_{\ell,32} & 7.117 & 0.183 &0.557& 0.586& 0.141(12) &[2.923,3.634] \\
        C_{\ell,32} & 8.684 & 0.065& 0.333 &0.339 &0.115(4)& [1.953,2.205]\\
        D_{h,32} & 10.555 & 0.037& 0.141 &0.146 &0.094(1) &[1.326,1.785]\\
        E_\ell & 12.677 & 0.056 & 0.311 &0.316 & 0.079(2)& [1.176,1.302]\\
        \hline
      \end{array}$
      \caption{\label{cont_lim_table} Results for the decay rate $\rho$ and the corresponding
        errors: statistical $\delta_{\text{stat}}\rho$, systematic $\delta_{\text{syst}}\rho$, total
        $\delta\rho$. We also give the inverse decay rate (with its combined statistical and systematic
        error in parentheses) and the fit interval for the best fit.}
    \end{center}    }
\end{table}  

In Tab.~\ref{cont_lim_table}, we summarize the results obtained for the values of the decay
rate $\rho$, together with their statistical, systematic and total
errors. As already suspected from Fig.~\ref{individual} the total error is
dominated by the systematic part.

\begin{figure}[t!]
\begin{center}
\includegraphics[width=0.7\textwidth]{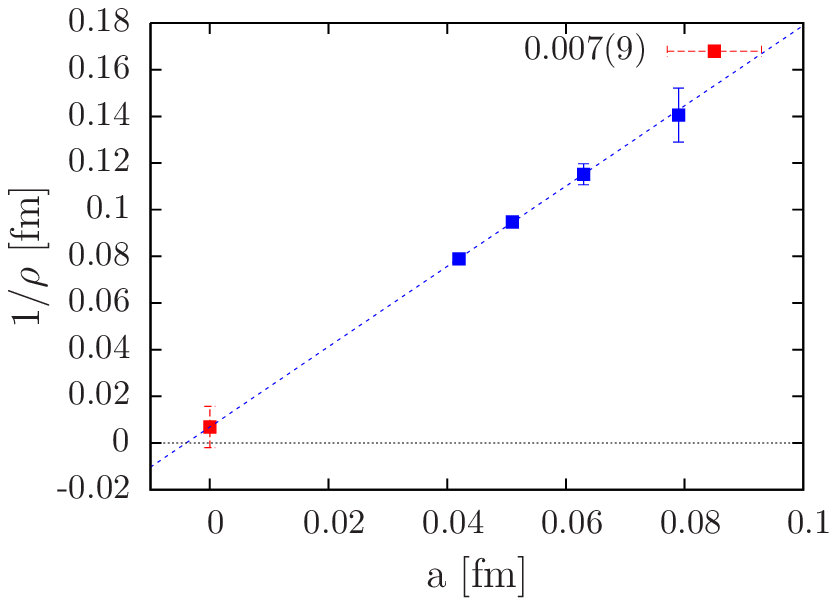}
\caption{Continuum limit scaling of the inverse decay rate $1/\rho$.}
\label{cont_lim}
\end{center}
\end{figure}

Using data from Tab.~\ref{cont_lim_table}, we perform the continuum limit scaling test of the
inverse decay rate (locality radius) $1/\rho$. The values at non-zero lattice spacings can be
interpreted as the physical length that correspond to the slowest decrease of the Dirac operator norm.
The value of the locality radius extrapolated to
$a=0$ is 0.007(9) fm, i.e. compatible with zero. Thus, in the
continuum limit, point-like locality is indeed restored. If we consider the lattice
artefacts and therefore perform a fit adding a term quadratic in $a$, instead of a
simple linear fit, we obtain a compatible result: -0.002(37) fm, but with a much larger
error, indicating that we are not sensitive to $\mathcal{O}(a^2)$ cut-off effects in the effective
decay rates.

\subsection{Comparison of decay rates and hadron masses} 

The relevance of the decay rate $\rho$ of the overlap operator
can be quantified when compared to the size of a hadron. The
locality radius, $r_{\textrm{loc}}=1/\rho$, is expected to be
sufficiently smaller than the hadron size in order to extract the
hadron mass without being affected by non-locality problems. To
illustrate this point, we consider the approximation in which the
size of the hadron is related to the inverse of its mass
$m$. There is a similarity of the asymptotic exponential decay of zero-momentum correlation functions
(with Euclidean time) and the likewise asymptotic exponential 
decay of the overlap Dirac operator norm (with taxi-driver distance). 
At large Euclidean time or
taxi-driver distance one finds (barring effects from periodic boundary conditions):
\begin{equation}
\begin{array}{c}
\norm{D(\norm{n})}=A\e^{-\hat\rho\norm{n}}\\
\displaystyle\hat\rho_\text{eff}=\ln\frac{\norm{D(n)}}{\norm{D(n+1)}}
\end{array}
\quad\longleftrightarrow\quad
\begin{array}{c}
C(t,{\textbf{p}}=0)=\tilde A\e^{-mt}\\
\displaystyle m_\text{eff}=\ln\frac{C(t)}{C(t+1)}\,.
\end{array}
\end{equation}
where $m$ corresponds to the mass of the lightest hadron with given quantum
numbers and $\hat\rho_\text{eff}$ denotes the effective decay rate.

In this simplified picture, in order to have an extraction of a
hadron mass which is not influenced by a possibly too slow decay rate of
the overlap operator, the decay rate is expected to be larger
than the measured hadron mass: $\hat\rho>ma$.

In terms of the 
Compton wavelength of the hadron $\lambda_C=1/m$ and 
the locality radius
$r_{\textrm{loc}}=1/\rho$ we want $\lambda_C>r_\textrm{loc}$. 
When the condition 
$\lambda_C>r_\textrm{loc}$ is met, once we have reached the asymptotic 
Euclidean time region to extract masses, the norm of the overlap 
operator has decayed already sufficiently strongly such as not to 
influence the hadron mass measurement. 

We now show that the condition $\lambda_C>r_\textrm{loc}$ is indeed satisfied in our mixed action simulations,
considering the masses of the lightest particles in the meson and baryon sectors. 
We take the unitary values of
the pion and nucleon mass, given in Ref.~\cite{Alexandrou:2009qu}. 
Since we match the pion
mass in our mixed action setup, the values of the pion mass are the same as in the unitary setup, by
definition, whereas the masses of the nucleon can differ from the unitary ones by $\Oasq$ effects.
These turn out to be very small in practice -- see Sec.~\ref{section-baryons} for details. 
We summarize the values for the pion mass, the nucleon mass, the decay rate 
and the ratios of the masses and the decay rate in Tab.~\ref{masses_table}.

\begin{figure}[t!]
\begin{center}
\includegraphics[width=0.509\textwidth]{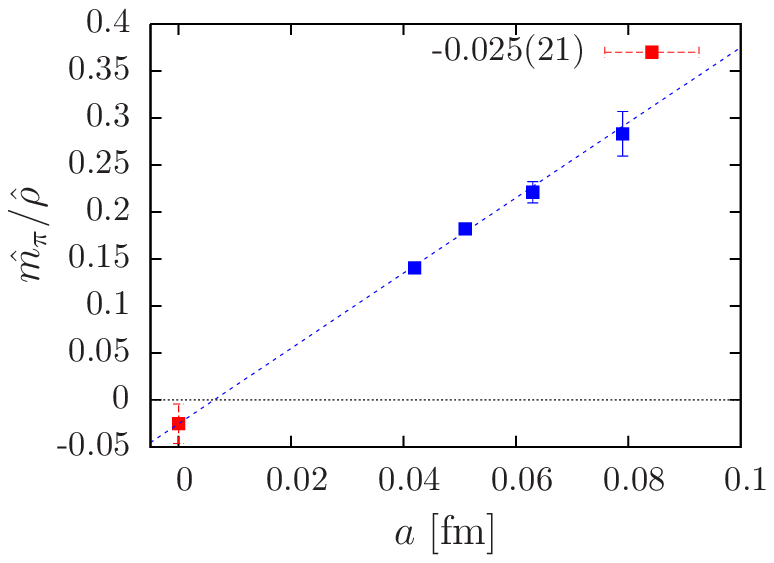}
\hspace*{-0.5cm}
\includegraphics[width=0.509\textwidth]{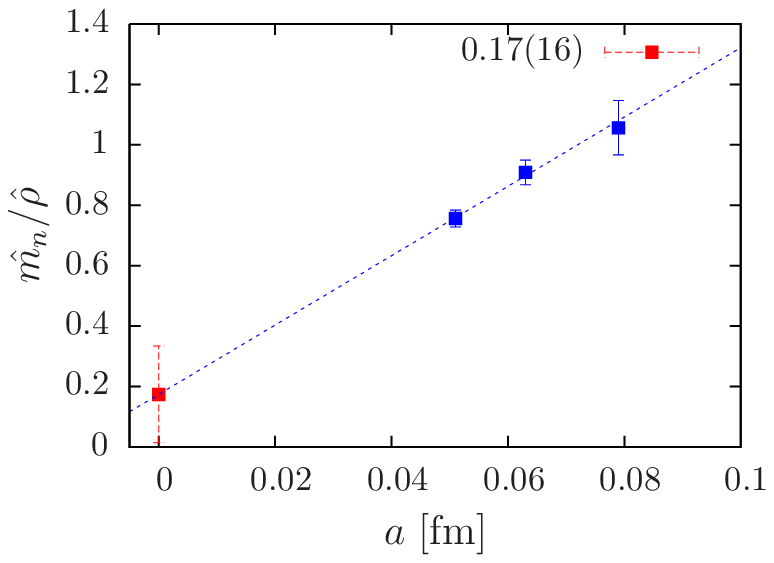}
\caption{Continuum limit scaling of $am/\hat\rho$ for the pion (left)
  and nucleon (right) masses. In these plots only the linear fits are shown.}
\label{pion_nucleon}
\end{center}
\end{figure}

In Fig.~\ref{pion_nucleon}, we show the ratio $am/\hat\rho$ as a function of the lattice
spacing $a$ in a fixed physical situation, see above. 
Both for the cases of the pion and the nucleon masses, the values of the ratio are below one (slightly
above one in the case of the nucleon at $\beta=3.9$)
and the continuum limit value is compatible with zero. This suggests that the overlap Dirac
operator is sufficiently local in our mixed action setup to allow for 
a clean computation of these masses. However, working at a coarser lattice spacing
or considering heavier hadrons, one could enter the regime where the size of the hadron of
interest is smaller than the locality radius. When using overlap
fermions, it is thus crucial to monitor that the locality radius
is sufficiently smaller than the hadron size for all the values
of the lattice spacing which are being considered. 

\begin{table}[t!]
  \begin{center}
    $\begin{array}{c c c c c}
      \hline
      $Ensemble$ & a\mps & a\mps/\hat\rho
      & am_N  & am_N/\hat\rho \\
      \hline
   B_\ell  & 0.159(2)  &0.283(24)   & 0.511(6)  &1.056(90)\\
      C_\ell  &  0.121(4) &0.221(11) & 0.409(6)  &0.908(41)\\
      D_\ell  & 0.098(2)  &0.182(4) & 0.305(4)  &0.756(28)\\
      E_\ell  & 0.075(2)  & 0.141(5)   &    -          &     -  \\
      $cont.limit $       &      -        &  -0.025(21) &   -  &0.17(16)\\   
      $cont.limit (quadratic)$       &    -        &  -0.10(9) &   -  &-0.3(1.2)\\   
    \hline
    \end{array}$
    \caption{\label{masses_table} The pion masses $\mps$ and nucleon masses $m_N$ in lattice
units and the ratios of these masses divided by the decay rate of the overlap Dirac
operator $a\mps/\hat\rho$ and $am_N/\hat\rho$. The continuum limit
values quoted in the table correspond to a linear extrapolation in the first
case and in the  second case we added a quadratic term to take into
account possible higher order lattice artefacts.}
  \end{center}
\end{table}

\subsection{Dependence on the $s$ parameter}
The overlap Dirac operator is constructed from the Wilson Dirac operator taken at large negative
mass shift, depending on the $s$ parameter (see Eq.~\eqref{overlap-A}), which has to fulfill
$\abs{s}<1$, as shown in Ref.~\cite{Jansen:1992tw}. 
In this subsection, we analyze the influence of the $s$-parameter on the decay rate $\rho$ 
of the overlap operator norm. This allows to choose the optimal value of $s$ for simulations, i.e.
the one that gives the highest value of $\rho$. We compare the case when our gauge field
configurations were HYP smeared (1 iteration) -- which is the case for the results 
in the previous sections -- with the one of un-smeared configurations. Since HYP
smearing brings the plaquette values closer to unity and lifts the low-lying eigenvalues
of the kernel Wilson-Dirac operator, improved locality can be expected and is one
of the motivations of using HYP smearing in our mixed action setup.

\begin{figure}[t!]
\begin{center}
\includegraphics[width=0.509\textwidth]{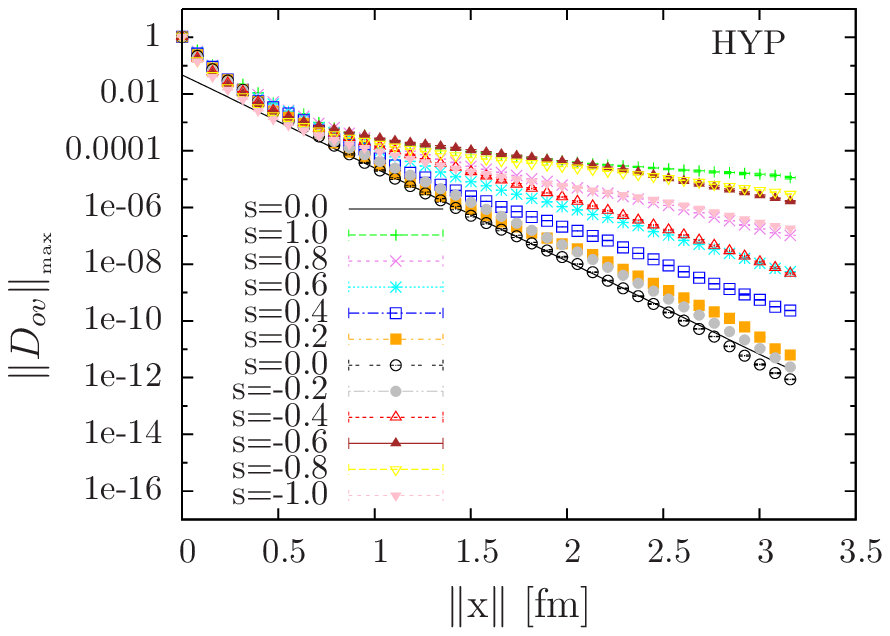}
\hspace*{-0.5cm}
\includegraphics[width=0.509\textwidth]{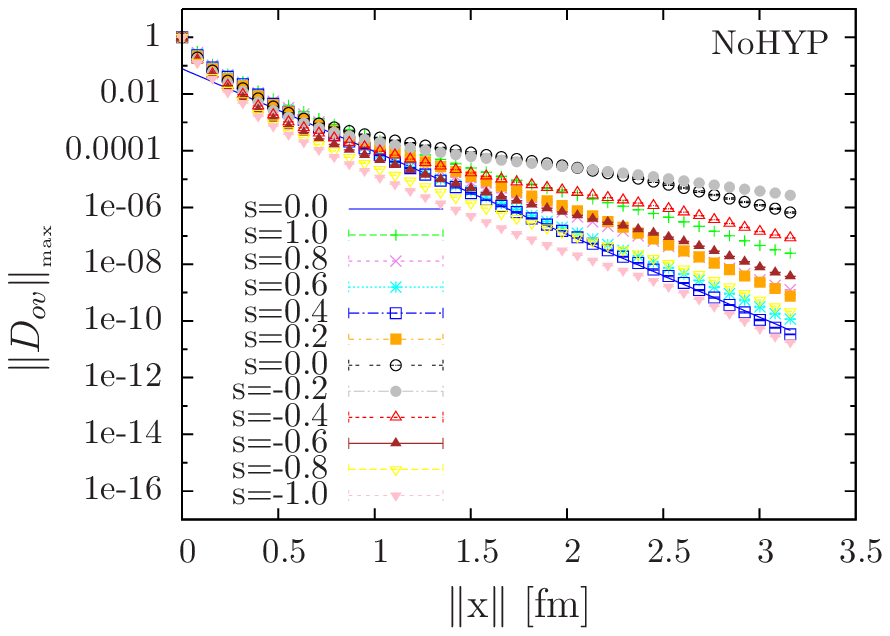}
\caption{\label{norm_HYP_NoHYP} Exponential decay of the norm of the 
overlap Dirac operator (normalized by the value at $\|x\|=0$) for
several values of the $s$ parameter as a function of the taxi-driver distance. 
The graph represents our results for the Ensemble $B_\ell$. 
(left) 1 iteration of HYP smearing was
applied to our gauge field configurations. (right) No HYP smearing applied.}
  \end{center}
\end{figure}

\begin{figure}[h!]
\begin{center}
\includegraphics[width=0.7\textwidth]{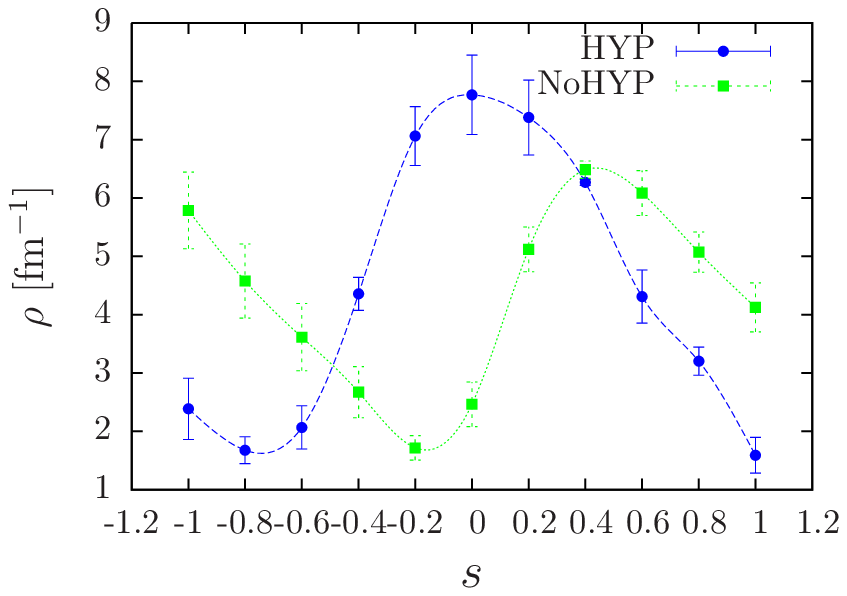}
\caption{\label{HYPNoHYP} The dependence of the decay rate $\rho$ on the value of the $s$
parameter for HYP and non HYP smeared configurations. The lines are just to guide the eye.}
\end{center}
\end{figure}

In Fig.~\ref{norm_HYP_NoHYP}, we plot the dependence of the norm of the
overlap Dirac operator on the taxi-driver distance, for the ensemble 
with the coarsest lattice spacing 
$B_\ell$. We compare several values of the $s$ parameter for HYP smeared and original, 
non-smeared 
configurations. In all cases, we observe an exponential decay, 
with a significant dependence on the $s$
parameter. The resulting values of the decay rate $\rho$, which are 
extracted using the strategy described above, are shown in
Fig.~\ref{HYPNoHYP}. For the case of HYP smeared configurations, the optimal value of $s$ is the
free-field optimal value $s=0$, in accordance with 
Ref.~\cite{Bar:2006zj}, while for the original configurations $s=0.4$. 
In general, we see a strong dependence of $\rho$ on the $s$-parameter and 
it seems from 
Fig.~\ref{HYPNoHYP} that the main difference between HYP-smeared and non-smeared 
gauge field configurations is solely a shift 
in the $s$-parameter. In particular, within the errors, there seems to be
no clear gain of using smeared or non-smeared configurations for the
effective decay rate. Nevertheless, we do observe that in the HYP-smeared case the norm of the
Dirac operator $\|D_{ov}\|$ reaches lower values for the same
values of the taxi driver distance than in the Non-HYP smeared case, as
is shown in Fig.~\ref{norm_HYP_NoHYP}.

\subsection{Finite Volume Effects}
\label{sec:locality_fve}
Since our locality analysis was performed for different physical volumes (from ca. 1.3 to 2.5 fm),
it is natural to ask whether some of these volumes are not too small for a reliable 
calculation of the decay rate. In order to see whether finite size effects affect 
the results for the decay rate, we extended 
our analysis for ensembles $B_{\ell,32}$, $C_{\ell,32}$, $D_{\ell,32}$ and $E_{\ell}$ to the
corresponding ensembles with smaller physical volumes: $B_{\ell,8}$, $B_{\ell}$, $B_{\ell,20}$,
$C_{\ell}$ and $D_{\ell}$.

\begin{figure}[t!]
\begin{center}
\includegraphics[width=0.509\textwidth]{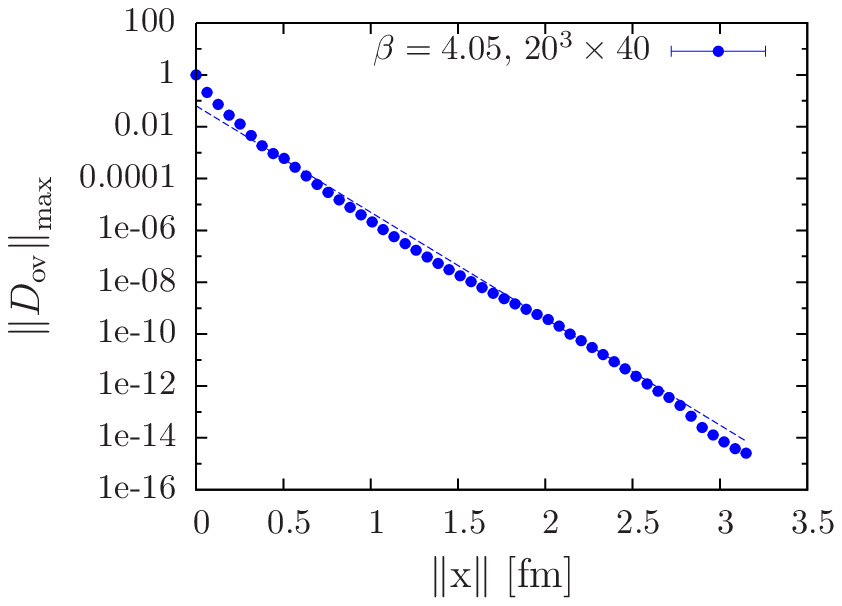}
\hspace*{-0.5cm}
\includegraphics[width=0.509\textwidth]{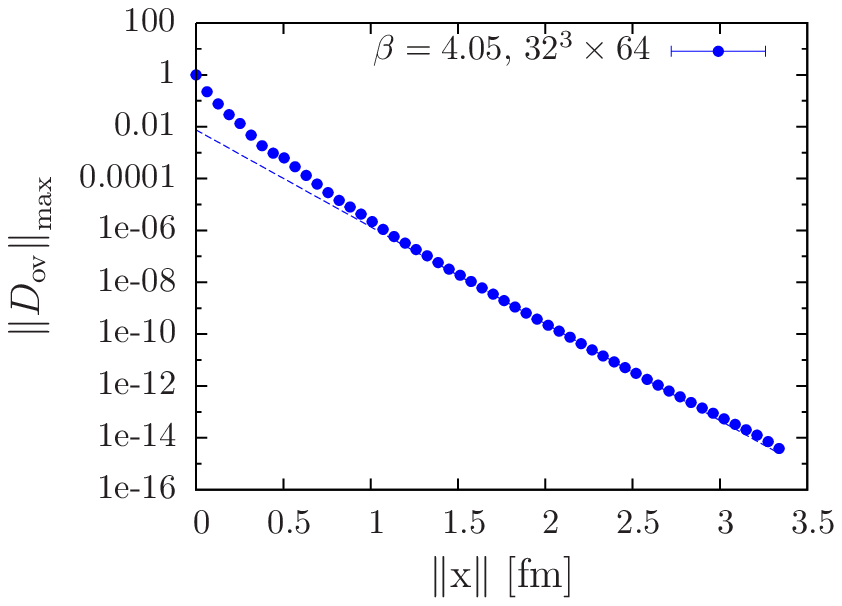}
\includegraphics[width=0.509\textwidth]{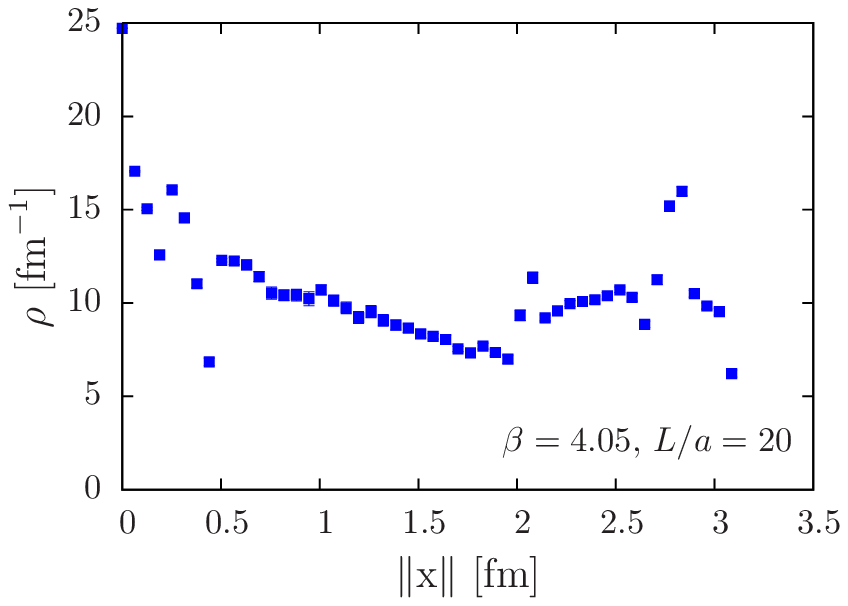}
\hspace*{-0.5cm}
\includegraphics[width=0.509\textwidth]{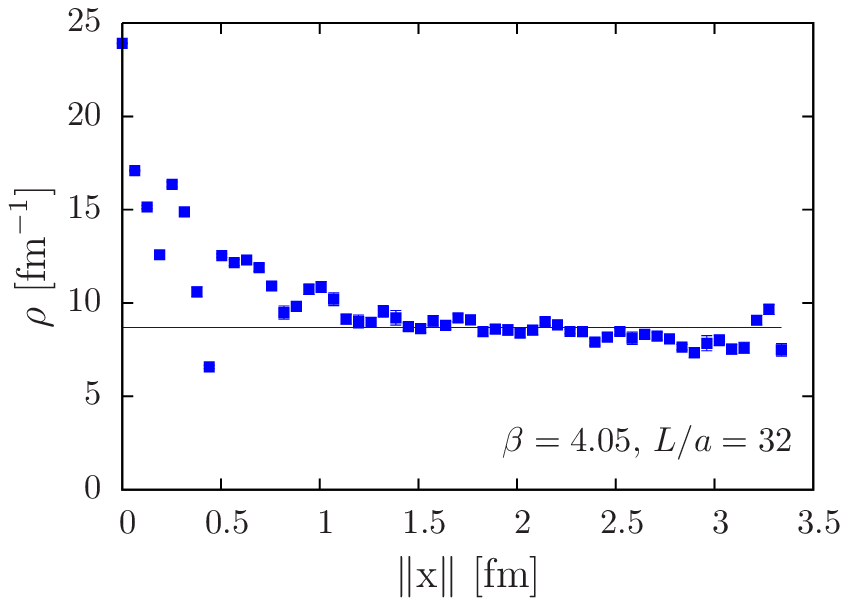}
\caption{\label{fin_vol2} 
(upper) Exponential decay of the norm of the overlap Dirac operator (normalized by the value at
$\|x\|=0$) 
and the effective decay rate $\rho$ (lower) as a
  function of the  taxi driver distance [fm] 
for $\beta=4.05$, $L/a=20$ (left) and $L/a=32$ (right).
In the lower left plot, we observe no plateau. In the right plot, the plateau is present and hence we
can perform meaningful fits.} 
\end{center}
\end{figure}

We illustrate in Fig.~\ref{fin_vol2} the outcome of this analysis by comparing results for ensembles
$C_{\ell}$ and $C_{\ell,32}$ (although they are practically the same at other values of $\beta$).
The upper plots show that in both cases we observe an exponential decay of the overlap Dirac operator
norm.
However, as the lower plots show, for the case of $L/a=20$, we observe no plateau of the effective
decay rate and
hence we cannot perform meaningful fits and extract a reliable value of the decay
rate\footnote{If
we insist on extracting a value, it is compatible with the one of $L/a=32$, but within a large
systematic error -- it therefore shows that our way of extracting the systematic error is reliable.
We emphasize that even in the case of $L/a=32$, we have to take into account systematic effects,
which dominate over statistical errors.}.
Further investigation of the problem in free-field theory for several values of $L/a$ indicates the
presence of large hypercubic artefacts, reaching taxi-driver distances as large as
$\norm{x}/a\approx20$.
Moreover, it is natural to expect that at $\norm{x}/a\approx L/a$ finite volume effects start to
become large. Hence, the plateau region of the effective decay rates is located approximately between
$\norm{x}/a=20$ and 32, as observed in the lower right plot of Fig.~\ref{fin_vol2}. This
conclusion is valid for all values of $\beta$, as can be seen in Fig.~\ref{rhos} and it motivated
our choice of $L/a=32$ for all lattice spacings, rather than keeping the physical extent $L$ fixed.
Indeed, we observe that the decay rate does not depend in a significant way on the volume once the
lattice size is sufficiently large. Therefore, for the range of lattice spacings considered in this
work, we expect that the finite size effects in the decay rate are negligible when working on $L/a =
32$ lattices. 
We have also checked that
performing the continuum limit with the physical volume fixed to 1.3 fm yields a compatible result,
but with much worse precision, due to large hypercubic artefacts for $L/a=16$ ($\beta=3.9$) and
$L/a=20$ ($\beta=4.05$).

\section{Pion decay constant and the role of the zero modes}
\label{sec:scaling}

In Ref.~\cite{arXiv:1012.4412}, we have reported the results of a 
continuum limit scaling
test of the pion decay constant $\fps$, using maximally twisted mass (MTM) sea fermions and two
kinds of
valence quarks: MTM (unitary setup) and overlap (mixed action setup). 
In the mixed action setup, both kinds of fermion actions were 
matched, using the pion mass or by employing alternative matching conditions. While in the matched
mixed action setup the value of $\fps$ (or any other observable) at finite lattice spacing might be
different for two different fermion discretizations, one expects that this
difference vanishes in the continuum limit, provided that the matching 
conditions were chosen properly. Let us shortly summarize the findings of Ref.~\cite{arXiv:1012.4412}.

In order to compare the continuum limit of $\fps$ for the unitary and the mixed
action setups, we matched them by employing several matching
conditions. 
\begin{enumerate}
 \item Naive matching condition -- we extracted the pion mass from the
correlator constructed from the pseudoscalar interpolating field (PP
correlator $C_{PP}(t)$) for both MTM and overlap valence quarks. We defined the
matching point as the overlap valence quark mass that yields the same PP 
correlator pion
mass as in the unitary case.
 \item Improved matching condition -- the unitary pion mass was extracted from
the PP correlator, but the overlap valence pion mass was extracted from a
correlator defined as $C_{PP-SS}(t)=C_{PP}(t)-C_{SS}(t)$. In such a correlator
(PP-SS correlator), the effects of zero modes exactly cancel, since zero modes
couple in the same way to the pseudoscalar and scalar correlators. We defined
the matching point as the overlap valence quark mass that yields the same
PP-SS correlator pion mass as the PP correlator pion mass of the unitary case.
\item Alternative matching condition -- matching was performed using the PP
correlator, but at a heavier quark mass, using partially quenched MTM data. In
this way, one also suppresses the effects of zero modes, since their
leading contribution is proportional to the squared inverse quark mass.
\end{enumerate}

For the regime of parameters that we had chosen, we found that using the naive matching condition
leads to incompatible continuum
limits for $r_0\fpstm$ and $r_0\fpsov$, if both are extracted from the PP
correlator. This was attributed to the effects of zero modes that couple
strongly to this correlator when overlap valence quarks are used. 
These zero mode contributions are not matched by the MTM sea quark action, 
at least not at the values of the lattice spacing used in Ref.~\cite{arXiv:1012.4412}.
To test this
hypothesis, we employed the improved matching condition and extracted $\fpsov$
from the PP-SS correlator. In this way, we found compatible continuum limits.
By investigating the three matching conditions described above, we finally
concluded that each of them can lead to compatible continuum limits of $\fpstm$
and $\fpsov$, provided that the latter is extracted from the PP-SS correlator,
which does not have the contribution of zero modes.

\begin{figure}[t!]
\begin{center}
\includegraphics[width=0.4\textwidth,angle=270]
{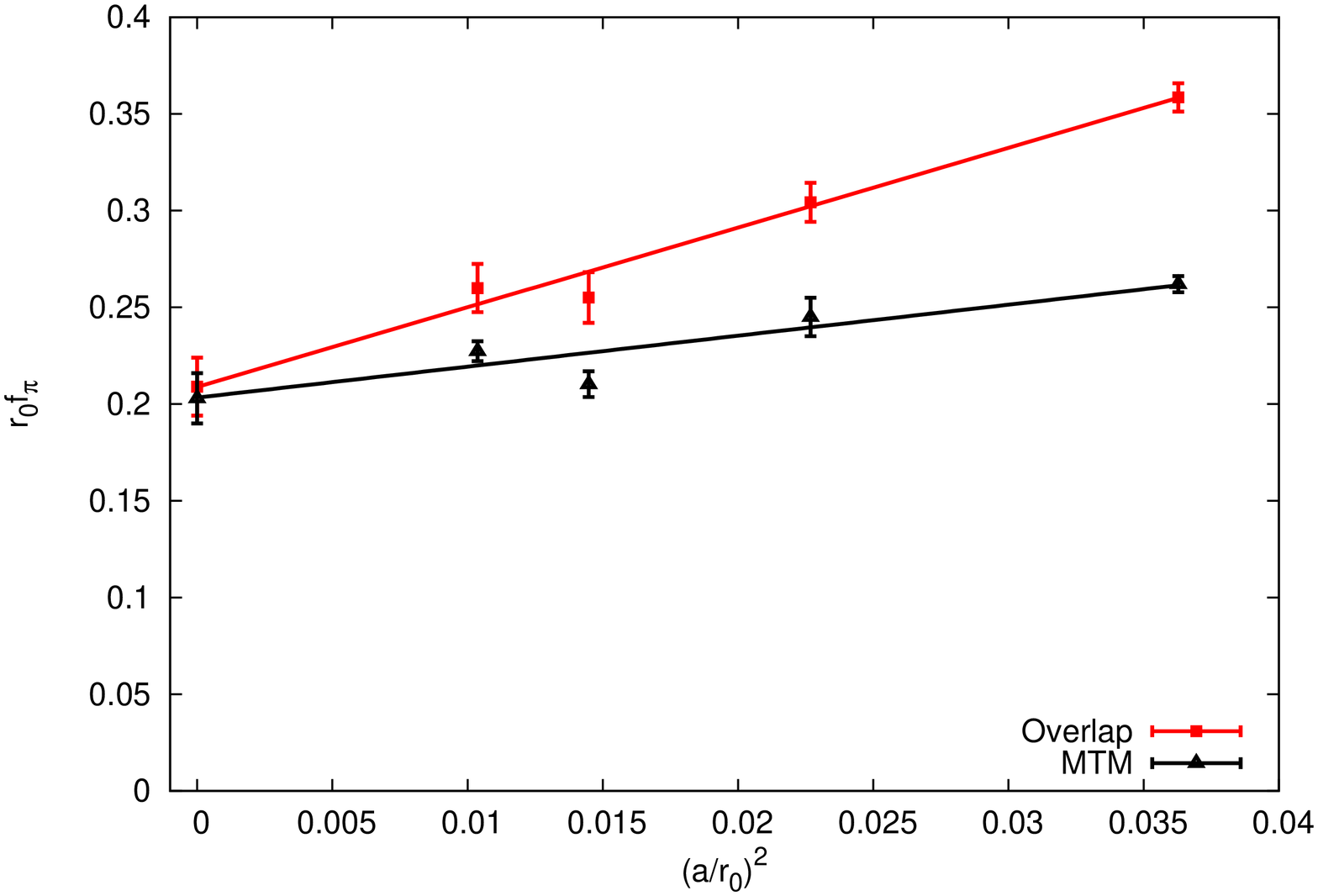}
\end{center}
\caption{Continuum limit scaling of $r_0\fps$. We used the improved matching
condition and extracted $\fpsov$ from the PP-SS correlator. The solid line
represents a linear extrapolation in $a^2$ to the continuum limit.}
\label{fig:scaling}
\end{figure}

\begin{table}[t!]
  \centering
  \begin{tabular*}{0.8\textwidth}{@{\extracolsep{\fill}}lcccc}
    \hline\hline
    \multirow{2}{*}{Ensemble}& \multirow{2}{*}{$a\mps$} & $a\mov^{\rm match}$ & $a\mov^{\rm
match}$ & $a\mov^{\rm match}$\\
    & & naive & improved & alternative\\
    \hline\hline
    $B_\ell$  & 0.1592(24) & 0.007 & 0.011 & 0.009\\
    $C_\ell$          & 0.1209(40) & 0.005 & 0.006 & 0.006 \\
    $D_\ell$          & 0.0980(19) & 0.002 & 0.004 & 0.004 \\
    $E_\ell$ 	& 0.0749(16) & 0.004 & 0.004 & -- \\
    \hline\hline\\
  \end{tabular*}
  \caption{The pseudoscalar masses and overlap matching quark masses for different ensembles
and different matching conditions. Results for ensembles $B_\ell$, $C_\ell$ and $D_\ell$ are taken
from Ref.~\cite{arXiv:1012.4412}. The ensemble $E_\ell$ at a very fine value 
of the lattice spacing has been newly analyzed in this work.}
  \label{tab:matching}
\end{table}

The aim of the remainder of this subsection is to update this analysis with the
fourth lattice spacing (ensemble $E_\ell$ at $\beta=4.35$). We followed the
same procedure as in the case of the coarser lattice spacing ensembles
($B_\ell$, $C_\ell$, $D_\ell$) -- we extracted the quark mass dependence of the
overlap pion mass from the PP-SS correlator and found the matching mass,
employing thus the improved matching condition. The values of the matching mass,
employing different matching conditions are summarized in Tab.~\ref{tab:matching}. 
At this
matching mass, we
computed the pion decay constant $\fpsov$ from Eq.~(\ref{fpsov}) and compared it
to the unitary value $\fpstm$, computed from Eq.~(\ref{fpstm}). The results for
our 4 lattice spacings are presented in Fig.~\ref{fig:scaling}. 

We obtained a result which is fully consistent with our finding in
Ref.~\cite{arXiv:1012.4412}. Both the unitary setup and the mixed action setup
lead to compatible continuum limits (compatible between themselves and with the values found in
Ref.~\cite{arXiv:1012.4412}). Let us emphasize again that the continuum
limits are compatible only when $\fpsov$ is extracted from the PP-SS
correlator, i.e. when the contribution of zero modes is removed. 
The necessity of using proper matching conditions as found in our work 
should, in our opinion, be taken into account for any calculation 
in a mixed action setup which uses chiral invariant fermions in the quark 
sector. 

In order to further corroborate this conclusion we also tried the 
naive matching condition, using solely the PP correlator, 
for a larger volume with $L\approx 2\,\mathrm{fm}$ and a pion mass of 
$m_\pi\approx 480\, \mathrm{MeV}$, i.e. the ensemble $B_s$ of 
Tab.~\ref{tab:setup}. 
For this ensemble, we could observe a significant drop of the 
difference between $\fpsov$ and $\fpstm$ from about 40\% 
for the small volume and small pion mass to about only 8\% 
for the ensemble $B_s$. 
Details of this analysis are presented in Appendix~\ref{sec:app}.

\section{Low energy constants of mixed action chiral perturbation theory}
\subsection{Pseudoscalar meson masses and $\Delta_{Mix}$}

The chiral Lagrangian including $\Oasq$ corrections 
for a mixed action with twisted mass
sea and Ginsparg-Wilson valence quarks can be derived from the twisted
mass chiral
Lagrangian~\cite{Bar:2002nr,Bar:2003mh,Sharpe:2004ny,Munster:2003ba,Scorzato:2004da}. 
In this section we will use the expressions of 
Ref.~\cite{furchner}. They contain the 
Wilson chiral perturbation theory ($\chi$PT) low energy constants (LECs)
$W_0$, $W'_{6,7,8}$ as well as an additional LEC, $W_M$,  appearing in a mixed action. 
A dependence on the twist angle $\omega_0$, defined by 
$\mathrm{atan}(\omega_0)=\mu/m_0$ is also present.

Considering the power-counting $a^2 \sim m$ leads to the following
leading order (LO) expressions for sea and valence 
quark mass dependence of the light pseudoscalar meson
masses~\cite{hep-lat/0611003,arXiv:0706.0035,Chen:2009su,furchner,Bar:2010ix,Hansen:2011kk}
(we set
$\omega_0 = \pi/2$):
\begin{eqnarray}
  \mss^2 &=& 2 B_0 \mu \,,
\label{eq:mSSc}\\
  M_{\rm SS,0,conn}^2 &=& 2 B_0 \mu - \hat a^2\, \frac{32}{f^2}\, W'_8 \,,
\label{eq:mSSn}\\
  \mvv^2 &=& 2 B_0 \mov \,,
  \label{eq:mVVtm}\\
  \mvs^2 &=& B_0 (\mov + \mu) + \hat a^2\, \frac{4}{f^2} \, W_M  -
  \hat a^2\,  \frac{8}{f^2}\,  W'_8  \,,
\label{eq:mVStm}
\end{eqnarray}
where the convention for the pion decay constant in the chiral limit $f$ is such
that $f_\pi$ is 132 MeV, $B_0$ is a low energy constant related to the chiral condensate:
$B_0=-2\langle0|\bar uu|0\rangle/f^2$ \cite{Gasser:1983yg}, $\hat a= 2 W_0 a$, $\mss$ is the charged
sea-sea (SS) meson mass, $\mssz$
the neutral sea-sea meson mass computed solely from the quark-connected contributions and $\mvv$,
$\mvs$ are
valence-valence (VV) and valence-sea (VS) meson masses, respectively.

Rearranging the above expressions, one can find the relation between the
dimensionless combination of LECs $(r_0^6W_0^2)(W_M-2W_8')$ and the SS, VV
and VS meson masses:
\begin{equation}
\label{WM-2W8}
 \left(r_0^6W_0^2\right)\left(W_M-2W_8'\right) =
r_0^2\left(\mvs^2-\frac{\mvv^2+\mss^2}{2}\right)
\frac{(r_0 f)^2}{16}\left(\frac{r_0}{a}\right)^2.
\end{equation} 
This equation will allow us to determine the combination
$(r_0^6W_0^2)(W_M-2W_8')$ from our lattice data for the masses $\mvv$, $\mss$ and $\mvs$.

An alternative parametrization of the VS meson mass
\cite{hep-lat/0611003,arXiv:0706.0035,Chen:2009su} is:
\begin{equation}
 \mvs^2=B_0(\mov+\mu)+a^2 \Delta_{Mix}.
\end{equation} 
The relation between $(W_M-2W_8')$ and $\Delta_{Mix}$ is the following:
\begin{equation}
\Delta_{Mix}=\frac{16W_0^2(W_M-2W_8')}{f^2}.
\end{equation} 

\begin{figure}[t!]
\begin{center}
\includegraphics[width=0.3\textwidth,angle=270]
{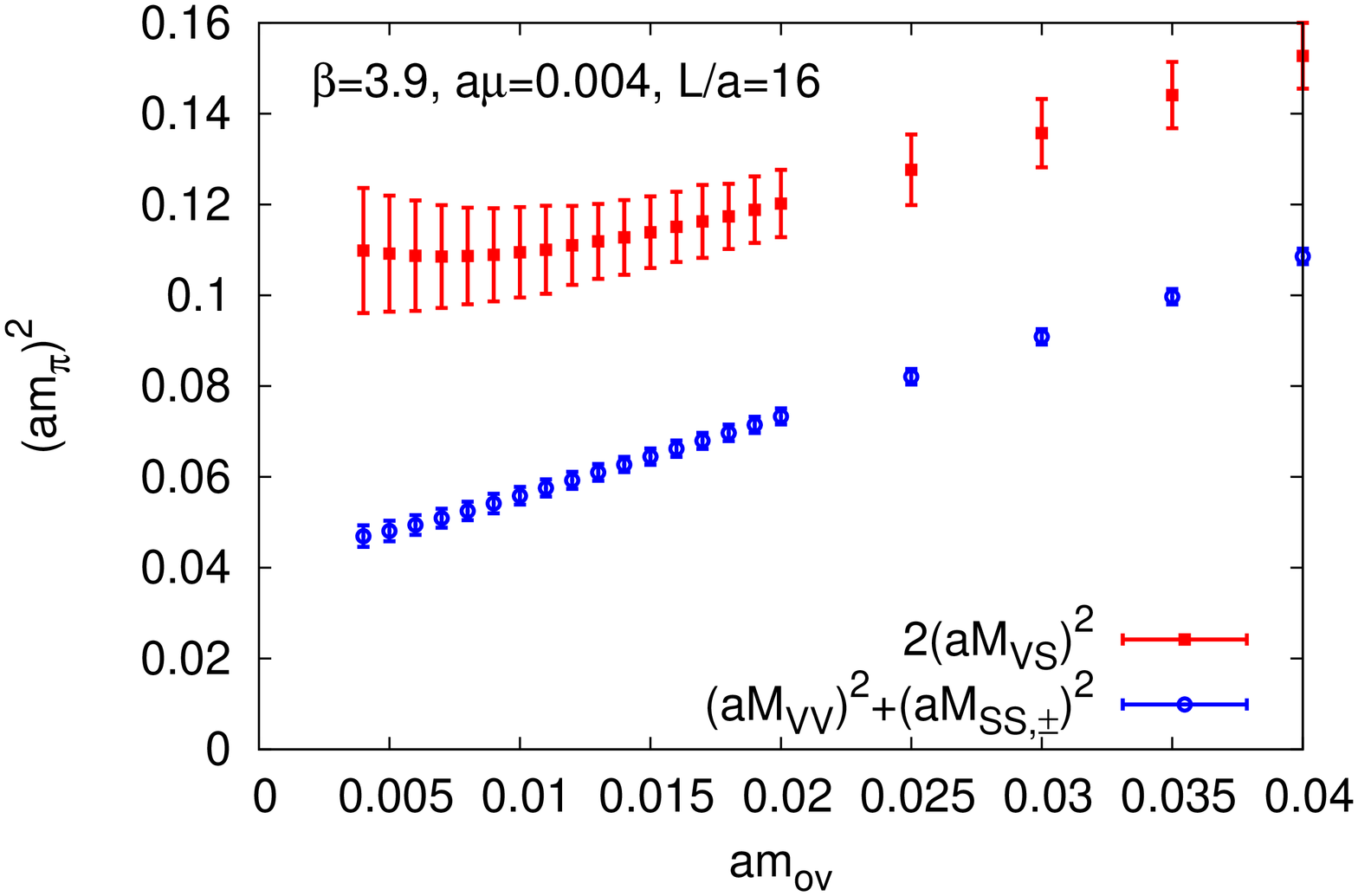}
\vspace*{-0.4cm}
\includegraphics[width=0.3\textwidth,angle=270]
{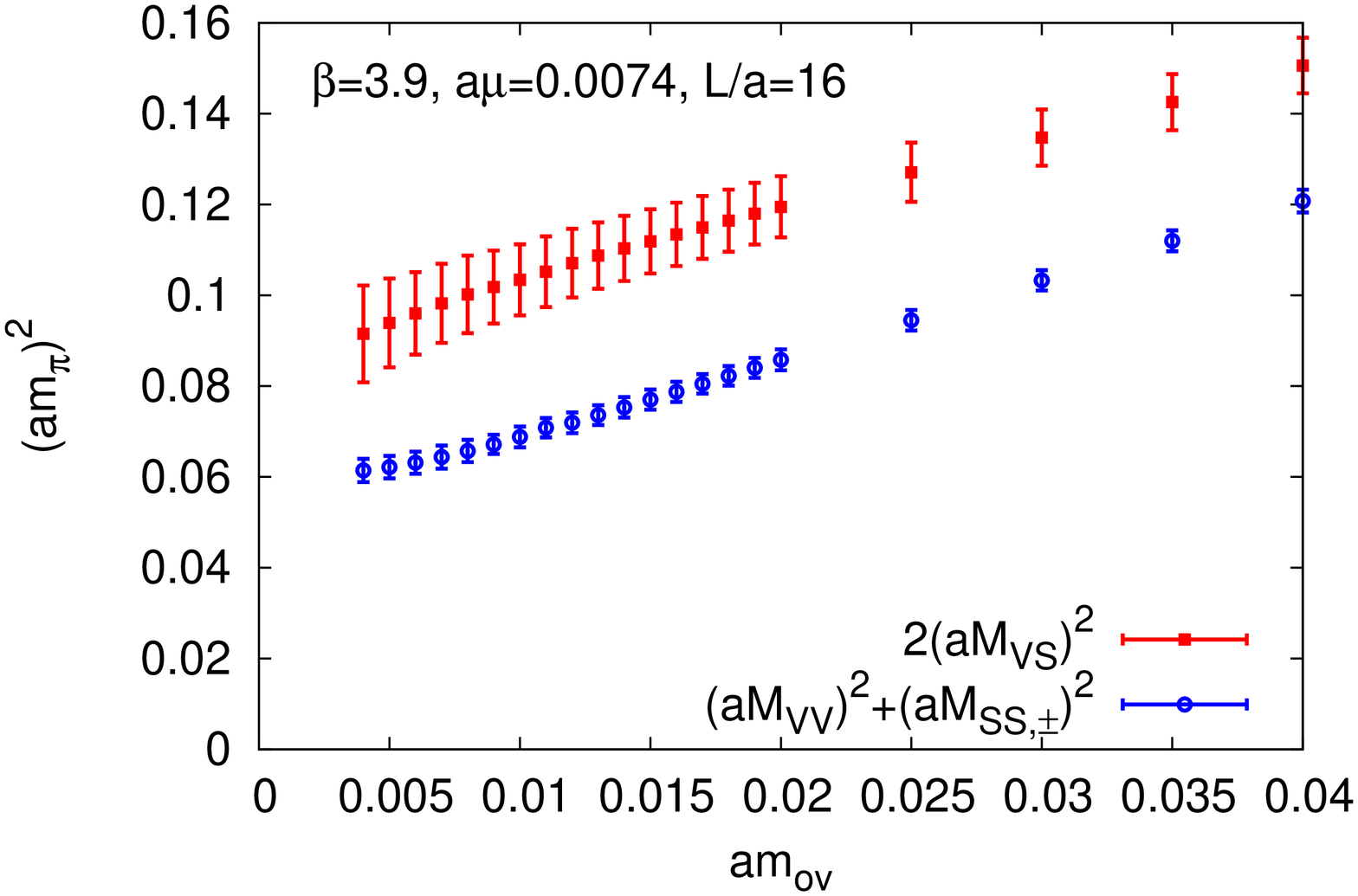}
\includegraphics[width=0.3\textwidth,angle=270]
{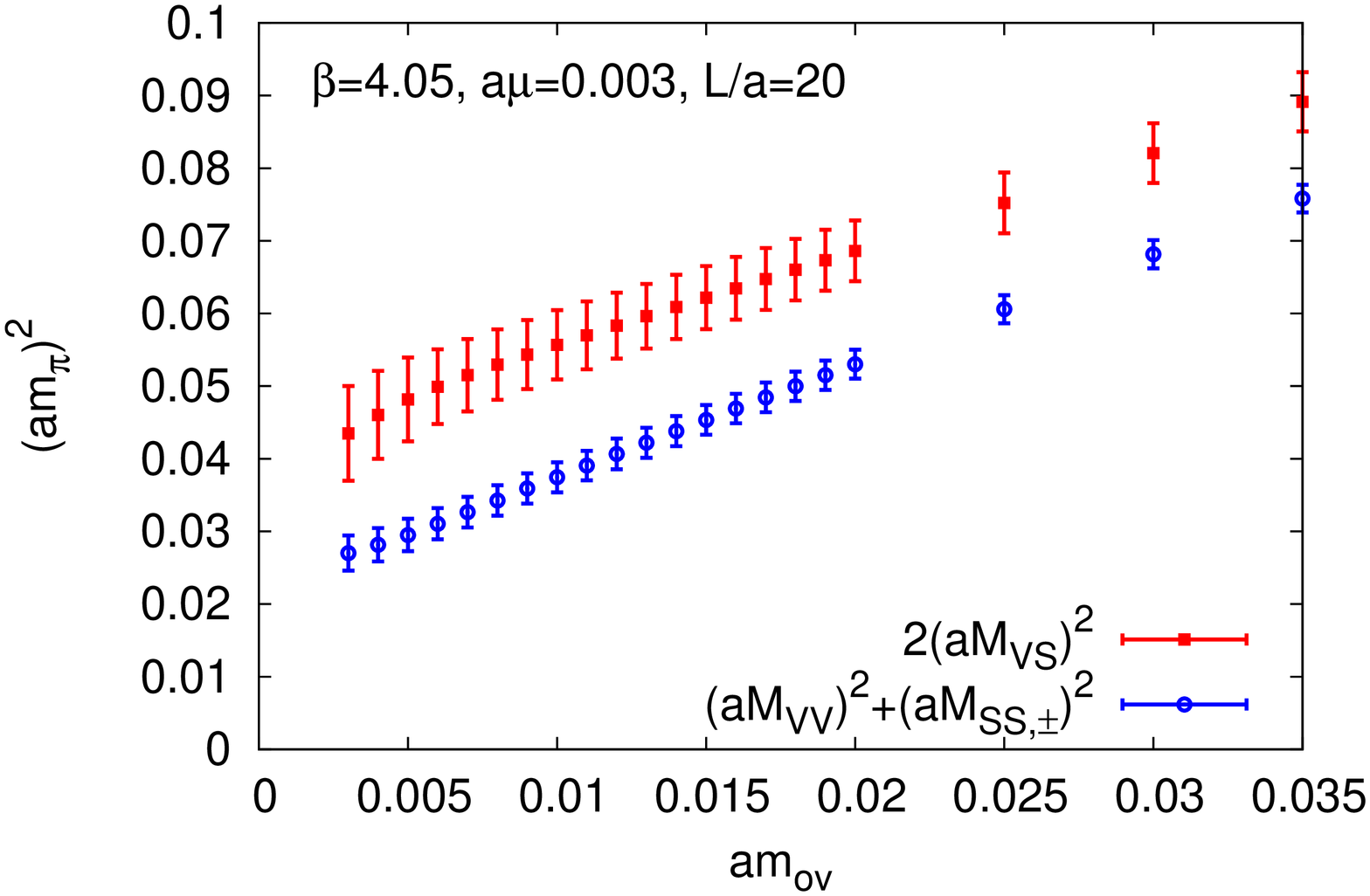}
\vspace*{-0.4cm}
\includegraphics[width=0.3\textwidth,angle=270]
{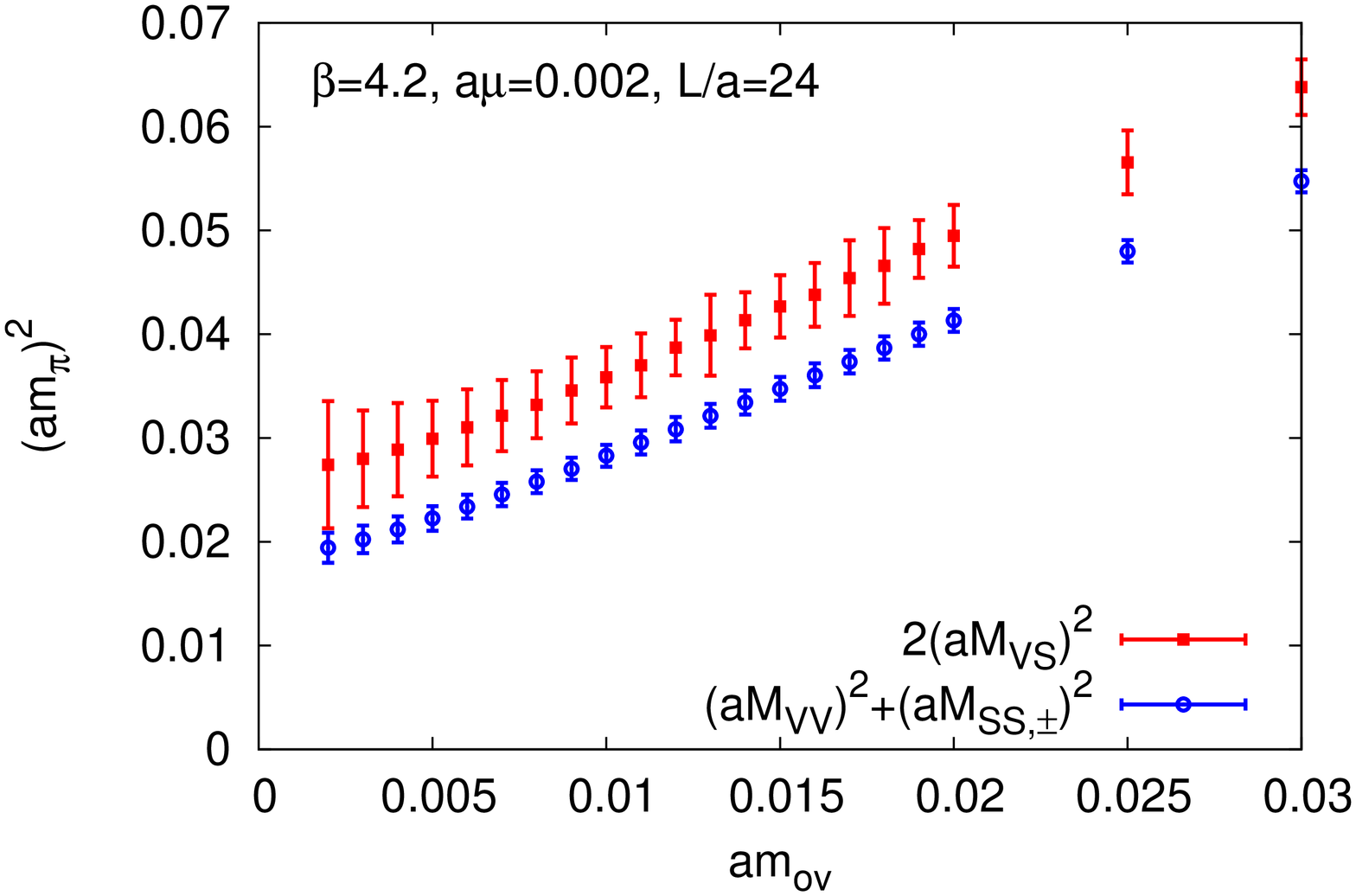}
\end{center}
\caption{The quark mass dependence of the mixed (VS) pseudoscalar meson mass and
the overlap (VV) pseudoscalar meson mass for the ensembles of Tab.~\ref{tab:setup}. 
The latter has been
shifted by the unitary MTM (SS,$\pm$) pseudoscalar meson mass, which  does 
not depend on $m_{\mathrm{ov}}$. The quark mass independent vertical distance between
$2(a\mvs)^2$ and $(a\mvv)^2+(a\mss)^2$ determines the value
of $\Delta_{Mix}$, which should in the LO of $\chi$PT considered here
not depend on the quark mass.} 
\label{fig:lec1}
\end{figure}

\begin{figure}[t!]
\begin{center}
\includegraphics[width=0.3\textwidth,angle=270]
{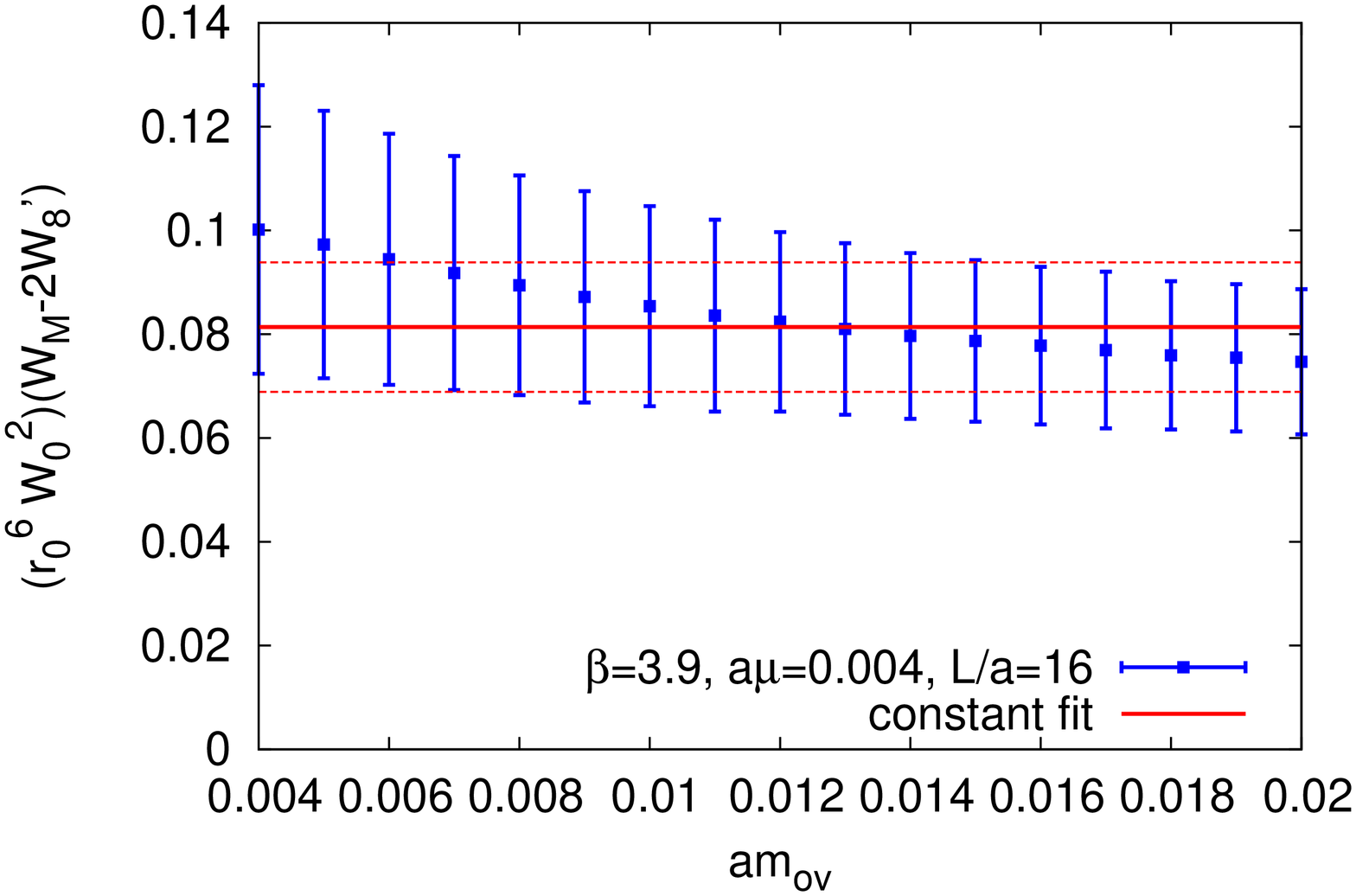}
\vspace*{-0.4cm}
\includegraphics[width=0.3\textwidth,angle=270]
{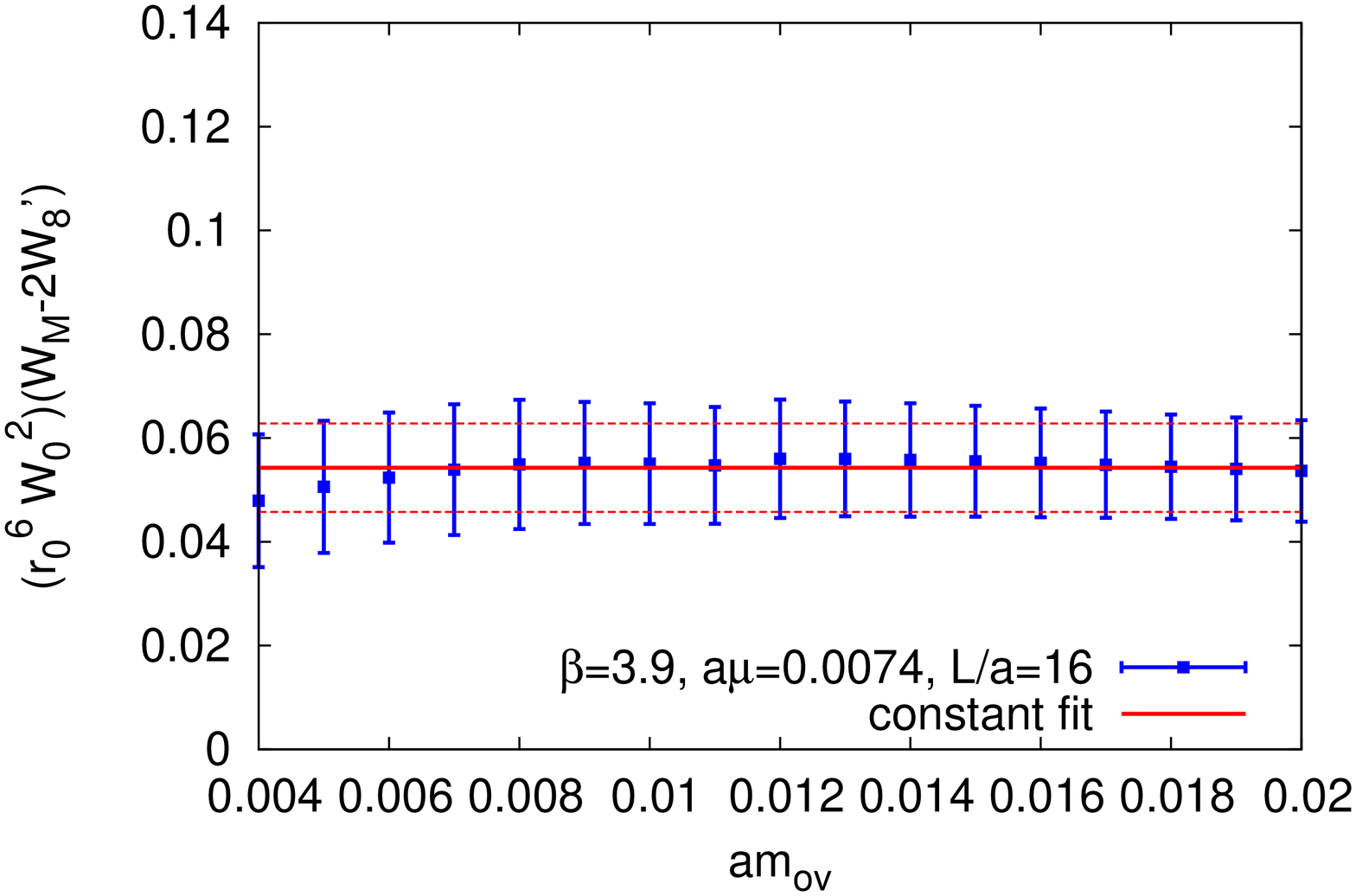}
\includegraphics[width=0.3\textwidth,angle=270]
{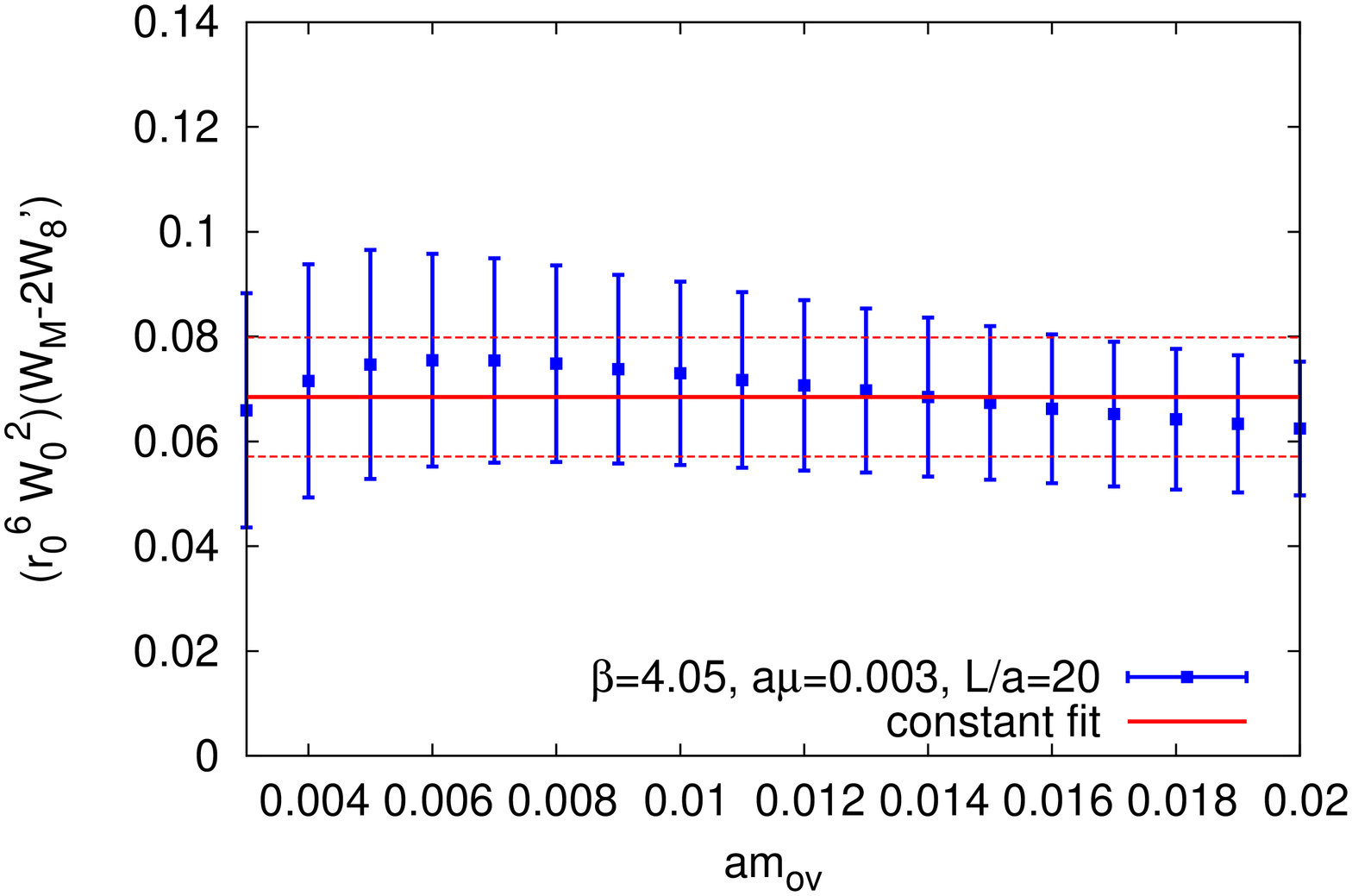}
\vspace*{-0.4cm}
\includegraphics[width=0.3\textwidth,angle=270]
{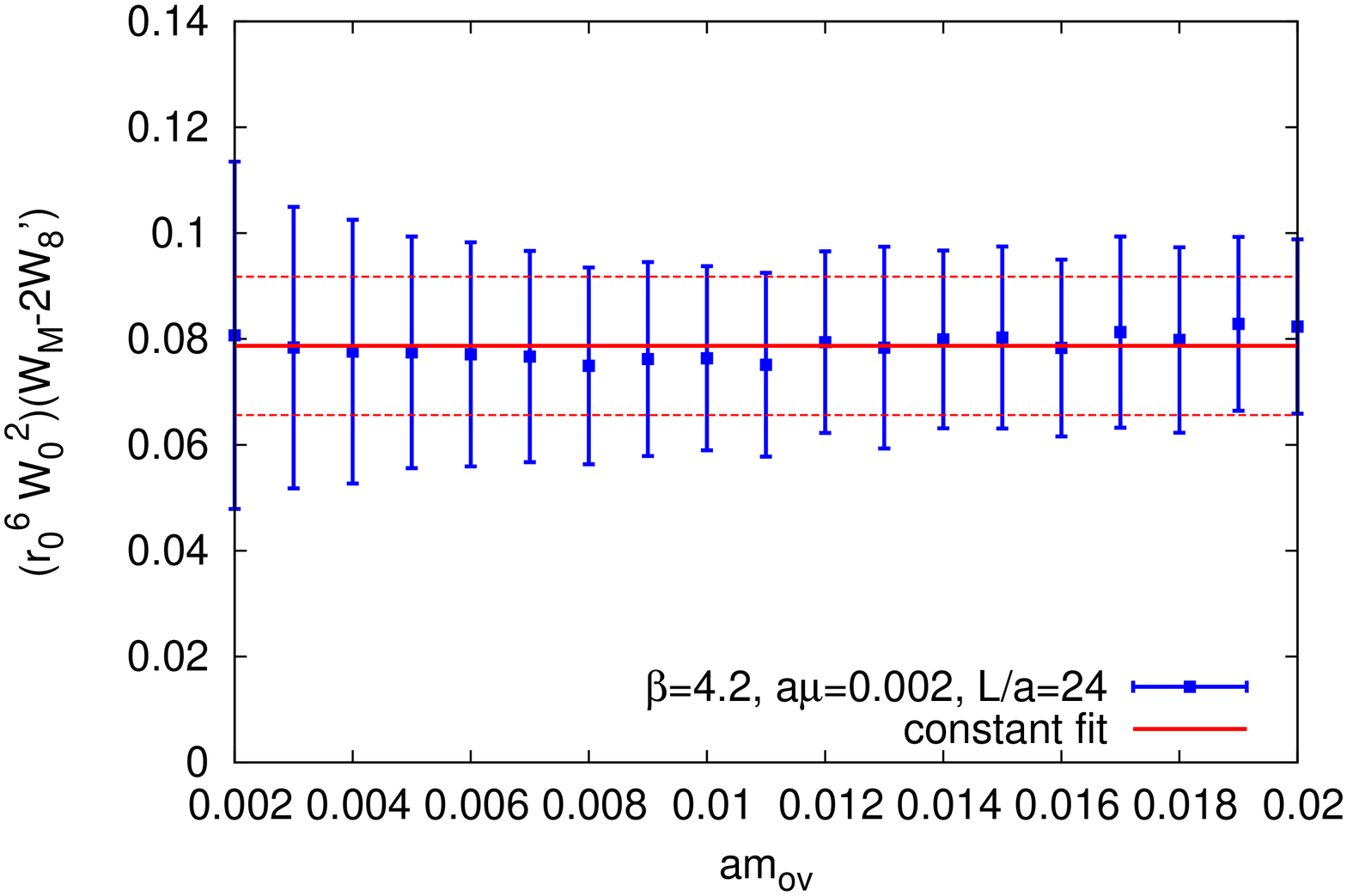}
\end{center}
\caption{The quark mass dependence of the combination
of LECs $(r_0^6W_0^2)(W_M-2W_8')$ for the ensembles of Tab.~\ref{tab:setup}. 
The vertical scale is the
same on all plots, indicating that the values extracted for all ensembles are
compatible with each other.}
\label{fig:lec2}
\end{figure}

\begin{table}[t!]
\begin{center}
\caption{The extracted values of the combination of LECs  $\left(r_0^6
W_0^2\right)\left(W_M-2W_8'\right)$. We show the values at the matching point
$\mvv=\mss$ (the matching quark mass is given in the column $a\mov$) and from the
constant fit to $2(a\mvs)^2-(a\mvv)^2+(a\mss)^2$
for the whole quark mass range shown in Fig.~\ref{fig:lec2}.
The last column is $\Delta_{Mix}^{1/4}$ in MeV, for which the first error is
statistical and the second one the systematic error of conversion to MeV, coming from
uncertainties in the values of lattice spacings.}
\vspace{0.5cm}
\begin{tabular}[b]{ccccc}
\hline
\hline 
\multirow{2}{*}{Ensemble} & matching & \multicolumn{2}{c}{$\left(r_0^6
W_0^2\right)\left(W_M-2W_8'\right)$} & $\Delta_{Mix}^{1/4}$ \\
& $am_{ov}$ & at matching mass & constant fit & [MeV]\\
\hline 
$B_\ell$ & 0.007 & 0.092(23) & 0.081(12) & 997(37)(21)\\
$C_\ell$ & 0.005 & 0.075(22) & 0.068(11) & 951(38)(21)\\
$D_\ell$ & 0.002 & 0.081(33) & 0.079(13) & 968(40)(21)\\
$B_h$ & 0.015 & 0.056(11) & 0.054(9) & 901(38)(20)\\
\hline
\hline
\end{tabular}
\label{tab:lec1}
\end{center}
\end{table}

For our ensembles $B_\ell$, $C_\ell$, $D_\ell$ and $B_h$, we have computed the
pseudoscalar correlation functions for mesons constructed from two MTM quarks
(SS,$\pm$), two overlap quarks (VV) and a combination of one MTM and 
one overlap quark (VS). We will
call the latter mixed correlators. From each of these correlators, we have
extracted pseudoscalar meson masses. The dependence of these masses ($\mss$,
$\mvv$ and $\mvs$, respectively) on the quark mass is shown in
Fig.~\ref{fig:lec1}. Since the combination of LECs $(r_0^6W_0^2)(W_M-2W_8')$ and
hence also $\Delta_{Mix}$ are determined by the difference between $\mvs^2$ and
the average of $\mvv^2$ and $\mss^2$, we use the so determined meson 
masses in the graph.
At the considered order of $\chi$PT, this difference should not depend on the overlap
quark mass $\mov$ and the twisted quark mass $\mu$, i.e. the data for 
$\mvs^2$ and $\mvv^2 + \mss^2$ should be parallel as a function of the quark mass.

As the Fig.~\ref{fig:lec1} shows, this is indeed the case for all 
ensembles we have investigated. In Tab.~\ref{tab:lec1}, we also
show the extracted values of the combination $(r_0^6W_0^2)(W_M-2W_8')$ and the
corresponding values of $\Delta_{Mix}$. The overlap quark mass dependence of these 
LECs is
compatible with a constant behaviour and hence we have used constant fits 
of the difference $2\mvs^2 - (\mvv^2 + \mss^2)$ to
extract $(r_0^6W_0^2)(W_M-2W_8')$ for each ensemble, see also Fig.~\ref{fig:lec2}. 
We find that the extracted
values are compatible for all ensembles, as they should at his order 
of $\chi$PT. 
In fact, the spread in $(r_0^6W_0^2)(W_M-2W_8')$ (or in $\Delta_{Mix}$)
between our light sea quark
ensembles, labeled with the subscript $\ell$, is very small, suggesting that
lattice spacing effects are rather small, at least within the precision 
we could reach here. 

As our final values for $(r_0^6W_0^2)(W_M-2W_8')$ and $\Delta_{Mix}^{1/4}$, we
take the weighted averages of the values from Tab.~\ref{tab:lec1} (with the
inverse statistical error squared as the weight). This yields:
\begin{displaymath}
 (r_0^6W_0^2)(W_M-2W_8')=0.067(6)(14),
\end{displaymath}
\begin{displaymath}
 \Delta_{Mix}=[951(21)(50)\,\mev]^4,
\end{displaymath}
where the first error is statistical (error of the weighted average)
and the second one is systematic, which we take as one half of the spread between
the values for the ensembles $B_\ell$ and $B_h$. 
This implies an approx. 140 MeV splitting between the VS and VV meson masses at the
matching mass, for ensemble $B_\ell$.
Note that our result for $W_M-2W_8'$ is compatible with the constraint $W_M-2W_8'\geq0$ \footnote{In
Ref.~\cite{Bar:2010ix}, the constraint takes the form $2W_M-W_8'\geq0$. However, $W_M$ in our
notation receives a factor of 4 compared to the one defined in Ref.~\cite{Bar:2010ix}.}, found in
Ref.~\cite{Bar:2010ix}.

The value of $\Delta_{Mix}$ that we have determined is comparable to, but
somewhat larger in comparison with previous studies with different
mixed action setups:
$\Delta_{Mix}=[706(4)\,\mev]^4$ \cite{arXiv:0705.0572} and [678(13) MeV]$^4$ 
(with a finer lattice spacing)
\cite{Aubin:2008wk} for domain wall fermions on a staggered sea, [769(77) MeV]$^4$ and [861(90)
MeV]$^4$ (for a heavier pion mass) 
\cite{Durr:2007ef} for overlap fermions on a clover sea and
-[427(338) MeV]$^4$ \cite{Li:2010pw} or [416(27) MeV]$^4$  \cite{Lujan:2012wg} for
overlap
fermions on a
domain wall sea\footnote{The
former value has been obtained using a different (indirect) method -- by examining a
meson state that wraps around the time boundary, instead of a direct extraction
from mixed correlators.}.

\subsection{Unitarity violations in the scalar correlator}
\subsubsection{Theoretical predictions}
Mixed actions, even when related to each other by suitable 
matching conditions, violate unitarity at any non-vanishing value of the 
lattice spacing. Unitarity is then only restored in the continuum limit. 
So far, we have looked at unitarity violations in the pion decay constant at the matching mass and
in mixed correlators.
Here, we investigate yet another way of studying unitarity violations, through a quantity which is
known to depend strongly on them and, in principle, allows to quantify their size -- the flavour
non-singlet scalar correlator.
The reason is that within $\chi$PT, 
unitarity violations appear as unphysical 
double poles in the propagator, affecting strongly 
the flavour non-singlet scalar correlator 
even at the matching point, i.e. if the valence-valence (VV)
pseudoscalar meson mass is matched to the sea-sea (SS) meson
mass~\cite{hep-lat/0106008,hep-lat/0307023,hep-lat/0407037,Golterman:2005xa}.
In fact, the scalar correlator can turn negative in a mixed action 
setup which is clearly an unphysical behaviour. 

The flavour non-singlet scalar correlator can be approximated by:~\cite{furchner}
\begin{eqnarray}
\label{eq:CssVtm}
  C_{\rm sca}^{\rm VV}(t) &=& 
- 2 \left( \mss^2 - \mvv^2  + \hat
a^2\,
    \frac{16}{f^2}\, W'_8 \right)
  \cdot B_{\rm DP}(T, L, t,\mvv,\mvv) \nn\\
  &\,& -2B_{\rm SP}(T, L,
  t,\mvv,\mvv) + 2B_{\rm SP}(T, L, t,\mvs,\mvs) \, \nn\\
  &\,& +Ae^{-m_{a_0}t}+\mathrm{excited\;states}\, ,
\end{eqnarray}
where $B_{\rm SP}$ and $B_{\rm DP}$ are the single and double pole
bubble functions including finite volume effects~\cite{furchner} and the last terms describe
contributions from scalar mesons (the lightest of which is $a_0$). The
label VV in $C_{\rm sca}^{\rm VV}(t)$ reminds that overlap valence
quarks are used in the scalar correlator.

Assuming matching of pseudoscalar meson masses $\mvv=\mss$, in
the limit $T \to \infty$, and at large Euclidean times $t$, the
bubble contributions can be simplified to yield:
\begin{eqnarray}
\label{eq:CssTinftm}
  C_{\rm sca}^{\rm VV}(t) \  \to \  \frac{B_0^2}{2L^3}\, \left[ \frac{e^{-2
M_{\rm VS}
        t}}{M_{\rm VS}^2} - \frac{e^{-2 M_{\rm VV}
        t}}{M_{\rm VV}^4} \left( M_{\rm VV}^2 +
      \hat a^2 \,\frac{16}{f^2} \,W'_8 ( 1 +   M_{\rm VV} \,t ) \right)
\right].
\end{eqnarray}
Identifying the LEC $\gamma_{\rm SS}$ in Ref.~\cite{Golterman:2005xa} with 
$\gamma_{\rm SS}=\frac{16}{f^2}\, W_0 W'_8$, one
recovers, in the large $t$ limit, the corresponding expression of Eq.~(23) of
Ref.~\cite{Golterman:2005xa}.

\begin{figure}[t!]
\begin{center}
\includegraphics[width=0.57\textwidth,angle=270]
{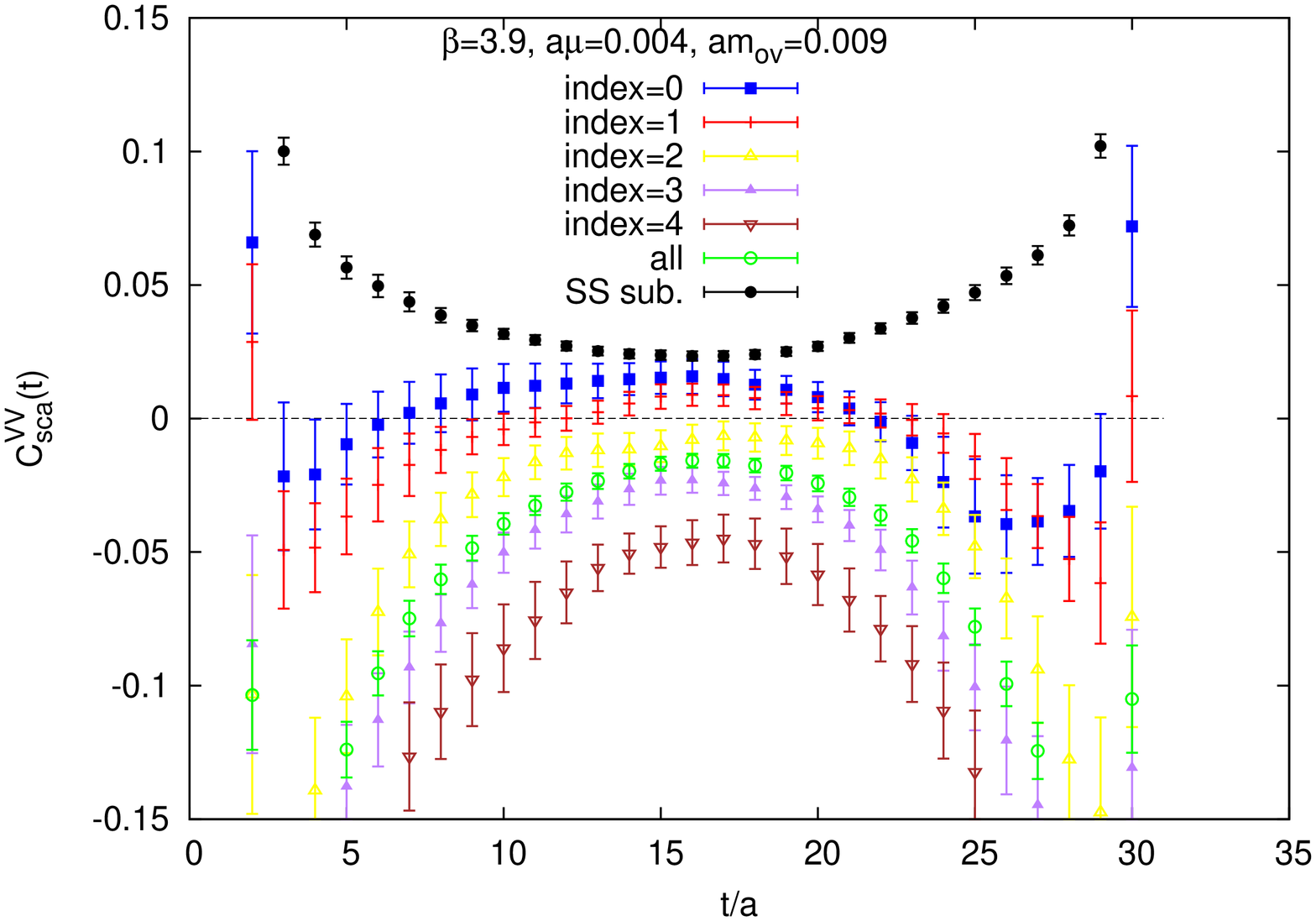}
\includegraphics[width=0.57\textwidth,angle=270]
{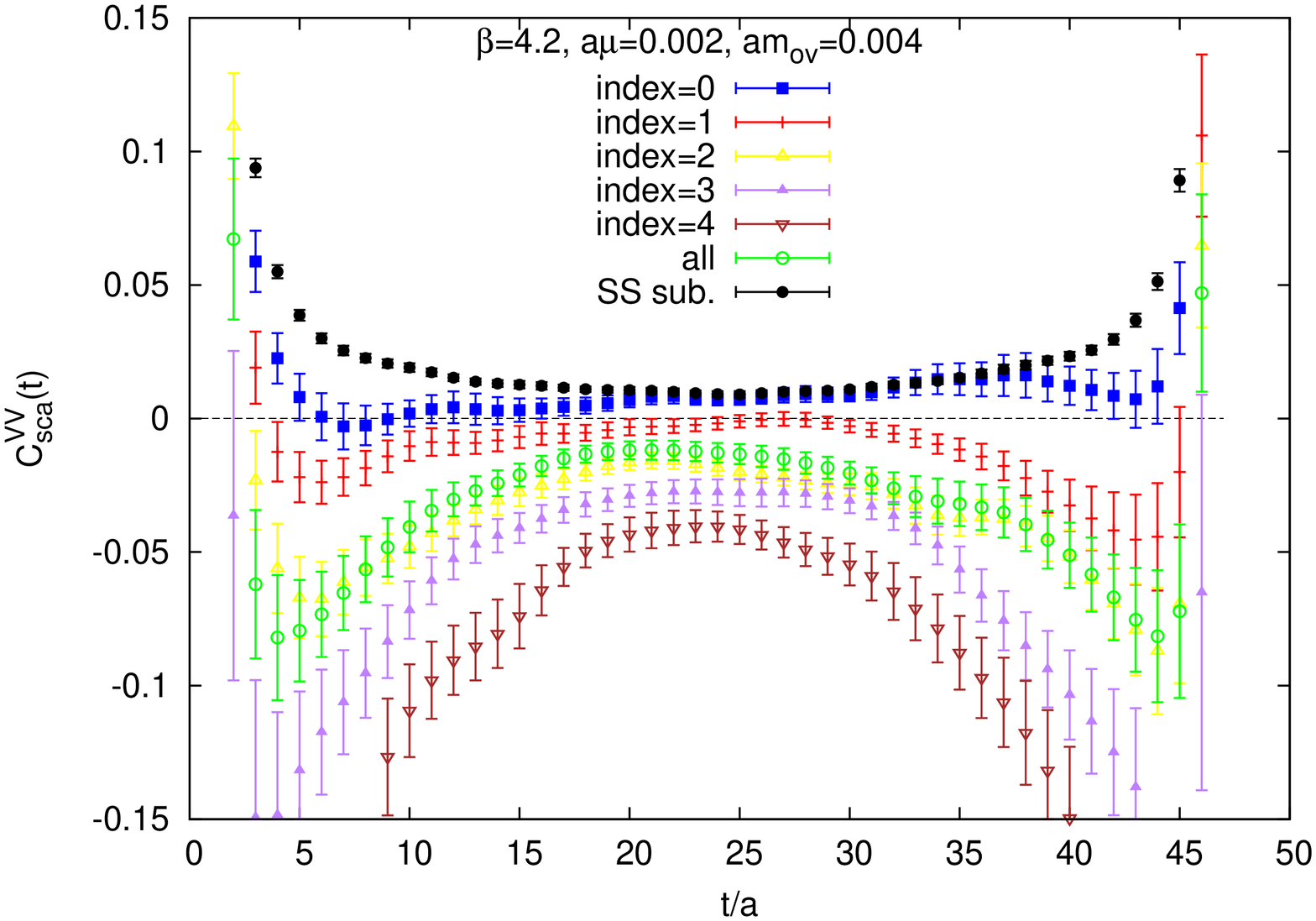}
\end{center}
\caption{The scalar correlator $\css(t)$ in fixed topological
charge sectors (index 0,1,2,3,4) and from summing over all topological charge 
sectors (all). In addition, we show the results for the scalar 
correlator with explicitly subtracted zero
modes (SS sub.). Upper: ensemble $B_l$, $\beta=3.9$, $L/a=16$, $am_s=0.004$,
$a\mov=0.009$. Lower: ensemble $D_l$, $\beta=4.2$, $L/a=24$, $am_s=0.002$,
$a\mov=0.004$.}
\label{fig:compare_index}
\end{figure}

\begin{figure}[t!]
\begin{center}
\includegraphics[width=0.5\textwidth,angle=270]
{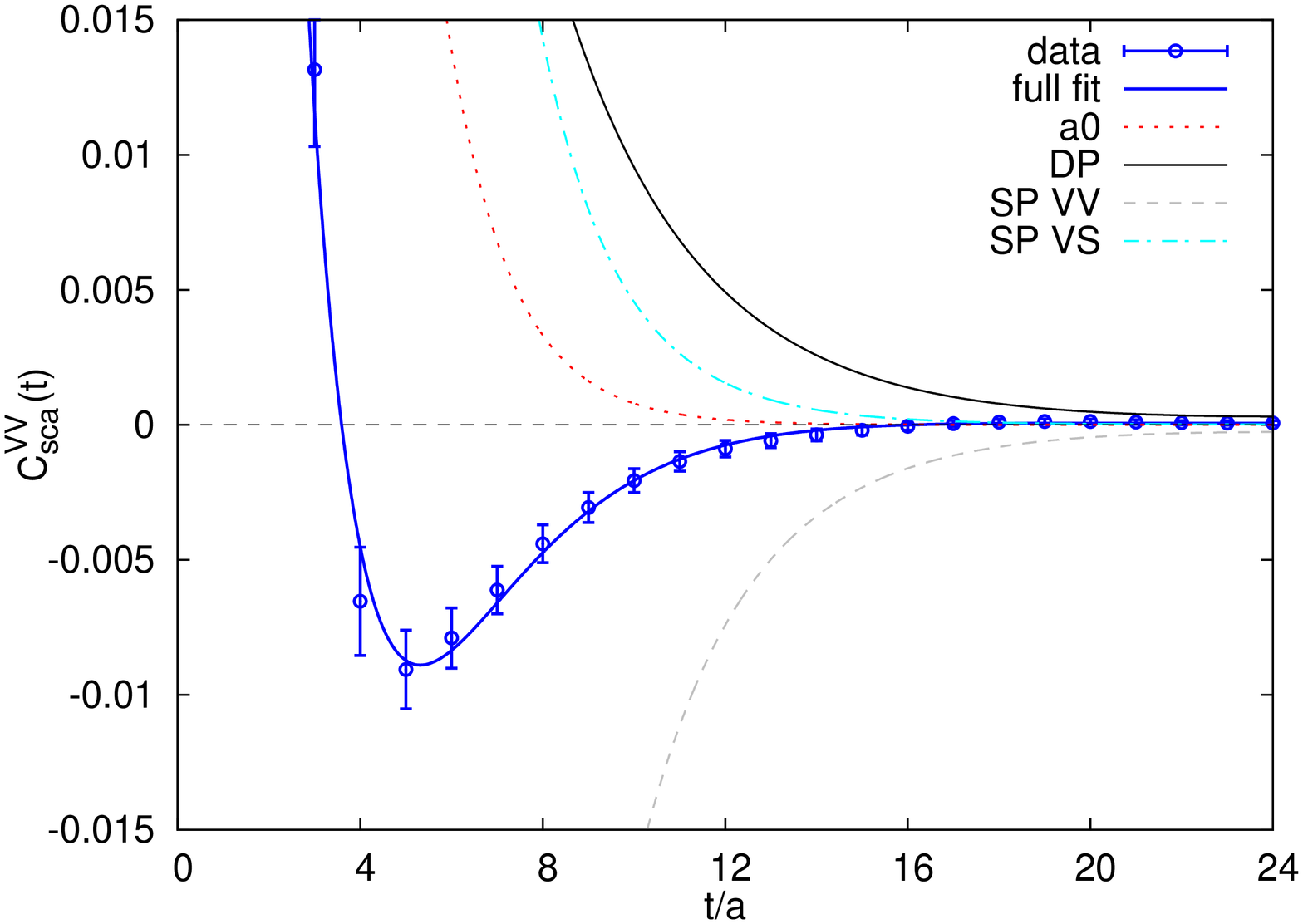}
\end{center}
\caption{Fits of Eq.~\eqref{eq:CssVtm}, using ensemble
$B_s$, for which the zero modes effects are very much suppressed. Blue solid line is the full fit
of Eq.~\eqref{eq:CssVtm}, which is a sum of 4 contributions: double pole (DP) -- black solid line,
single pole containing $M_{VV}$ (SP VV) -- gray dashed line, single pole containing $M_{VS}$ (SP
VS) -- cyan dash-dotted line, scalar $a_0$ meson (a0) -- red dotted line.}
\label{fig:fit_Bs}
\end{figure}

\subsubsection{Lattice results}
Due to the negative coefficients appearing in 
(\ref{eq:CssTinftm}) it can be that the SS correlator
becomes negative at large Euclidean times thus indicating a 
lack of unitarity. 
However, in order to isolate the different contributions to the violations of unitarity, it is
useful to disentangle the effect of zero modes, which can have a potentially very
large contribution to the SS correlator. 
This contribution is actually equal to the zero modes contribution to the PP correlator. As previously
discussed, this is behind the choice of the improved matching condition, which allowed us to extract
the pseudoscalar meson mass and decay constant by computing the PP-SS correlator, where the effect of
zero modes is exactly canceled. 
In Fig.~\ref{fig:compare_index}, we show the scalar correlator for 
ensembles $B_\ell$ and $D_\ell$. 
We have computed the index of the overlap Dirac operator for each
configuration and we have calculated the SS correlator for all configurations
with a given index. For large Euclidean times (i.e. in the region where
Eq.~(\ref{eq:CssTinftm}) should be applicable), the SS correlator 
has a tendency to become more and more negative for configurations with an increasing number of zero
modes, i.e. in higher topological charge sectors. When the
correlator is evaluated on all configurations, the SS correlator is clearly
negative, which signals unitarity violations.

In order to determine 
the low energy
constant $W_8'$ appearing in Eq.~\eqref{eq:CssVtm}, it is much safer to use larger volumes and quark
masses, i.e. 
a situation when the
contribution of zero modes is highly suppressed.
Among our ensembles, $B_s$ has the desired properties and will be used to extract $W_8'$.

Another possibility is to restrict oneself to topologically trivial
configurations. However, it is clear that the number
of such configurations is rather small for each ensemble (between 10 and 20\%
of all configurations) and hence the signal is of poor quality, as can also be 
inferred from Fig.~\ref{fig:compare_index} directly. Moreover,
taking only topologically trivial configurations means that 
the results of the ensemble average will differ from the full result (i.e. from all topological
sectors) by power-like finite volume effects \cite{Aoki:2008tq,Brower:2003yx}.
Therefore, such an analysis will be used only as a cross-check and as an estimate of a systematic
error related to our computation using ensemble $B_s$.

Yet another way to proceed is to explicitly subtract the contribution of zero modes
at the level of propagators. This is a potentially dangerous procedure with hard-to-control
systematic effects. The SS correlator
with explicitly subtracted zero modes (which we call \emph{SS subtr.}) is shown
in Fig. \ref{fig:compare_index}. 
However, given the fact that the zero mode subtraction is a rather doubtful procedure, we will 
quote the value obtained with this method, but we will not
take it into account in our analysis leading to the final value of $W_8'$.

We now describe our strategy to extract $W_8'$ from ensemble $B_s$, using ensembles $B_\ell$,
$B_h$, $C_\ell$ and $D_\ell$ to estimate the systematic error. We fit Eq.~\eqref{eq:CssVtm}, which
has 3
fitting parameters: the LEC of interest -- $W_8'$, the amplitude of $a_0$ scalar meson contribution
and its mass $m_{a_0}$ (however, we do not attempt to extract the $a_0$ contribution
quantitatively). In addition, we add a fourth fitting parameter -- the overall normalization
of the correlator $N$, into which we absorb unknown renormalization factors $Z_S$.
We work with symmetrized correlators and take into account the finite extent of the system by
adding in Eq.~\eqref{eq:CssVtm} the terms with $t\rightarrow T-t$.

\begin{figure}[t!]
\begin{center}
\includegraphics[width=0.3\textwidth,angle=270]
{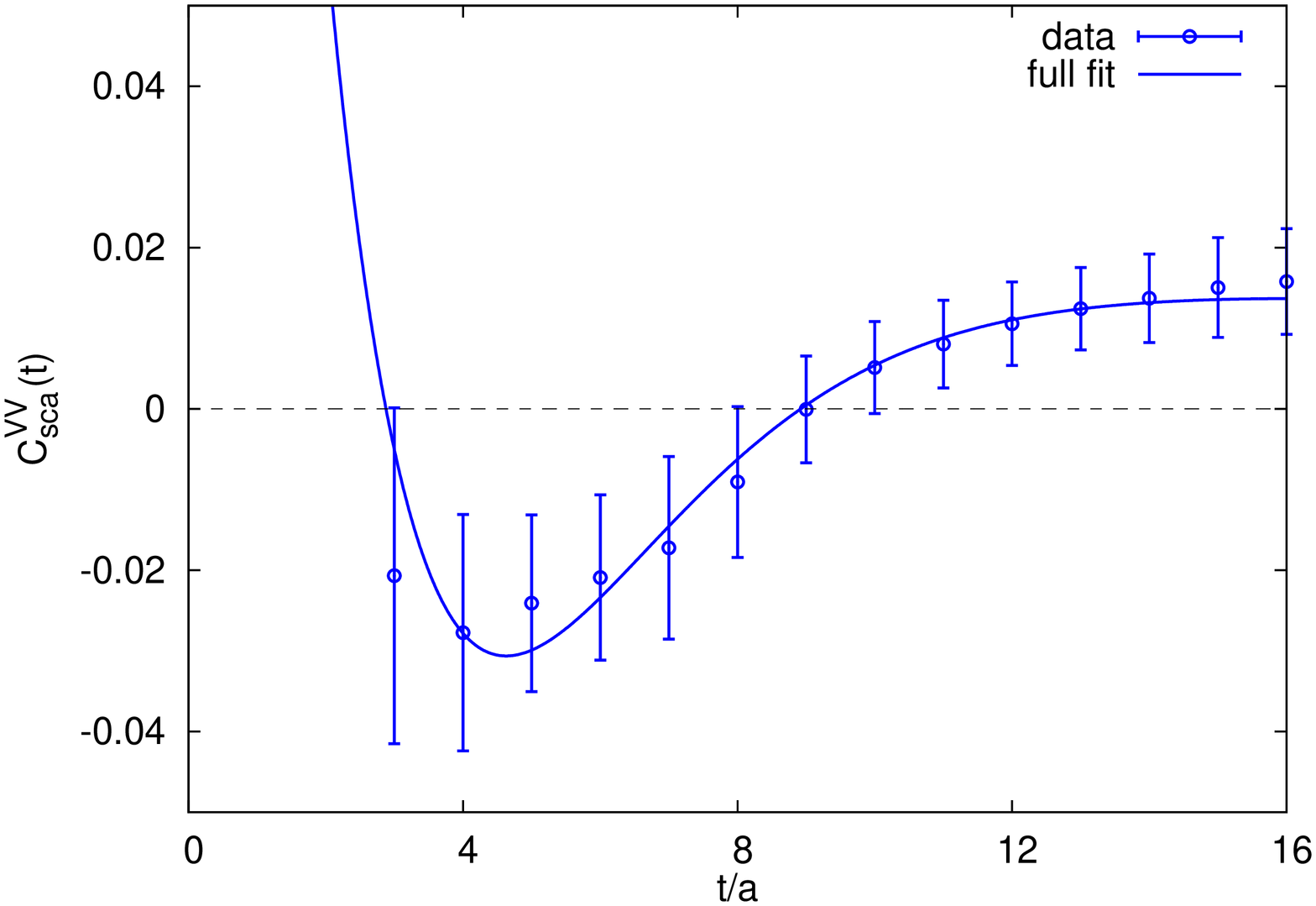}
\includegraphics[width=0.3\textwidth,angle=270]
{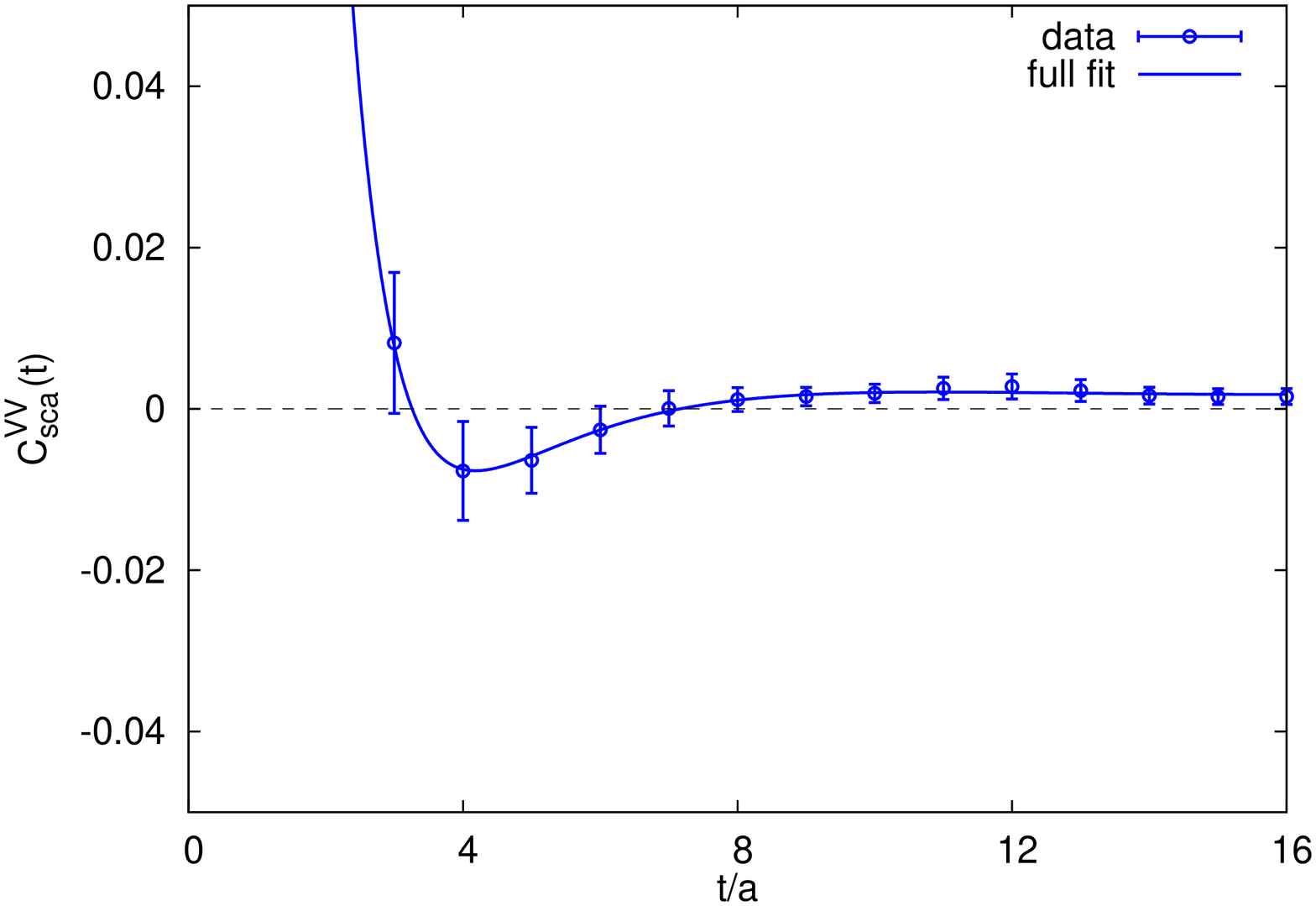}
\includegraphics[width=0.3\textwidth,angle=270]
{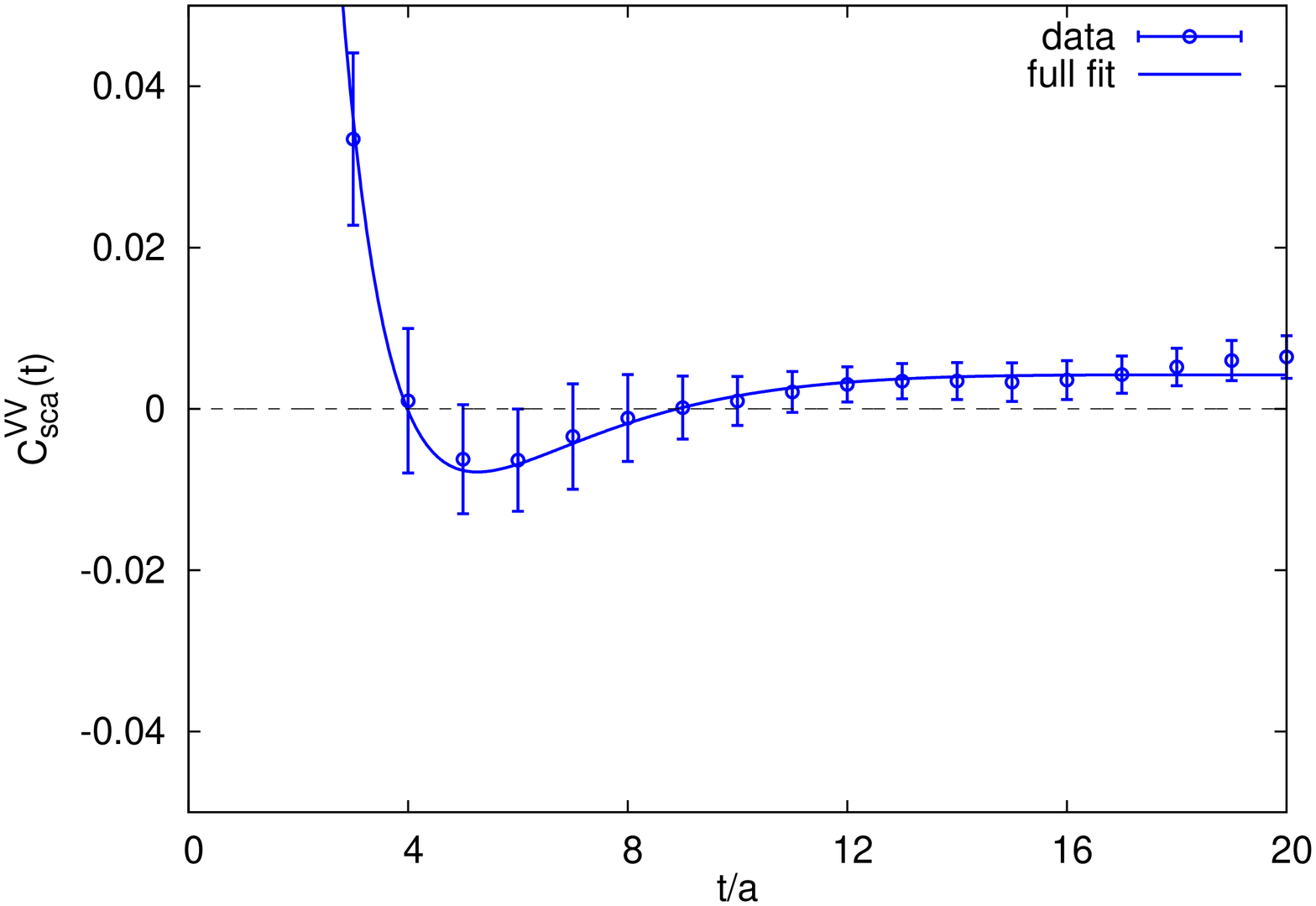}
\includegraphics[width=0.3\textwidth,angle=270]
{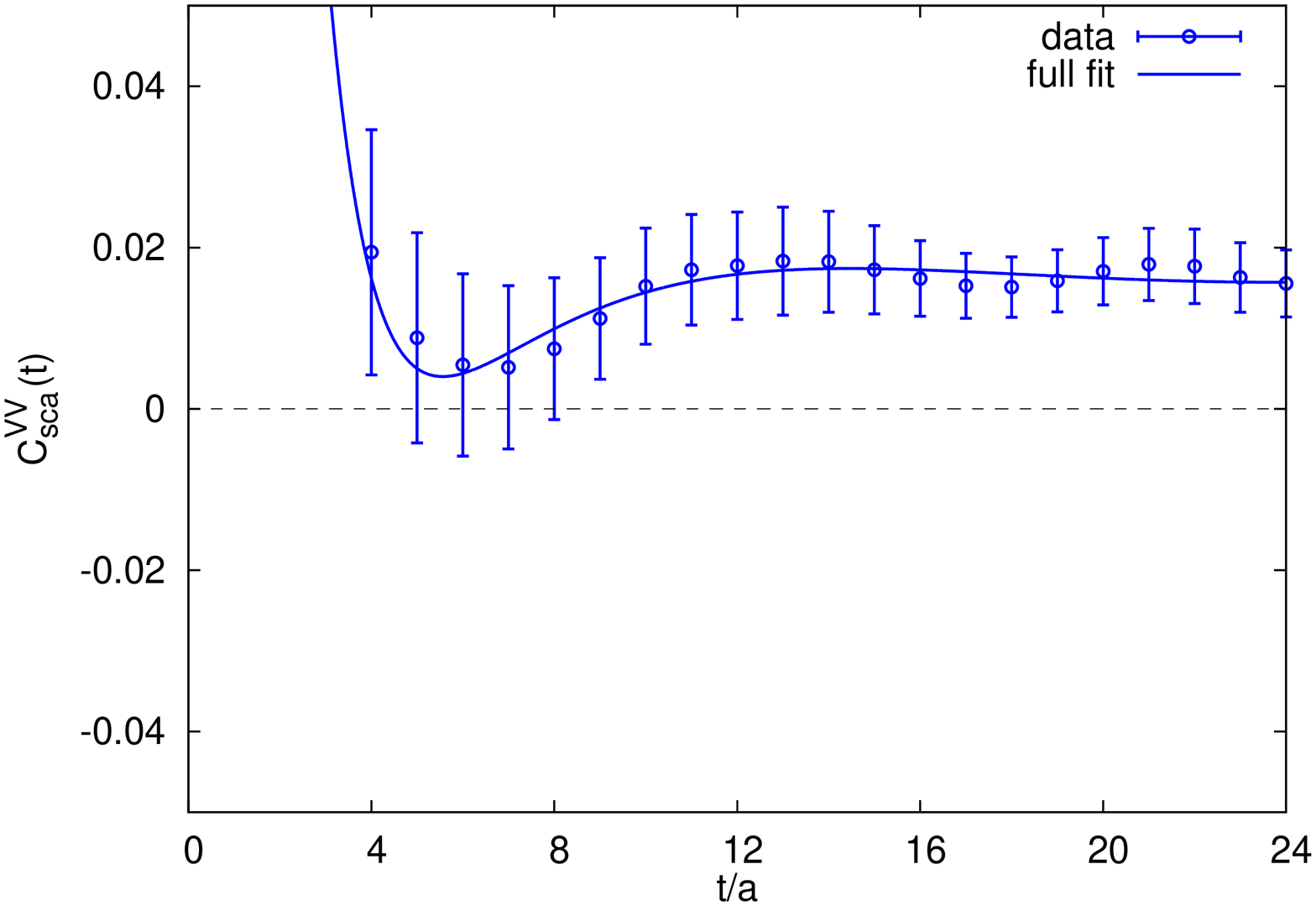}
\end{center}
\caption{Fits of Eq.~\eqref{eq:CssVtm} to the SS correlators for the small volume ensembles
$B_\ell$ (upper left), $B_h$ (upper right), $C_\ell$ (lower left), $D_\ell$ (lower right).}
\label{fig:fit_Bl}
\end{figure}

The results of our fits for ensemble $B_s$ are shown in Fig.~\ref{fig:fit_Bs}. The dimensionless
combination of LECs $r_0^6W_0^2W_8'$ determined from our best fit is $-0.00642(15)$, where the
error is statistical only. We now want to determine the systematic error, related to:
residual discretization
effects, finite volume effects, choice of the Euclidean time fit range and finite precision of the
input
parameters of our fits: $a\mvv=0.1918(12)$ at the matching mass $a\mov=0.017$ (see
Appendix~\ref{sec:app}), $a\mss=0.19403(50)$\footnote{Note that $a\mvv$ is not exactly equal to
$a\mss$, which we take into account by including the difference $\mss^2-\mvv^2$ in the fits of
Eq.~\eqref{eq:CssVtm}.}
\cite{Boucaud:2008xu}, $\Delta_{Mix}$ (taken from the previous subsection), $r_0/a=5.25(2)$,
$r_0f=0.259(11)$ \cite{Baron:2009wt}. 

We address the errors related to discretization effects, finite volume effects and sea quark
mass using our small-volume ensembles at 3 lattice spacings: $B_\ell$, $C_\ell$, $D_\ell$ (at a
pion mass of about 300 MeV) and $B_h$ (approximately 450 MeV). We illustrate fits for these
ensembles in Fig.~\ref{fig:fit_Bl}.
The remaining systematic errors were estimated using ensemble $B_s$ only and varying the respective
quantities, in particular the fitting range.
We have also checked the stability of our results with respect to including or excluding the 2
fitting parameters related to the scalar meson ($a_0$, $A$).
This gives fully compatible results (within the systematic error from the choice of the fitting
range), however we are then limited to larger values of $t/a$.
The summary of all systematic errors is presented in Tab.~\ref{tab:errors}.

\begin{table}[t!]
\begin{center}
\caption{Error budget for our estimates of the combination of LECs $r_0^6W_0^2W_8'$. The last
column shows which ensembles were used to determine the given type of systematic error (or central
value). The total systematic error comes from combining the individual ones in quadrature.}
%\vspace{0.5cm}
%\hspace*{-1.2cm}
%\begin{footnotesize}
\begin{tabular}{lcc}
%begin{tabular}[b]{\extracolsep{-5mm}}{ccccccc}
\hline
\hline
central value & -0.0064 & $B_s$\\
\hline
fit range & 0.0015 & $B_s$\\
$\Delta_{Mix}$ & 0.0004 & $B_s$\\
$aM_{VV}$ & 0.0002 & $B_s$\\
$aM_{SS}$ & 0.0001 & $B_s$\\
including/excluding $\mss^2-\mvv^2$ & 0.0003 & $B_s$\\
discretization effects & 0.0008 & $B_\ell$, $C_\ell$, $D_\ell$\\
finite volume & 0.0009 & $B_s$, $B_h$\\
sea quark mass & 0.0012 & $B_\ell$, $B_h$\\
$r_0/a$, $r_0f$ & 0.0006 & Ref.~\cite{Baron:2009wt}\\
\hline
total systematic & 0.0024 & $B_s$, $B_h$, $B_\ell$, $C_\ell$, $D_\ell$\\
statistical & 0.0002 & $B_s$\\
\hline
\end{tabular}
%\end{footnotesize}
\label{tab:errors}
\end{center}
\end{table} 

The final value that we quote is:
\begin{displaymath}
r_0^6W_0^2W_8'=-0.0064(2)(24), 
\end{displaymath}
where the first error is
statistical and the second systematic.
This value can be compared to the recent determination (with twisted mass fermions) \cite{ref:W8'} 
from the connected contribution to the neutral pion correlator (Eq.~\eqref{eq:mSSn}), which
gives $r_0^6W_0^2W_8'=-0.0106(11)$ (ensemble $B_\ell$ at a larger volume with $L/a=24$ instead of
16) or $r_0^6W_0^2W_8'=-0.0096(22)$ (ensemble $B_\ell$).
We also give here as a cross-check the value obtained from explicit subtraction of zero
modes: $r_0^6W_0^2W_8'=-0.0127(8)$ (where the error is purely
statistical). Although this value has the right order of 
magnitude, 
we emphasize again that the validity of the explicit zero modes subtraction procedure is doubtful
and hence this result should be interpreted with caution.
We also note that our finding for $W_8'$ is compatible with the constraint $W_8'\leq0$
found in Refs.~\cite{Hansen:2011kk,Sharpe:2006ia,Akemann:2010em,Splittorff:2012gp}.

Our results for the combination of LECs $r_0^6W_0^2W_8'$ and $r_0^6W_0^2(W_M-2W_8')$ can be
combined to isolate: 
\begin{displaymath}
r_0^6W_0^2W_M=0.054(6)(15), 
\end{displaymath}
where the meaning of the two errors is as above.

We also define quantities analogous to $\Delta_{Mix}$:
\begin{equation}
w_8' = \frac{16 W_0^2W_8'}{f^2}, 
\end{equation} 
\begin{equation}
w_M = \frac{16 W_0^2W_M}{f^2}.
\end{equation} 
With such definitions, the following relation holds: $\Delta_{Mix} = w_M - 2w_8'$. Our results imply
the values: $w_M=[901(21)(62)\,\textrm{MeV}]^4$ and $w_8'=-[529(4)(51)\,\textrm{MeV}]^4$.
The latter corresponds to approximately 125 MeV splitting between the charged and neutral
(connected\footnote{If one considers the full (connected and disconnected contributions) neutral
pion mass $M_{\rm SS,0}$, the splitting $M_{\rm SS,0}-\mss$ is governed by a combination of LECs
$W_6'$ and $W_8'$ (see e.g. Ref.~\cite{Scorzato:2004da}). The LEC $W_6'$ is not directly accessible
from the overlap scalar correlator.})
pion masses, for ensemble $B_\ell$.
For convenience, we summarize all our results for mixed action $\chi$PT LECs in Tab.~\ref{tab:lecs}.

\begin{table}[t!]
\begin{center}
\caption{The values for different combinations of LECs determined in this
work. The first error is statistical and the second one systematic.}
%\vspace{0.5cm}
\begin{tabular}[b]{cc|cc}
\hline
\hline 
$\left(r_0^6
W_0^2\right)\left(W_M-2W_8'\right)$ & 0.067(6)(14) & $\Delta_{Mix}$ &
 $[951(21)(50)\,\textrm{MeV}]^4$\\
$\left(r_0^6
W_0^2\right)W_8'$  & -0.0064(2)(24) & $w_8'$ & 
$-[529(4)(51)\,\textrm{MeV}]^4$\\
$\left(r_0^6
W_0^2\right)W_M$  & 0.054(6)(15) & $w_M$ & 
$[901(21)(62)\,\textrm{MeV}]^4$\\
\hline
\hline
\end{tabular}
\label{tab:lecs}
\end{center}
\end{table}

\section{Light baryon masses}
\label{section-baryons}

In order to investigate the effects of the mixed action setup in other observables, we have
computed the nucleon and $\Delta$ 
baryon masses in the mixed action and the unitary setup. Given the numerical
cost of the inversion of the overlap operator, we performed the analysis on the following subset of
gauge ensembles generated on only small lattice sizes up to 
$24^3\cdot 48$,  $B_{\ell}$ ($\beta=3.9$), $C_{\ell}$ ($\beta=4.05$)  and $D_{\ell}$ ($\beta=4.20$)
that correspond to a fixed spatial extent of $\approx 1.3~\rm{fm}$ and a fixed 
pseudo scalar mass of
$\sim 300~\rm{MeV}$. Having three values of the lattice spacing available, 
allows us to study the continuum limit for the nucleon (N) and $\Delta$
masses in both setups.
The strategy followed to extract the baryon masses in this section follows closely the 
work on the light \cite{Alexandrou:2008tn} and strange baryon spectrum \cite{Alexandrou:2009qu}
by ETMC. For
consistency, we recall here the basic ingredients of the computation, the reader interested in more
details is referred to the two aforementioned references.

% interpolating fields
The masses of the nucleon and of $\Delta$ baryons are computed through the two-point correlators, using the
following interpolating fields: 
\begin{eqnarray}
\label{eq:interpol_ND}
J^N &=& \epsilon^{abc} \left( u^{a,T} C\gamma_5 d^b \right) u^c ,\\
J^{\Delta}_{\mu} &=& \epsilon^{abc} \left( u^{a,T} C\gamma_\mu u^b \right) u^c ,
\end{eqnarray}
 where $C$ is the charge conjugation matrix. The quark fields $u$ and $d$ refer to the two
degenerate flavours of quarks considered in this work.\footnote{In the unitary case, the $u$ and $d$
quarks refer to the fields in the so-called physical basis.}  While the operator $J^N$ carries spin
$1/2$ and couples only to states which have the quantum numbers of the nucleon, the operator
$J^{\Delta}_{\mu}$ couples both to spin $1/2$ and spin $3/2$ states. Note, however, that 
in practical lattice computations 
the spin $3/2$ dominates the correlator at large time \cite{Alexandrou:2008tn}
thus allowing for a clean extraction of the $\Delta$ baryon. As in 
nature, the spin $1/2$ partner is much heavier than the spin $3/2$ state.

% smearing
 In order to improve the overlap between the ground state and the interpolating fields, we employ
source and sink smearing of the quark fields. We use Gaussian smearing
\cite{Gusken:1989qx,Alexandrou:1992ti} of the quarks field and APE smearing \cite{Albanese:1987ds}
of the gauge links entering in the Gaussian smearing procedure. We use the smearing parameters
obtained in our previous works on baryon spectroscopy \cite{Alexandrou:2008tn,Alexandrou:2009qu}. 
Note that we use the same smearing
parameters both in the twisted mass and overlap fermion cases. 

% 2pt  and meff 
In order to investigate the consequences of the various matching conditions discussed 
in Sec.~\ref{sec:scaling}, we use several valence overlap 
quark masses.  The baryon masses are extracted from the asymptotic Euclidean time 
behaviour of the correlators: 
 
 \begin{equation}
 \label{eq:corr_ND}
 C^{N,\Delta}(t) = \sum_{\vec{x}} \frac{1+\gamma_0}{2} \langle J^{N,\Delta} (\vec{x},t)
\bar{J}^{N,\Delta} (0) \rangle -  \frac{1-\gamma_0}{2} \langle J^{N,\Delta}   (\vec{x},t-T)
\bar{J}^{N,\Delta} (0) \rangle, 
  \end{equation}
The statistical
errors are estimated using $1000$ bootstrap samples.  Concerning the topological finite
size effects, isolating the zero mode contribution after the fermionic integration, the leading
behaviour of Eq.~\eqref{eq:corr_ND} is:
  \begin{equation}
  \label{eq:zero_mode_contribution_ND}
  C(t)\sim \left( am_{\rm ov} \right)^{-3}
  \end{equation}
as discussed in Ref.~\cite{Gattringer:2003qx}.

\begin{figure}[t!]
\includegraphics[width=0.495\textwidth]{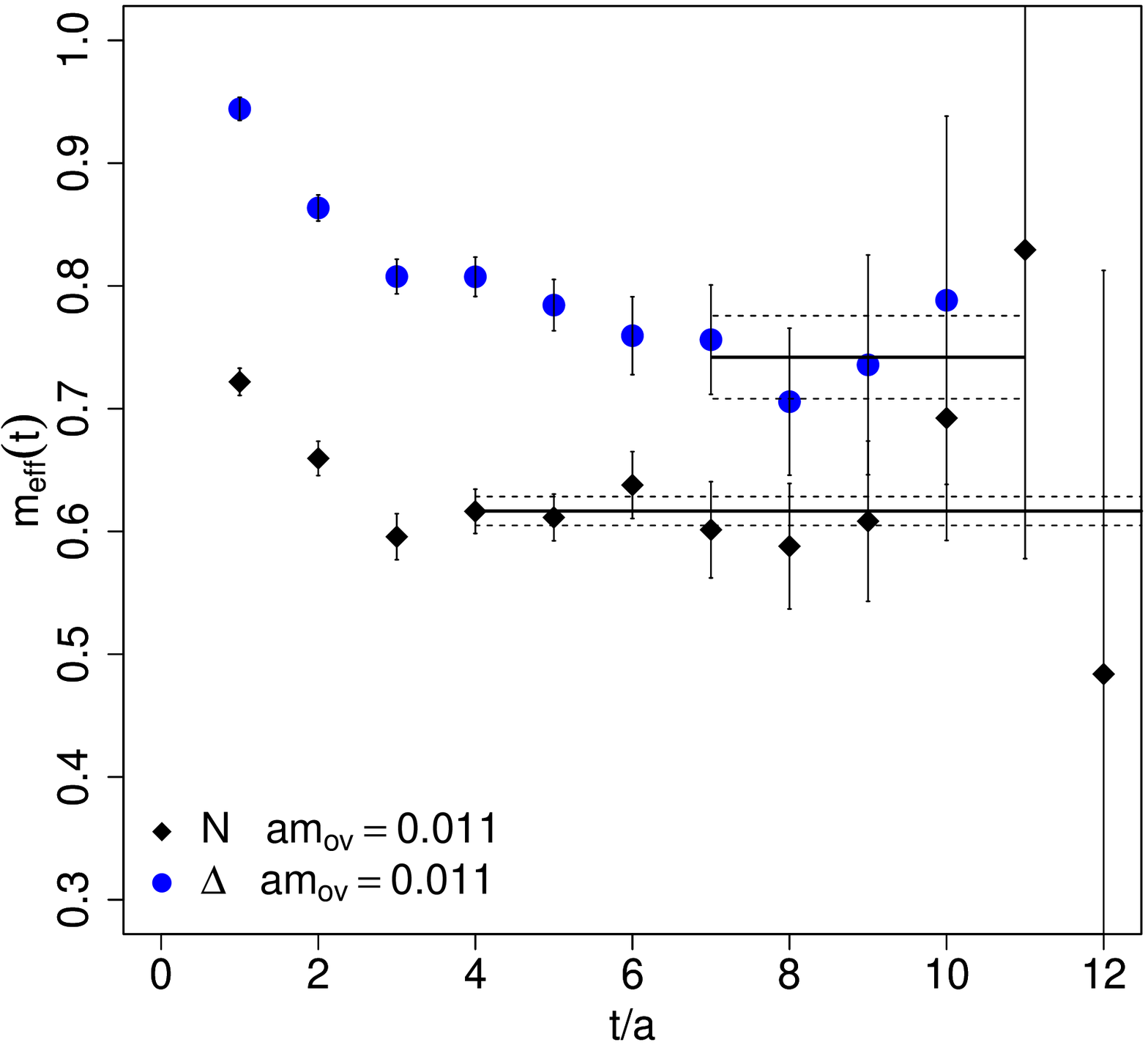}
\includegraphics[width=0.495\textwidth]{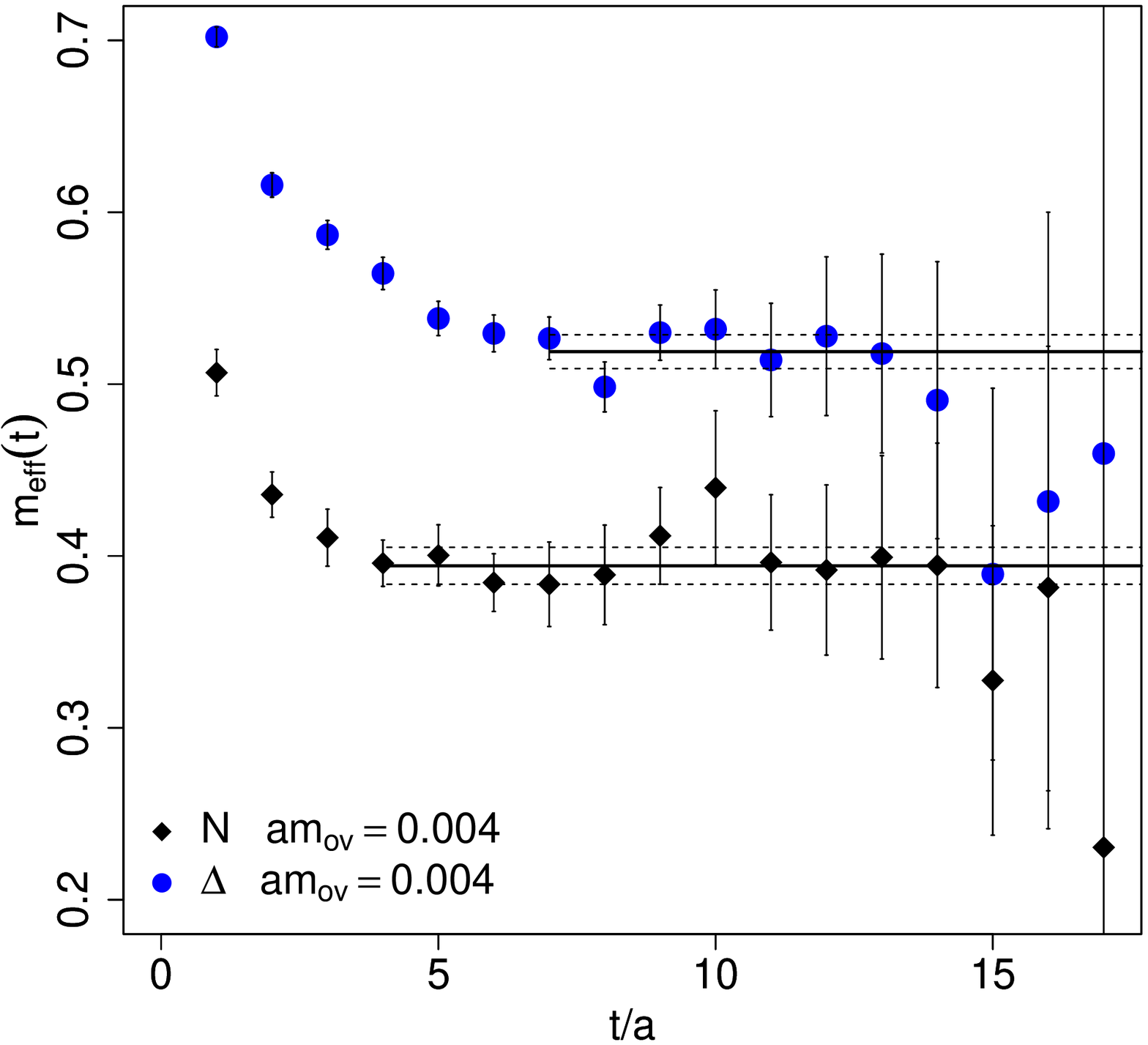}
\caption{\label{meff_ND} Nucleon and  $\Delta$ effective masses in the mixed action setup. The
valence quark mass is fixed to the (improved) matching overlap quark 
mass, see Tab.~\ref{tab:matching} for their values. In the left plot we show the results obtained
for the ensemble 
$B_\ell$ ($\beta=3.9$) and in the right for $D_\ell$ ($\beta=4.2$).}
\end{figure}

We use the effective masses defined by: 
\begin{equation}
m^{N,\Delta}_{\rm eff} (t) = - \log{ \frac{C^{N,\Delta} (t)}{C^{N,\Delta} (t-1)}}  = a m_{N,\Delta} + \mathcal{O}(e^{- \delta t})
\end{equation}
where $\delta$ is the mass difference  between the ground state and the first excited state.
In Fig.~\ref{meff_ND}, we show examples of effective masses of the nucleon and $\Delta$ in the
mixed action setup. The results are obtained at the 
coarsest and the finest lattice spacing used in this work, corresponding to the gauge ensembles 
$B_\ell$ (left) and $D_\ell$ (right). The valence quark masses
are fixed to the (improved) matching overlap quark masses summarized in Tab.~\ref{tab:matching}
We fit the effective masses to a constant in the Euclidean time region where
$m^{N,\Delta}_{\rm eff} (t)$ become time independent. The corresponding fits are shown in
Fig.~\ref{meff_ND}
where the fit result is represented by
horizontal black lines and their statistical errors by black dotted lines.
The effective mass plots in Fig.~\ref{meff_ND} show that we can identify 
a plateau region, allowing us a good determination of the nucleon 
and the $\Delta$ masses.

\begin{figure}[t]
\includegraphics[width=0.495\textwidth]{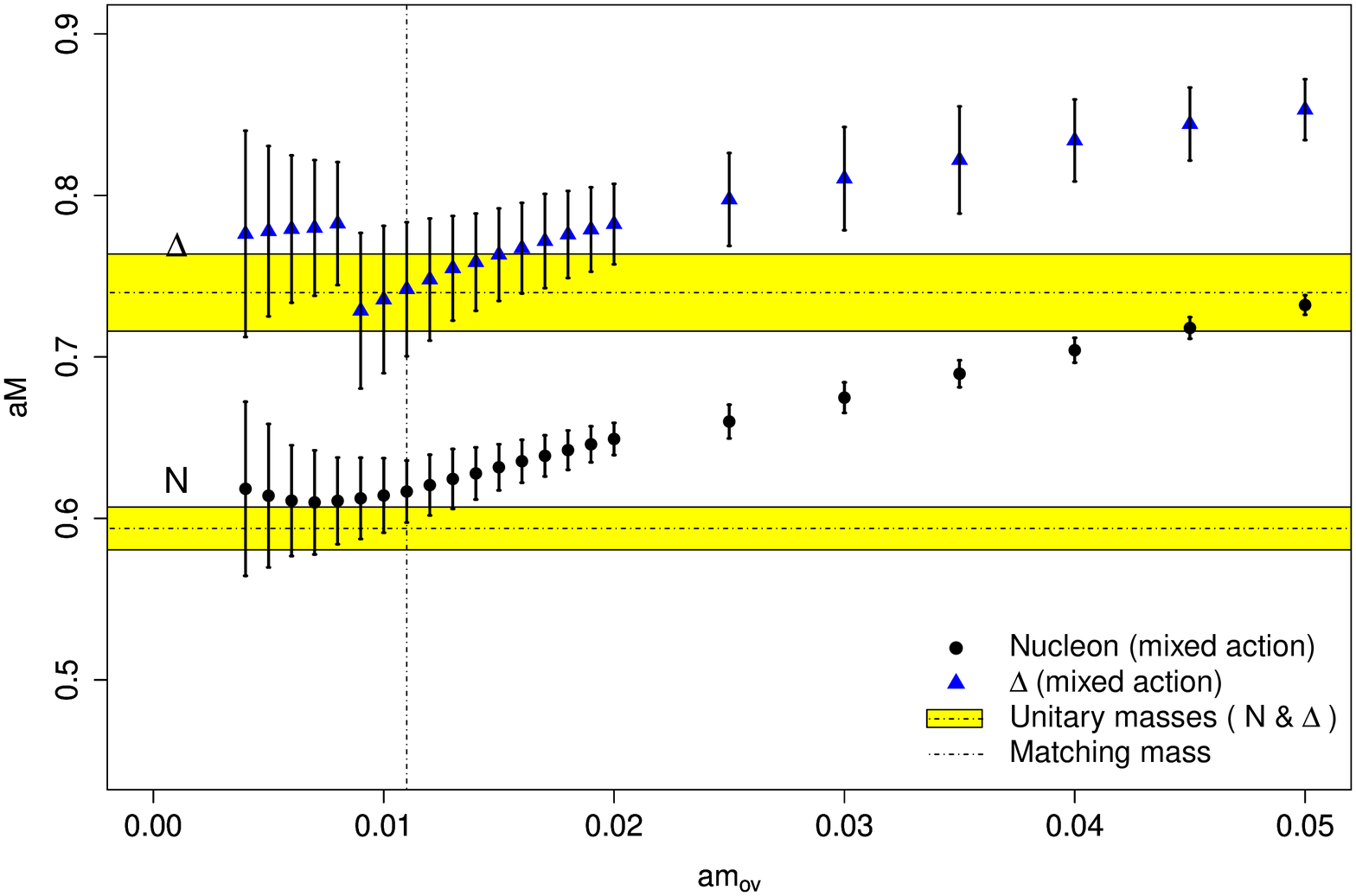}
\includegraphics[width=0.495\textwidth]{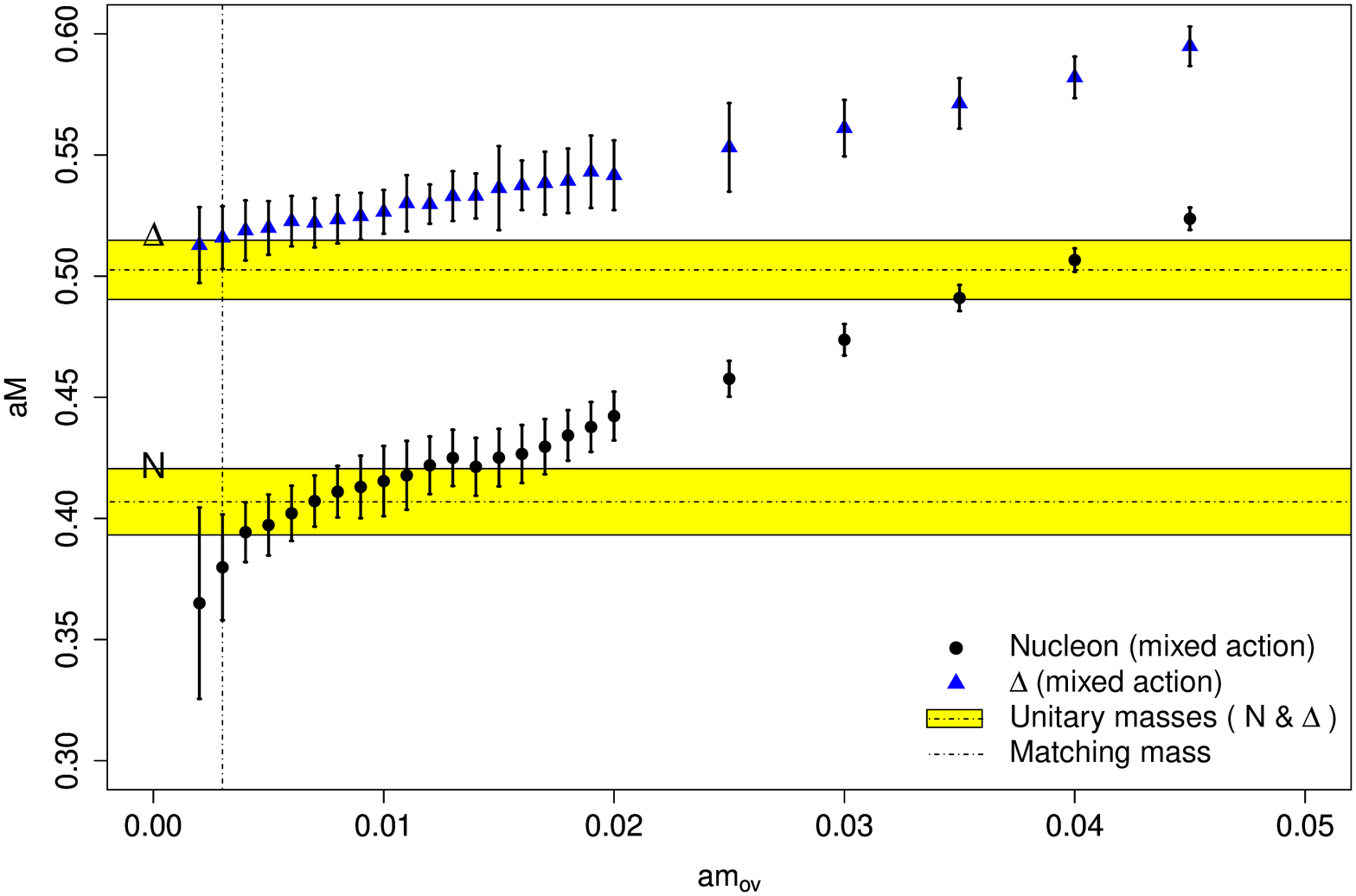}
\caption{\label{mov_dep_ND} Nucleon and Delta masses vs. $am_{\rm ov}$. The vertical dotted line
indicates the matching mass. The left plot shows results for ensemble $B_\ell$ ($\beta=3.9$) and the
right one for $D_\ell$ ($\beta=4.2$).}
\end{figure}

Results for the nucleon and $\Delta$ masses 
in the unitary setup and in the mixed action setup (for $a\mov$ set to the matching mass)
are reported in Tab.~\ref{tab:res_ND_tm}. We also show our results obtained for several valence
quark masses in Fig.~\ref{mov_dep_ND} for ensemble $B_\ell$ (left) and for $D_\ell$ (right) -- each
plot shows the overlap quark mass dependence of the nucleon mass (black
dots) and the $\Delta$ mass (blue triangles). The matching mass is indicated by a vertical dotted
line, while the results obtained in the unitary setup are indicated by a yellow band with a width
corresponding to the statistical errors. Note that for very low valence quark masses, 
below the matching quark mass, a plateau
region in the effective mass can hardly be found. This explains why statistical errors become
larger for very small overlap quark masses.
The uncertainty to extract reliably effective masses in this region is also 
responsible for the somewhat irregular behaviour in Fig.~\ref{mov_dep_ND}. 
As can be seen from the very good agreement of the baryon masses between 
the unitary and the mixed action setups, 
even in a regime where effects of chiral zero modes of the overlap operator are
sizable in the pion sector, in the nucleon or $\Delta$ mass case, the effects are negligible for
our definition of the (improved) matching mass.

\begin{table}[t!]
  \begin{center}
    \begin{tabular}{ccccccc}
      \hline
      Ensemble & $N^{\rm TM}$ & $N^{\rm ov}$ & $am_N^{\rm TM}$ & $am_N^{\rm ov}$ &
$am_\Delta^{\rm TM}$ & $am_\Delta^{\rm ov}$\\
      \hline
      \hline
      $B_\ell$ & 636 & 426 & $0.594(13)$ & $0.617(12)$ & $0.740(24)$ & $0.741(34)$ \\
      $C_l$ & 286 & 142 & $0.497(11)$ & $0.522(27)$ & $0.627(13)$ & $0.641(15)$ \\
      $D_\ell$ & 189  & 371 & $0.407(14)$ & $0.394(11)$ & $0.503(13)$ & $0.519(10)$ \\
      \hline
      \hline
    \end{tabular}
  \end{center}
  \caption{Summary of our results for for the nucleon masses $am_N$ and $\Delta$ masses
$am_\Delta$. The superscript ``TM'' denotes the unitary setup, while ``ov'' denotes the mixed
action setup. We also give the number of gauge field configurations $N$ used for 
the measurements on each ensemble.}
  \label{tab:res_ND_tm}
\end{table}

\begin{figure}[h!]
\begin{center}
\includegraphics[width=0.7\textwidth]{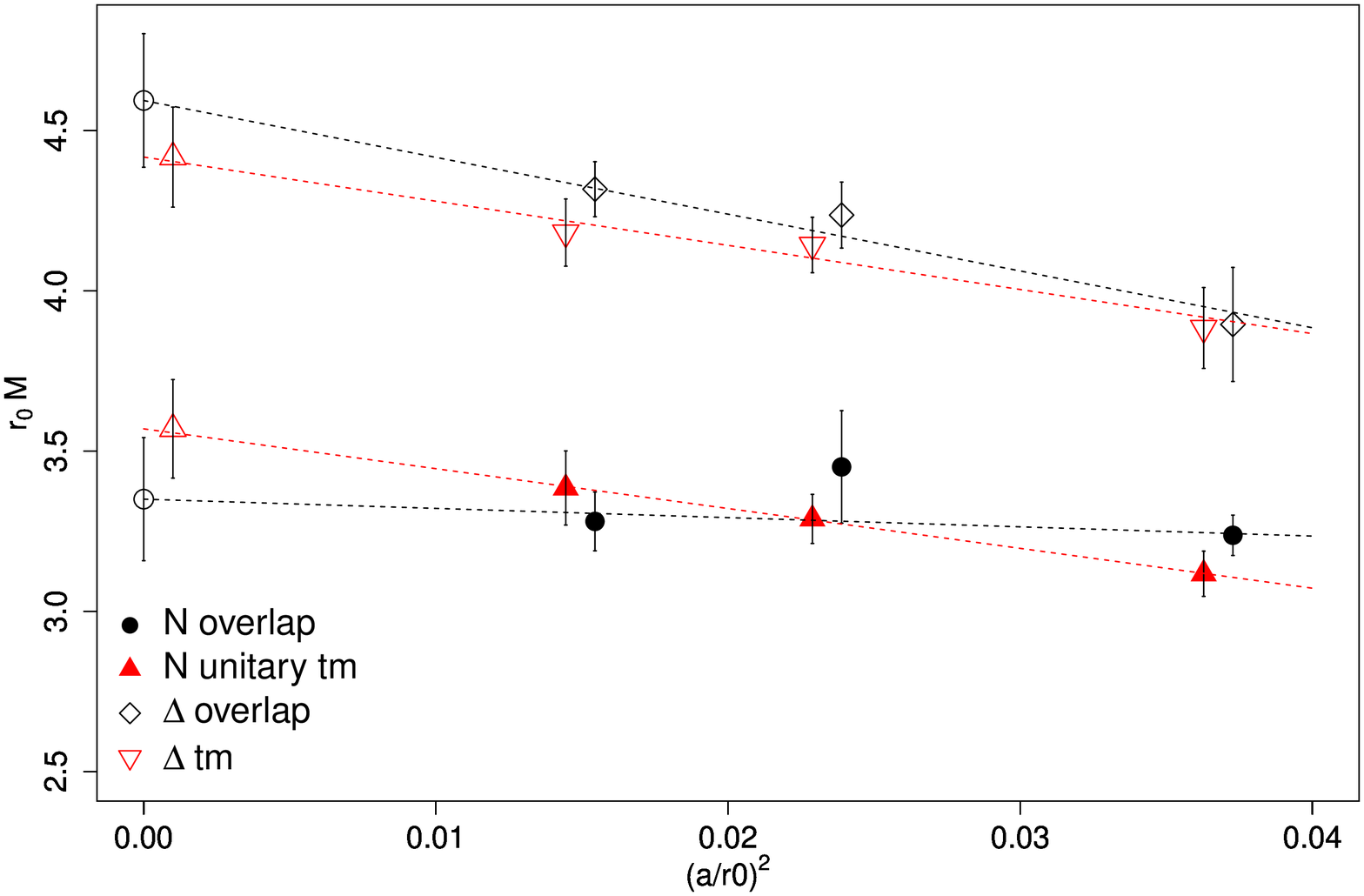}
\caption{The continuum limit scaling of the nucleon and $\Delta$ masses.}
\label{scaling_ND}
\end{center}
\end{figure}

Finally, we show in Fig.~\ref{scaling_ND} the continuum limit 
scaling plot of the two baryon masses
with three lattice spacings, in the unitary and mixed action case. All
three ensembles have pion masses of $\approx 300\mev$ and the same volume of $\approx1.3\fm$, so no
interpolation in volume or quark masses is needed within our statistical accuracy to compare our
results at different lattice spacings. 
For all ensembles, the results in the mixed action case and in the unitary case are compatible
within errors in the continuum limit. Furthermore, in the two setups, the results for the nucleon
and for the $\Delta$ mass show only a small dependence on the lattice spacings. We perform a linear
fit in $a^2$ represented by dotted lines (red in the unitary case and
black in the mixed action one). The continuum extrapolated values  are represented slightly shifted
for better readability.
 
\section{Summary and prospects}

In this paper, we have studied a particular example of a mixed action 
setup, namely valence overlap fermions on maximally twisted mass sea fermions. 
By adding a fourth and finer value of the lattice spacing 
than available to us earlier \cite{arXiv:1012.4412}, we were able to perform 
a continuum limit scaling test for the pion decay constant and to compare 
the unitary and mixed action theories. We found a continuum limit
value of the pion decay decay constant that is in full agreement 
between the two theories and confirmed thus 
the conclusion of Ref.~\cite{arXiv:1012.4412}
that with an improved matching condition the dangerous effects arising from the mismatch of the chiral
zero modes -- between the chiral invariant overlap Dirac operator in the valence sector and the
twisted mass operator in the sea -- can be avoided

Applying the same strategy for other observables for which we took 
the nucleon and $\Delta$ masses, we found that also these quantities are 
not affected by the presence of the zero modes, once the improved
matching condition is applied. This supports the interpretation reached 
in Ref.~\cite{arXiv:1012.4412} that the improved matching condition leads to a 
proper matching of the sea and valence theories.  
In general, in the context of mixed actions, we nevertheless advocate to check the possible effects
on physical observables of the mismatch of zero modes between valence and sea sectors, introduced
by the use of a valence overlap Dirac operator.

We also performed a continuum limit investigation of the locality 
properties of the overlap Dirac operator. While at any non-zero value of the 
lattice spacing the overlap Dirac operator is exponentially localized, 
our study, employing four values of the lattice spacing, strongly suggests
that a point-like localization is indeed recovered in the continuum limit. 
One essential element in our investigation was to compare the exponential decay rate 
of the norm of the overlap Dirac operator with 
hadronic scales, for which we took the pion and nucleon masses. 
We could demonstrate that for all values of the lattice spacing, the ratio of 
the hadronic masses over the decay rate
is smaller than one and assumes 
a value compatible with zero in the continuum limit. Thus, the exponential 
locality of the overlap Dirac operator is not expected to distort the evaluation
of these hadronic observables. We think that such kind of tests should be performed 
on hadronic physical quantities one is interested in, in order to monitor the possible distortions of
the results by a finite decay rate of the overlap Dirac operator. 

Finally, we compared our numerical computations of meson masses with predictions
of mixed action $\chi$PT and extracted in this way some of the LECs which 
parametrize the mixed action chiral effective Lagrangian. We also provided evidence 
that there are indeed unitarity violations in the mixed action setup and we quantified these
effects by computing the $\Delta_{Mix}$ and $W_8'$ parameters.

In summary, we believe that the results of Ref.~\cite{arXiv:1012.4412}, in combination with 
the investigations performed in this work, can serve as a basis for future 
mixed action computations employing chiral invariant fermions in the valence 
sector: 
we advocate to use the improved matching condition proposed in Ref.~\cite{arXiv:1012.4412}; 
we suggest to test whether the localization rate of the overlap Dirac operator is sufficiently
larger than the hadronic scale of interest; we further point to the possibility to use mixed action 
chiral perturbation theory to look at effects of a mixed action setup, allowing in particular
to isolate the effects of unitarity violations.

\vspace{0.3cm}
\noindent {\bf Acknowledgments} We thank the European
Twisted Mass Collaboration for generating ensembles of gauge field
configurations that we have used for this work and for a very enjoyable collaboration. 
We are grateful to O.~B\"ar for useful comments and suggestions.
Sec.~\ref{section-locality} of this work originated from a DESY summer student project carried out
by O.~Haas, E.~Panzer, S.~Tordjman.
The computer time for this project was made available to us by the Leibniz Rechenzentrum in Munich,
Pozna\'n Supercomputing and Networking Center in Pozna\'n, GENCI-CINES grant
2010-052271 and CCIN2P3 in Lyon. We thank these 
computer centers and their staff for all technical advice and help.
K.C. has been supported by Foundation for Polish Science fellowship ``Kolumb'' and Ministry of
Science and Higher Education grant nr. N N202 237437.
This work has been supported in part by the DFG Sonderforschungsbereich/Transregio SFB/TR9. 
G.H. acknowledges the support from the Spanish Ministry for
Education and Science project FPA2009-09017, the Consolider-Ingenio 2010
Programme CPAN (CSD2007-00042), the Comunidad Aut\'onoma de Madrid
(HEPHACOS P-ESP-00346 and HEPHACOS S2009/ESP-1473) and the European
project STRONGnet (PITN-GA-2009-238353).

\appendix

\section{Results for ensemble $B_s$}
\label{sec:app}

One of the main conclusions of Ref.~\cite{arXiv:1012.4412} were the values of the relevant
parameters -- physical volume and pion mass -- for which the role of the zero modes is strongly
suppressed. With such parameters, the simulation results are safe against the contribution of zero
modes, i.e. a continuum limit scaling test of e.g. the pion decay constant, computed with MTM and
overlap fermions would lead to a consistent behaviour (same continuum limits), even when both
$\fpstm$ and $\fpsov$ are computed from the PP correlator.

For this work, we have decided to test this expectation explicitly, employing an ensemble with
large enough volume and pion mass. We have not performed a full continuum limit scaling test, since
this is still very expensive computationally. Instead, we compare the values of $\fpstm$ and
$\fpsov$ at the matching mass, expecting only a small difference that can be attributed to $\Oasq$
cut-off effects. Our ensemble, labeled $B_s$ has the following parameters: $\beta=3.9$, $L/a=24$,
$a\mu=0.0085$, which corresponds to $L\approx2$ fm, $\mps\approx480$ MeV in infinite volume and
$\mps L \approx 4.7$. The latter is well above $\mps L = 4$, the minimal product of the pion mass
and spatial extent of the lattice that leads to negligible effects from the zero
modes \cite{arXiv:1012.4412}.

\begin{figure}[t!]
\begin{center}
\includegraphics[width=0.34\textwidth,angle=270]
{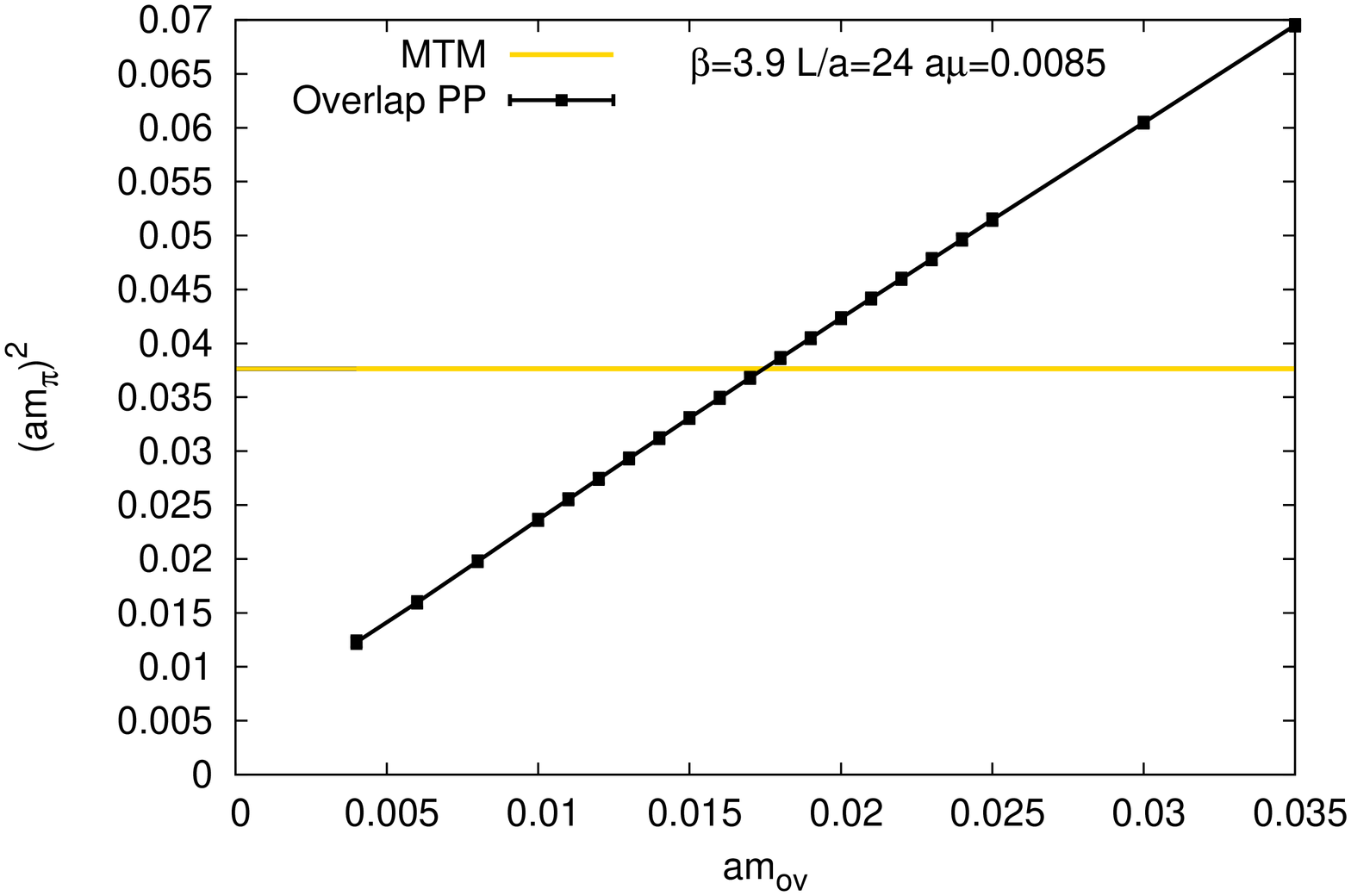}
\includegraphics[width=0.34\textwidth,angle=270]
{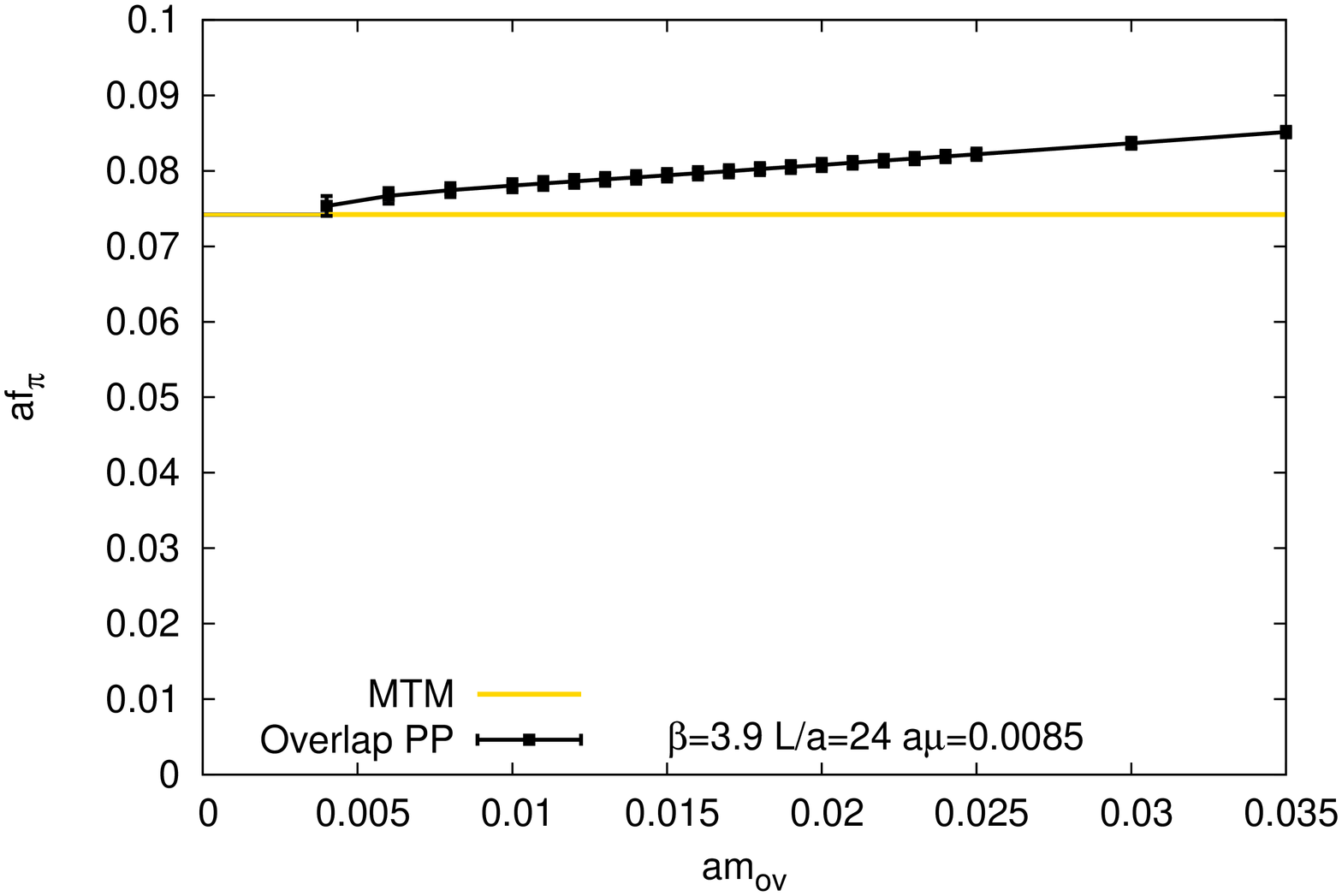}
\end{center}
\caption{(left) Matching the overlap and MTM pion mass. 
(right) The dependence of the overlap pion decay constant on the overlap quark mass $am_{ov}$. The
matching mass is around $a\mov=0.017$. Ensemble $B_s$ ($\beta=3.9$, $L/a=24$, $a\mu=0.0085$).}
\label{fig:safe_fps}
\end{figure}

In Fig.~\ref{fig:safe_fps}, we show the matching of the pion mass between the unitary setup and the
mixed action setup. We find equal pion masses ($\mpstm=\mpsov$) at around $a\mov=0.017$. At this
matching quark mass, we then compare the value of the pion decay constant. $\fpsov$ is greater than
$\fpstm$ by 7.7\% ($6.8\,\sigma$ away). This difference can be compared to the differences
between $\fpstm$ and $\fpsov$ for other ensembles at the same value of the lattice spacing: 40.0\%
($L\approx1.3$ fm, $a\mu=0.004$, $\mps L\approx2.5$), 27.0\% ($L\approx1.6$ fm, $a\mu=0.004$, $\mps
L\approx2.8$), 19.5\% ($L\approx2$ fm, $a\mu=0.004$, $\mps L\approx3.3$), 16.5\% ($L\approx1.3$ fm,
$a\mu=0.0074$, $\mps L\approx3.1$). Indeed, for the current ensemble of interest, the difference of
around 7.7\% is much smaller than for other ensembles at $\beta=3.9$, with either a smaller
physical volume, or a smaller pion mass (or both). While certainly this can not be treated as a
proof that the role of the zero modes is negligible for this ensemble, we believe that this
provides a clear hint that our interpretation of the role of the zero modes in our overlap/MTM
mixed action setup is correct.

\end{document}